\documentclass{article}

\usepackage{amsmath}
\usepackage{amscd}
\usepackage{amsthm}
\usepackage{amssymb} \usepackage{latexsym}
\usepackage{eufrak}
\usepackage{euscript}
\usepackage{epsfig}
\usepackage{tikz}
\usepackage{graphics}
\usepackage{array}
\usepackage{enumerate}
\usepackage{authblk}
\usepackage[colorlinks=true]{hyperref}
\usepackage{authblk}

\numberwithin{equation}{subsection}

\date{}

\begin{document}
\title{Communication on structure of biological networks}

\author[1]{\rm Krishanu Deyasi}
\author[1]{\rm Shashankaditya Upadhyay}
\author[1,2]{\rm Anirban Banerjee}
\affil[1]{Department of Mathematics and Statistics}
\affil[2]{Department of Biological Sciences}
\affil[ ]{Indian Institute of Science Education and Research Kolkata}
\affil[ ]{Mohanpur-741246, India}
\affil[ ]{\textit {krishanu1102@iiserkol.ac.in, shashankaditya@yahoo.co.in, anirban.banerjee@iiserkol.ac.in}}

\maketitle
\bigskip

\begin{abstract}
Networks are widely used to represent interaction pattern among the components in complex systems. 
Structures of real networks from different  domains may vary quite significantly. Since there is an interplay between network architecture and dynamics, 
structure plays an important role in communication and information spreading on a network. 
Here we investigate the underlying undirected topology of  different biological networks which support faster spreading of information and are better in communication. 
We  analyze the good expansion property by using the spectral gap and communicability between nodes. 
Different epidemic models are also used to study the transmission of information in terms of disease spreading through individuals (nodes) in those networks.
Moreover, we  explore the structural conformation and properties which may be responsible for better communication.
 Among all biological networks studied here, the undirected structure of neuronal networks not only possesses the small-world  property but  
 the same is expressed remarkably to a higher degree than any randomly generated network which possesses the same degree sequence. 
 A relatively high percentage of nodes, in neuronal networks, form a higher core in their structure.
 Our study shows that the underlying undirected topology in neuronal networks is significantly qualitatively different than 
 the same from other biological networks and that they may have evolved in such a way that they inherit a (undirected) structure which is excellent and robust in communication. 
\end{abstract}

\section{Introduction}
Over the past few years, network science has drawn attention from a
large number of researchers from diverse fields. Networks in which the underlying topology is a graph, are generic representations of the interactions among components of a complex
system.  
Biological networks provide an insight to analyze and understand various processes
that occur in several biological systems. These systems range
from intracellular protein interactions to inter-species interactions 
(see \cite{newman03} for details). Most networks have the function to transport or 
transfer entities like information, mass, energy etc. along their edges.
Structure of a network plays a crucial role in spreading of the above entities.
 In the last few years  different heuristic parameters (clustering coefficient, 
transitivity, average path length, betweenness, centrality etc. see \cite{newmanbook}
 for details) have been introduced for analyzing the network structure, and
various models (e.g. Erd\"{o}s-R\'{e}nyi's model random network model,
  Barab\'{a}si and Albert's scale free model,  Watts and Stogatz's small-world 
network model, duplication-divergence model etc. \cite{erdos59,barabasi99,sw98,IspolatovEtAl05}) have been proposed to represent the architecture of real networks. 
The qualitative properties of biological networks cannot be well captured by the heuristic parameters.
 However, spectral analysis is also used for elucidating the global property of a network (see \cite{farkas01,farkas02}). Features like
 ``good expansion'' and ``communicability'' are well quantified by
spectral analysis (see \cite{bollobas78,bollobas84}). 
\emph{Good expansion}
network can be thought of as a network in which a small subset of vertices has comparatively  large number of neighbours (see \cite{chung97,bollobas78,bollobas84})
and \emph{communicability} can be understood as networks in which 
information is ``capable of being easily communicated or transmitted in terms of 
passage or means of passage between the different nodes in a network"
(see \cite{estrada08}).

 Here, we study the underlying undirected structure of empirical biological networks from five different
 classes (neuronal, food web, protein protein interaction, metabolism and gene regulation) and explore the undirected topology that supports better communication and information spreading.
We observe that one class of biological networks has higher
\emph{good expansion} property and excellent \emph{communicability} than  the others, though most of the biological 
networks (see \cite{sw98}) have high
clustering coefficients (or transitivity) and low average shortest path lengths which
give them the liberty to reach from one node to another with a fewer number of 
steps. This property of a network is quite well known as the small-world property
(see \cite{sw98}). Here, we have also investigated the topological properties 
 that make a particular class of biological networks possess the excellent expansion property.


\section{Methods}

\subsection*{Spectral gap and Good expansion network:}

  Generally, sparsely populated and highly connected network topologies are contradictory 
  properties and hard to find in real-world networks. However, good expansion networks are known to possess
  those properties. There are extensive applications of good expansion networks in designing algorithms, 
  error correcting codes, extractors, pseudo-random generators (\cite{reingold02}) etc.
  Good expansion networks are also important as they show excellent communication 
  properties (\cite{estrada06}). The excellent spreading property or the good expansion property
  can be captured by the \textit{spectral gap} in a network (see \cite{tanner84, alon86}).
  
     A network can be represented as a simple graph $G=(V,E)$, where $|V|$ (=n) is the number
  of vertices or nodes and $|E|$ is the number of connections or edges between nodes.
  An unweighted and undirected network has the good expansion property, if any set $S \subset V$ with $|S| \leq |V|/2$
  satisfies 
                     $$|S^{'}| \geq c|S|,$$
  where $S^{'} \subset V\setminus S$ is the set of neighbours of $S$ and $c$ is a parameter
  called the expander constant (see \cite{estrada06,sarnak04}). The adjacency matrix $A$ = ($A_{ij}$) corresponding to a graph $G$
  is an $n\times n$ matrix with entries in \{0, 1\} such that $A_{ij}$ = 1, if there exists an edge in $G$ between the vertices $i$ and $j$, and 0 otherwise.
 The set of eigenvalues $\lambda_1\geq\lambda_2\geq\dots\geq\lambda_n$
 of $A$ is called the spectrum of the network. The larger the spectral gap $|\lambda_1|-|\lambda_2|$ is, the faster the random walk (on the graph) will converge to its steady-state.
 Observe that the largest eigenvalue is always positive. Thus, a network shows good information spreading 
 character if the largest eigenvalue of $A$ is much higher than the absolute value of the second largest eigenvalue (see \cite{tanner84,alon86}), i.e. if 
$$|\lambda_1|\gg|\lambda_2|.$$


\subsection*{Communicability:}
\hspace*{0.5 cm} Spreading of information, mass or entities on a network is a
common process and eventuates in most networks. The nature of information or 
entities varies depending on the type of network. In neuronal networks, information
spreading means electrical signal propagation. In food webs, it is regarded as mass flow from prey
to predator. In signal transduction networks, it is the signal which spreads, and so on.
Earlier, from a structural perspective, it was considered that the
communication (information, mass, entities spreading) between two nodes in a network can happen only through the 
shortest routes connecting them because it is the most economical way of communication.
But, communication between two nodes in a real network may not always only happen via the shortest routes.

Communicability can be thought of as transforming information easily 
between different nodes in a network. Communicability in a complex network is a broad generalization of 
the concept of the shortest path. To study the communicability, the above situations should be taken under consideration. Hence, communicability
can be thought as how effectively information can be propagated between a pair of nodes in a network.
We consider the communicability, introduced in \cite{estrada08}, to study which undirected (biological) network structures are more favourable for excellent communication.

The $(i,j)$-entry of the $k^{th}$ power of the adjacency matrix $A^{k}$ shows the 
number of walks of length $k$ between the vertices $i$ and $j$. The information in a network can 
flow back and forth several times before reaching the final destination, like particle
transversal through the graph. The communicability 
between any two nodes $p, q$ of a graph $G$ is defined by

\begin{equation} \label{equ1}
 G_{pq} = \sum_{j=1}^{n} \phi_j(p)\phi_j(q) e^{\lambda_j}, 
\end{equation}
         
 where $\phi_j(p)$ is the $p^{th}$ element of the $j^{th}$ orthonormal 
eigenvector corresponding to the eigenvalue
$\lambda_j$. A large $G_{pq}$  implies that the communicability between the nodes $p$ and $q$ is high (see \cite{estrada08} for details).

\subsection*{Epidemic spreading in network:}

Different epidemic models can be used to describe transmission of information in terms of disease spreading through individuals (nodes) in a
network.
 Here, we study the nature of information spreading in the empirical networks by using three epidemic models, 
 the SI (\emph{susceptible-infected}) model, the SIR (\emph{susceptible-infected-recovered}) model, and 
the SIS (\emph{susceptible-infected-susceptible}) model (see ~\cite{newmanbook} for details on these
models). We use these models for characterising the underlying undirected structures of biological networks which are more favourable in spreading information or disease.\\

{\bf The SI model:}
    The simplest mathematical model among all epidemic models is the SI model consisting 
    of two states, the \emph{susceptible} and the \emph{infected} individual. An individual who does not 
    have the disease yet, but can catch the disease from \emph{infected} individuals if in contact with them, is 
    treated as \emph{susceptible}. \emph{Infected} individuals are those who currently have
    the disease and can infect susceptible individuals (see ~\cite{newmanbook}).  
    
    Suppose that a disease is spreading in a population of $n$ individuals. Let $s(t)$  and $x(t)$ denote the fraction of susceptible
    and infected individuals respectively at time t. If one infected individual can 
    infect $\beta$ number of susceptible individuals per unit time, then the differential equations for the rate of change 
    of x and s become    
    
    \begin{equation}
    \left.\begin{aligned}
                      \frac{dx}{dt} &= \beta{sx},\\
                      \frac{ds}{dt} &= -\beta{sx}.
    \end{aligned}
    \right\}
    \end{equation}                  
We randomly
    choose a node as infected and an infected node can infect its neighbours with infection 
    probability 1.\\

{\bf The SIR model:}
   The SIR model unlike the SI model, consists of three states, namely, \emph{susceptible}, \emph{infected} and \emph{recovered}.
   Susceptible individuals are infected by the infected ones and the infected individuals are immunised. Immunised individuals are entered into the recovered state.
   Initially every individual is in the susceptible state except a small
number of individuals. At each time step, one individual can infect their neighbour. Infected individuals are 
entered into the recovered state by immunisation.   
   
   If $s(t)$, $x(t)$, and $r(t)$ denotes the fraction of susceptible, infected and recovered individuals respectively at time $t$, then the equations
   for the SIR model are
   \begin{equation}
   \left.\begin{aligned}
                         \frac{ds}{dt} &= -\beta{sx},\\
                         \frac{dx}{dt} &=\beta{sx}-\gamma{x},\\
                         \frac{dr}{dt} &=\beta{x},
   \end{aligned}
   \right\}
   \end{equation}
   
  where $s + x + r=1$. \\

{\bf The SIS model:} 
  Here, the individuals can have two states \emph{susceptible} and \emph{infected}, like in the SI model.
   The only difference is that \emph{infected} individuals after recovery, can become \emph{susceptible} again. 
   
   The governing equations for this model are
   \begin{equation}
    \left.\begin{aligned}
                       \frac{ds}{dt} &= \gamma{x}-\beta{sx},\\
                       \frac{dx}{dt} &= \beta{sx}-\gamma{x},
          \end{aligned}
   \right\}
   \end{equation}

                       with the condition $s+x=1$.

\section{Network construction and Data resources:}

{\bf Neuronal network:} The data for 
macaque visual cortex, macaque visual and sensorimotor area, macaque cortical connectivity, cat cortex (complete), and cat 
cortex connectivity that was used by Rubinov and Sporns in \cite{rubinov10} was downloaded from \url{https://sites.google.com/site/bctnet/Home}.
To construct a network from these data, we consider the cortical areas as nodes and large corticocortical
tracts as edges of the network. Neuronal connectivity data of {\it C. elegans} which was used by Watts and Strogatz in 
(\cite{sw98}) and by White et al. in (\cite{white86}) was downloaded from \url{http://www-personal.umich.edu/~mejn/netdata/}.
The nodes and edges of the network represent the neurons and the synaptic connections respectively.

{\bf Food web:} Here, different species in the ecosystem are considered as nodes and the prey-predator relationships are considered as the edges of the network.
The data was downloaded from \url{http://www.cosinproject.org/}. 

{\bf Protein-protein interaction network:} Here the nodes are proteins and we connect two proteins by an edge if they physically bind together. 
The {\it E. coli} data which was also used by Butland in (\cite{butland05}), was downloaded from \url{http://www.cosinproject.org}.  

{\bf Metabolic network:} Here metabolites are represented by nodes and an educt-product relation is represented by an edge.
The data was downloaded from \url{http://www3.nd.edu/~networks/resources.htm} (used in \cite{jeongEtAl2000}).

{\bf Gene regulatory network:} In this network, nodes are genes and if one gene regulates another we connect them by an edge.
The data of {\it E. coli} and
{\it S. cerevisiae}  were downloaded from \url{http://www.weizmann.ac.il/mcb/UriAlon/} (used in \cite{milo02}).

\section{Results and Discussion}

Here, we study the underlying undirected structure of five different classes of biological networks: neuronal 
networks, food webs, protein-protein interaction networks, metabolic networks, gene regulatory networks. 
To investigate which structure is better for communication or spreading of mass, information or entities, 
we explore the good expansion property (by using the spectral gap) of a network and study the communicability between every pair of nodes. 
Different epidemic spreading models are also used to investigate the same on these networks.\\

We observe that the underlying undirected topology of all neuronal networks
and a few food webs show the good expansion property (see Table~\ref{spectralgap}) unlike other biological networks.

\begin{figure}[!]
\begin{center}
\includegraphics[scale=.33]{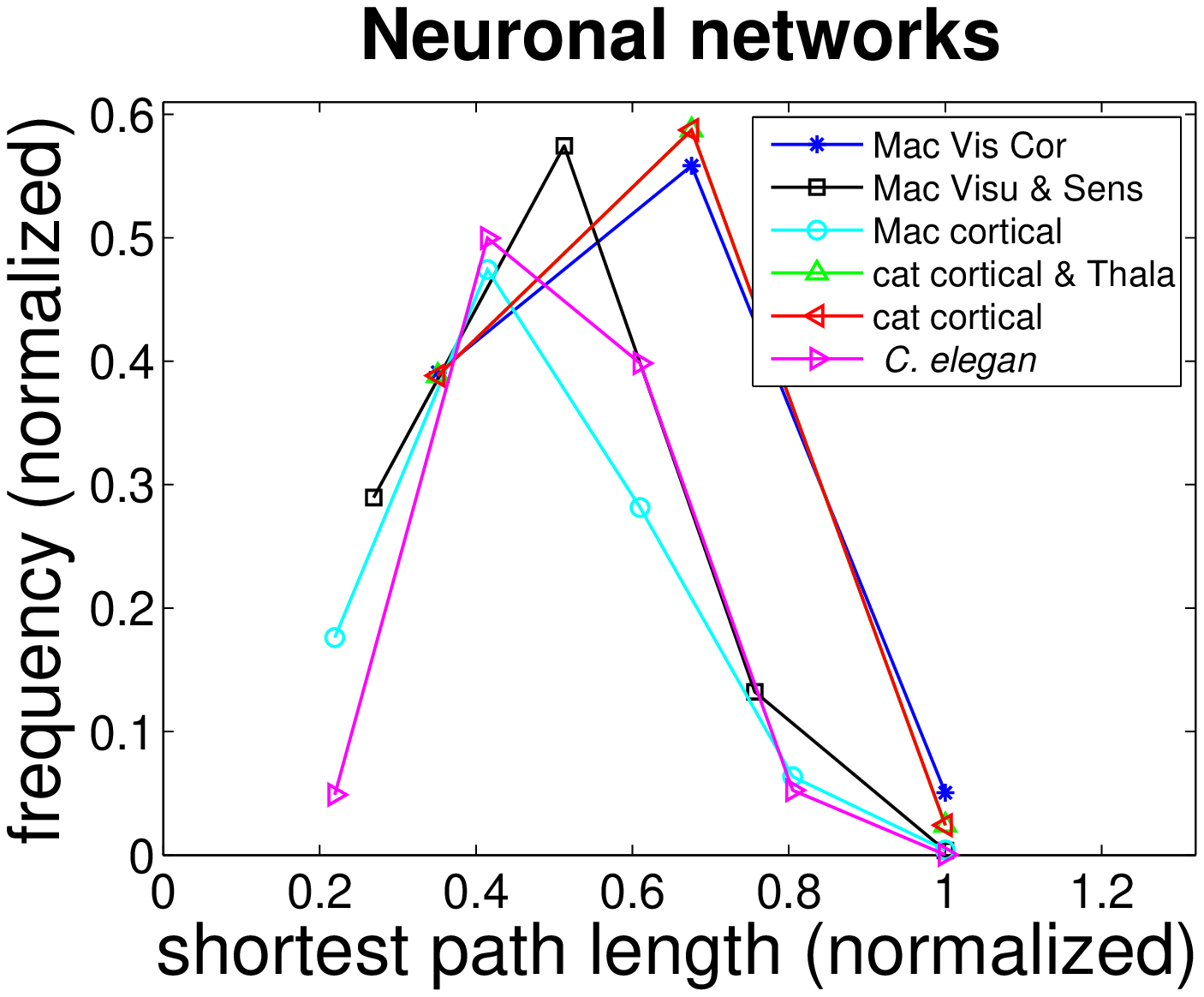}\includegraphics[scale=.33]{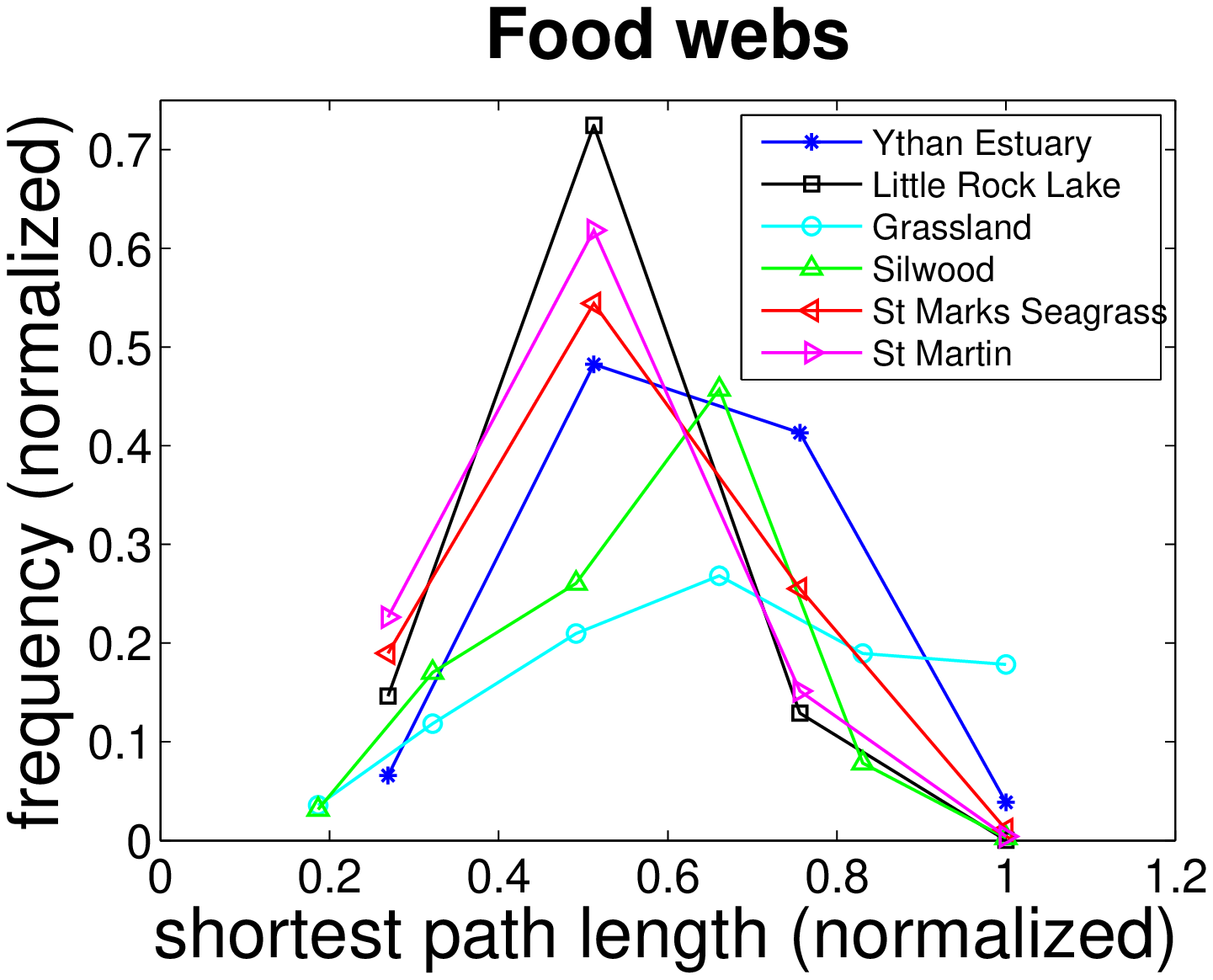}\includegraphics[scale=.33]{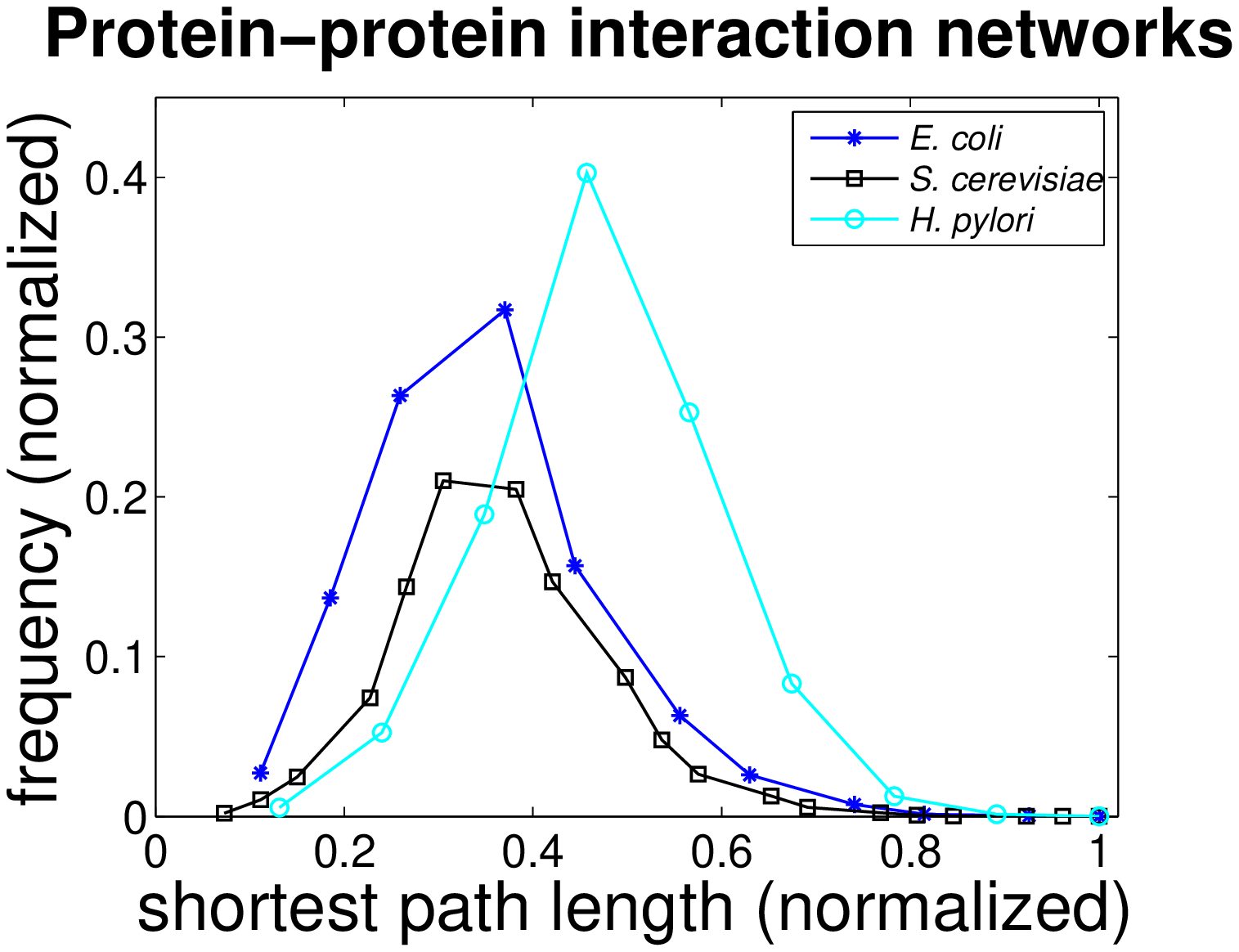}\\
\vspace{-3.45cm}
\hspace{.8cm}(a)\hspace{4.8cm}(b)\hspace{4.6cm}(c)\\
\vspace{3.2cm}
\includegraphics[scale=.33]{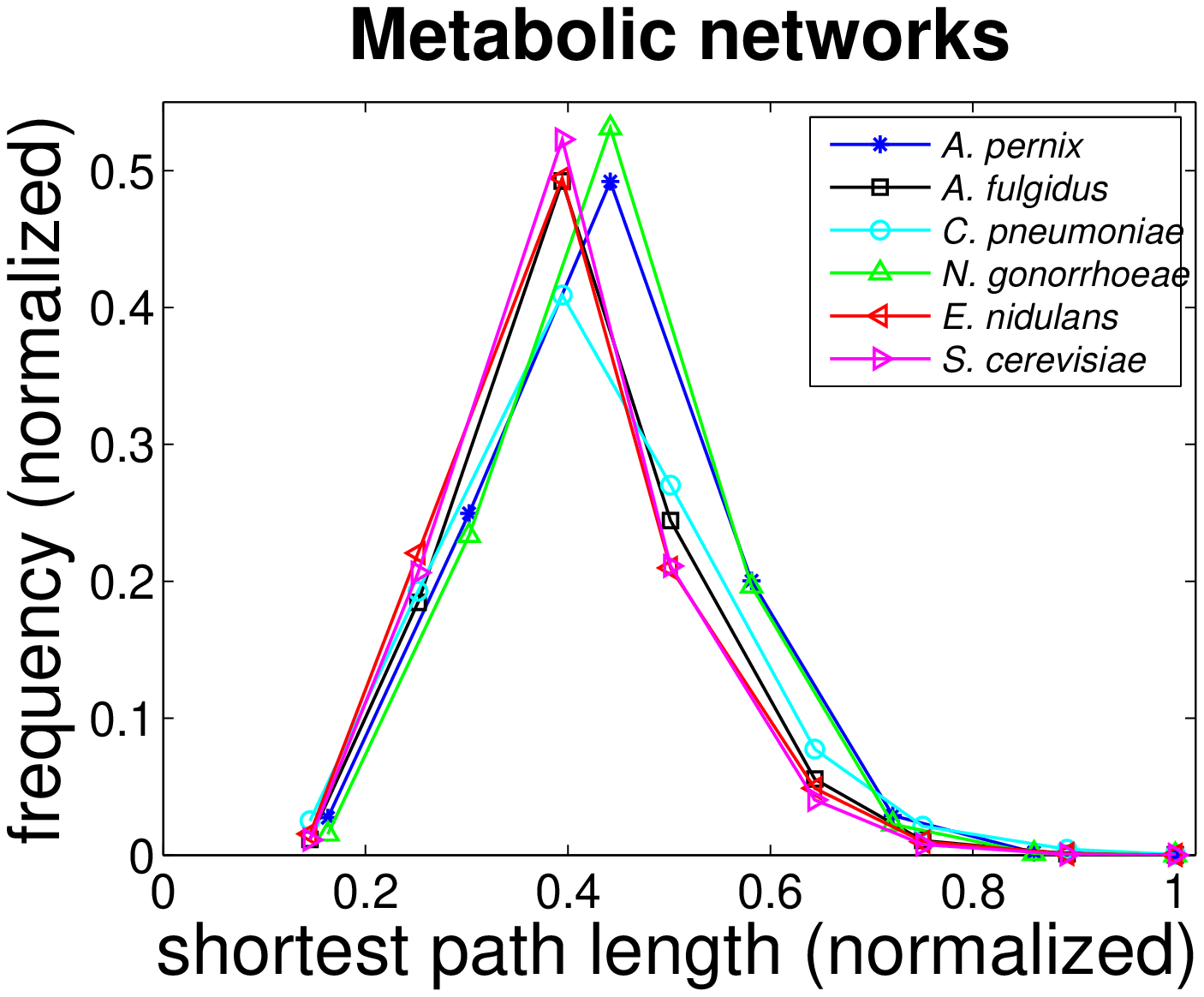}\includegraphics[scale=.33]{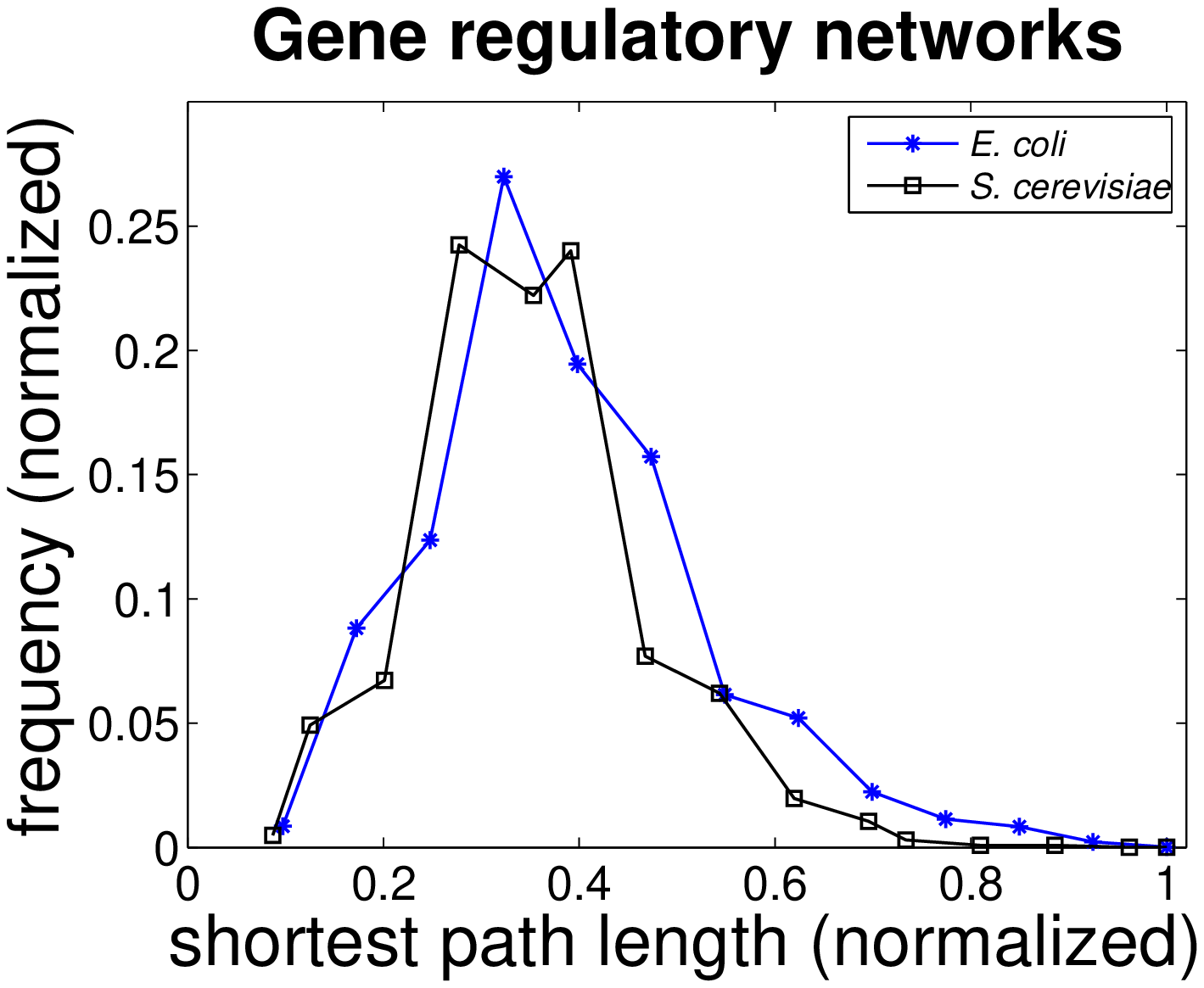}\\
\vspace{-3.4cm}
\hspace{-2.4cm}(d)\hspace{4.8cm}(e)\\
\vspace{2.4cm}
\end{center}
\caption{{\bf Histogram of shortest path length}. X-axis represents the (normalized) shortest path length
and Y-axis represents the (normalized) frequency. (a) {\bf Neuronal networks}: (macaque visual cortex area, 
macaque large-scale visual and sensorimotor area corticocortical connectivity, macaque cortical connectivity, 
cat cortical area, cat cortical and thalamic areas, {\it C. elegans}), (b) {\bf Food webs}: (Ythan Estuary, 
 Little Rock Lake, Grassland, Silwood, St Marks Seagrass, St Martin),
 (c) {\bf Protein-protein interaction networks}: ({\it E. coli}, {\it S. cerevisiae}, {\it H. pylori}), (d) {\bf Metabolic networks}:
         archaea ({\it A. pernix}, {\it A. fulgidus}), eukaryota ({\it E. nidulans}, {\it S. cerevisiae}), bacteria ({\it C. pneumoniae}, {\it N. gonorrhoeae}), 
(g) {\bf Gene regulatory networks}: ({\it E. coli}, {\it S. cerevisiae}).}
\label{sdneuronal}
\end{figure}

\begin{figure}[!] 
\begin{center}

\includegraphics[scale=.33]{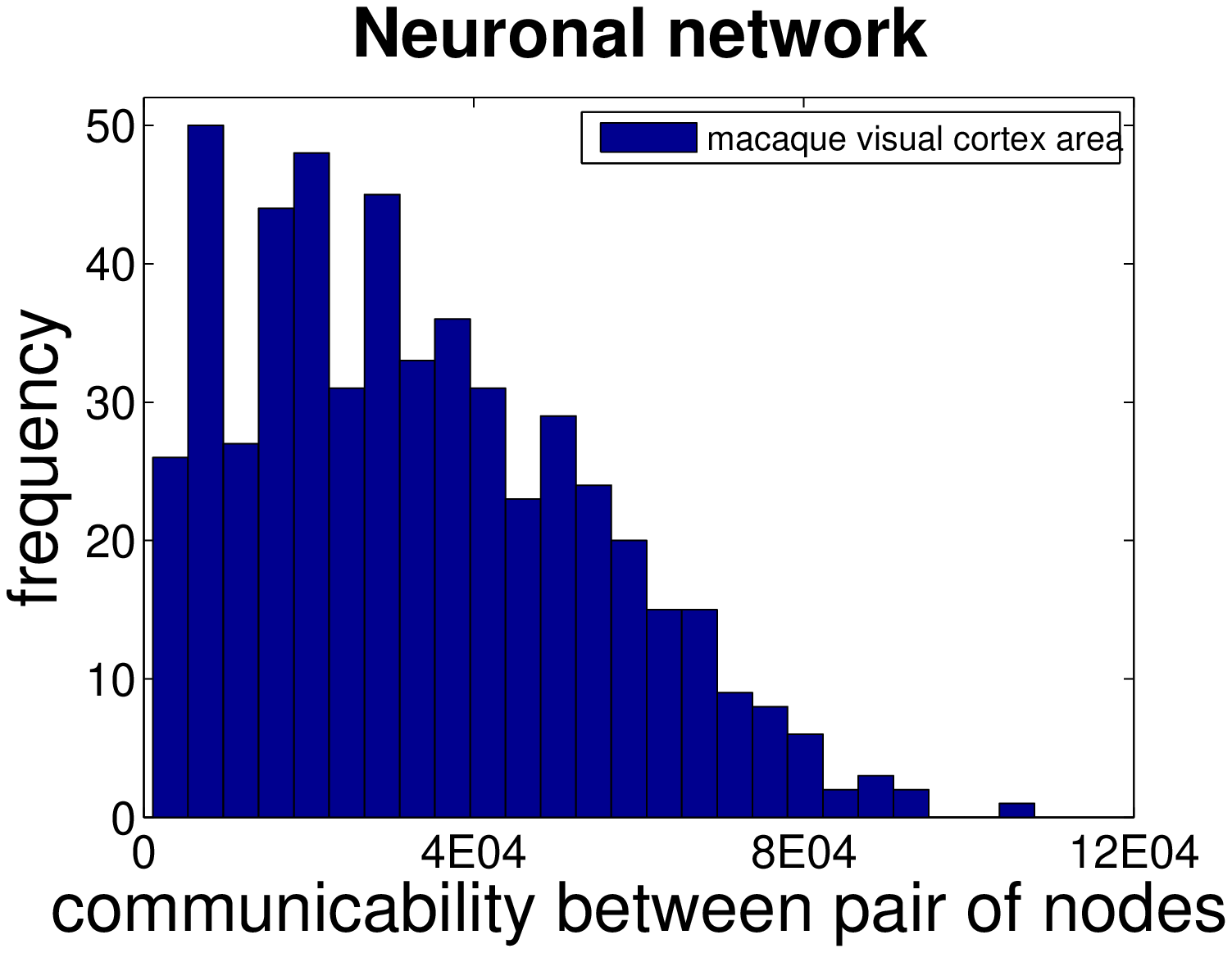}\includegraphics[scale=.33]{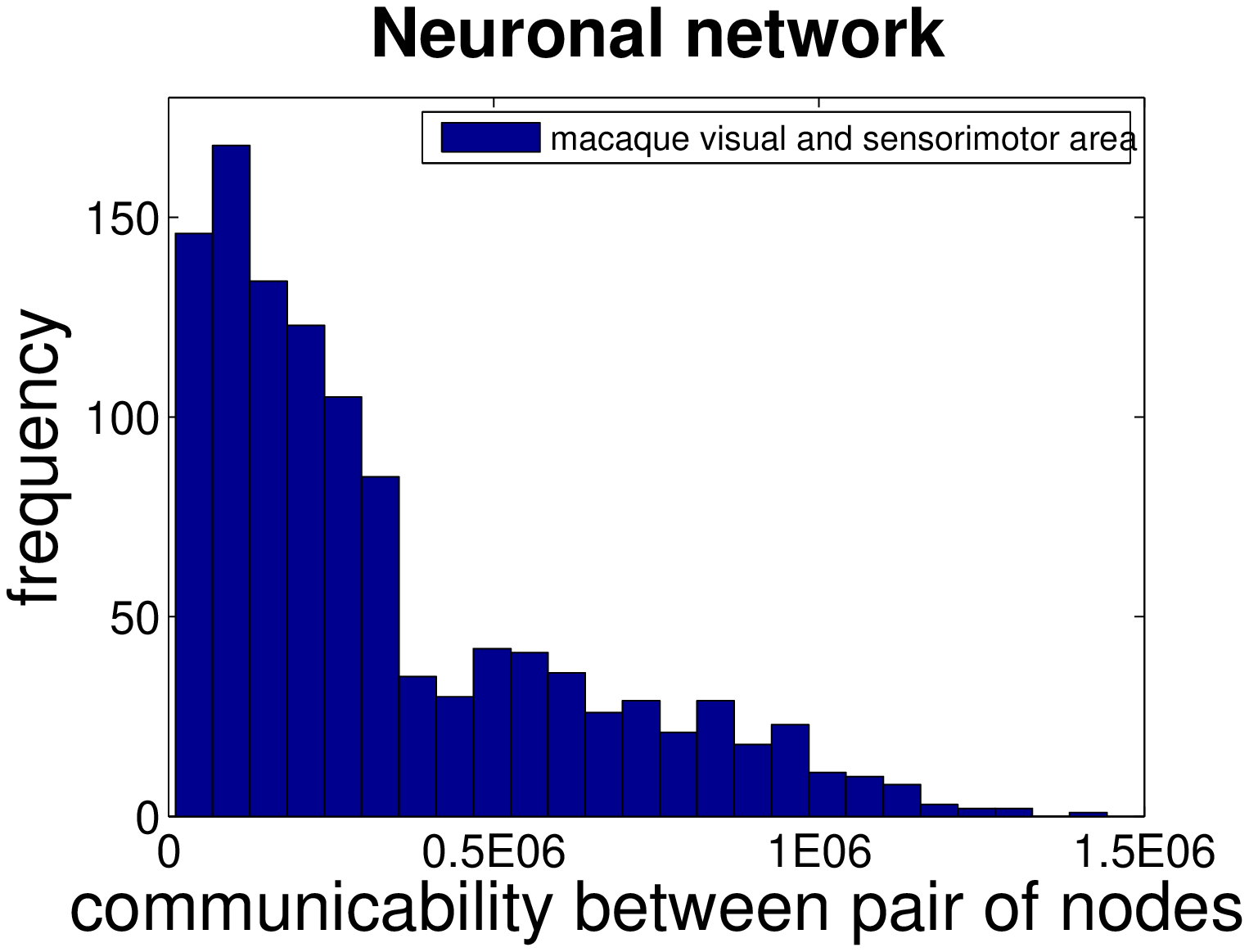}\includegraphics[scale=.33]{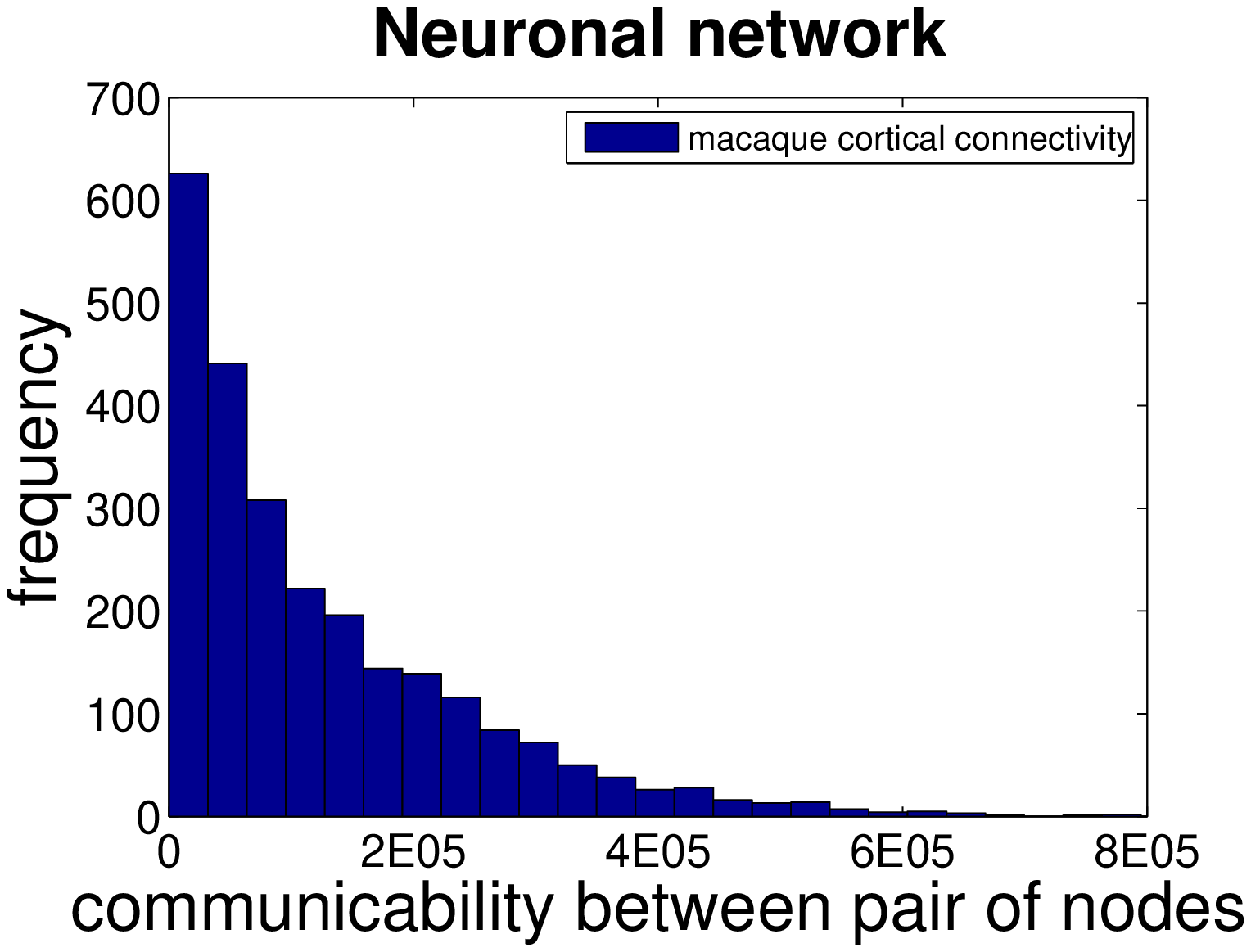}\\
\vspace{-3.2cm}
\hspace{2cm}(a)\hspace{4.4cm}(b)\hspace{4cm}(c)\\
\vspace{2.65cm}
\includegraphics[scale=.33]{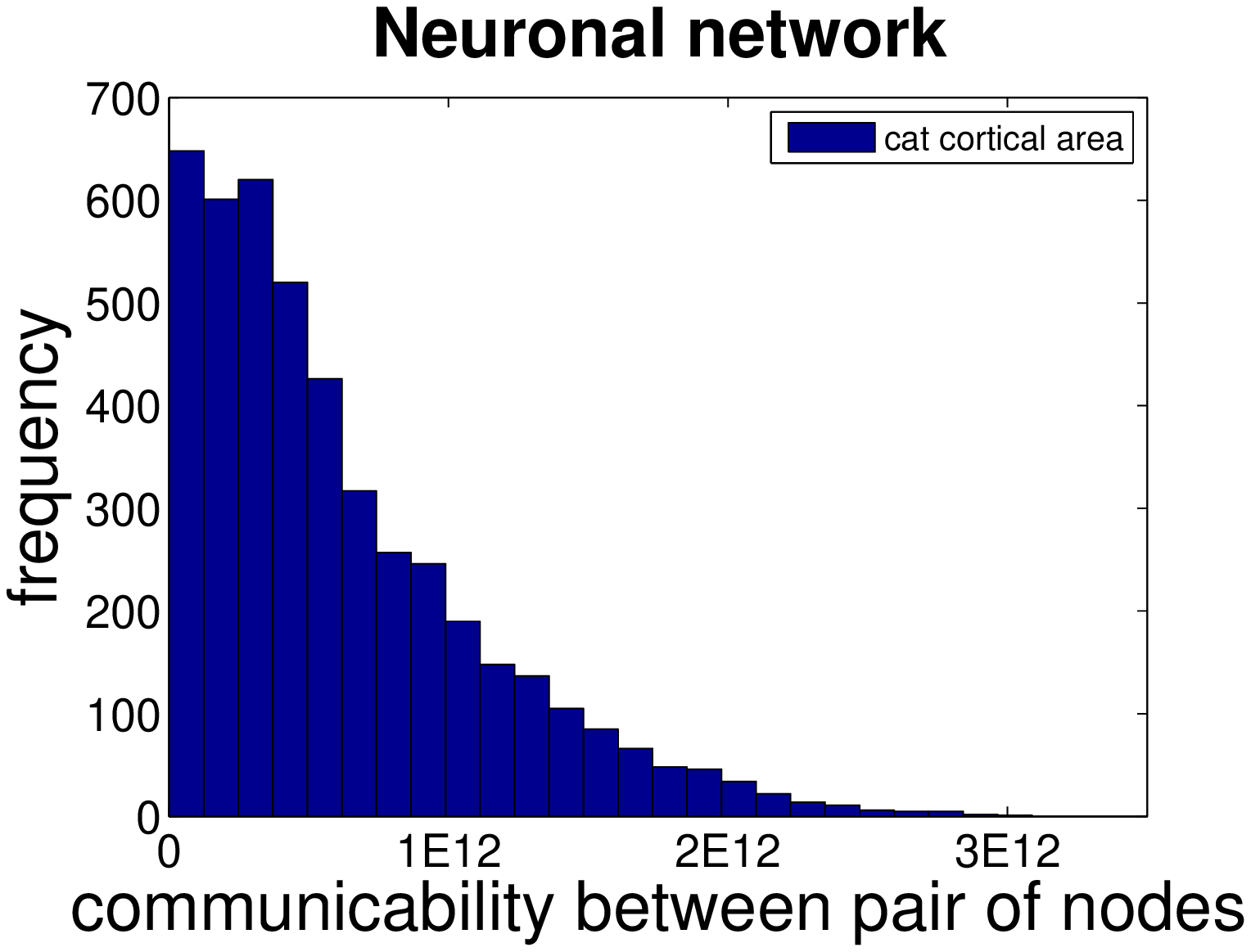}\includegraphics[scale=.33]{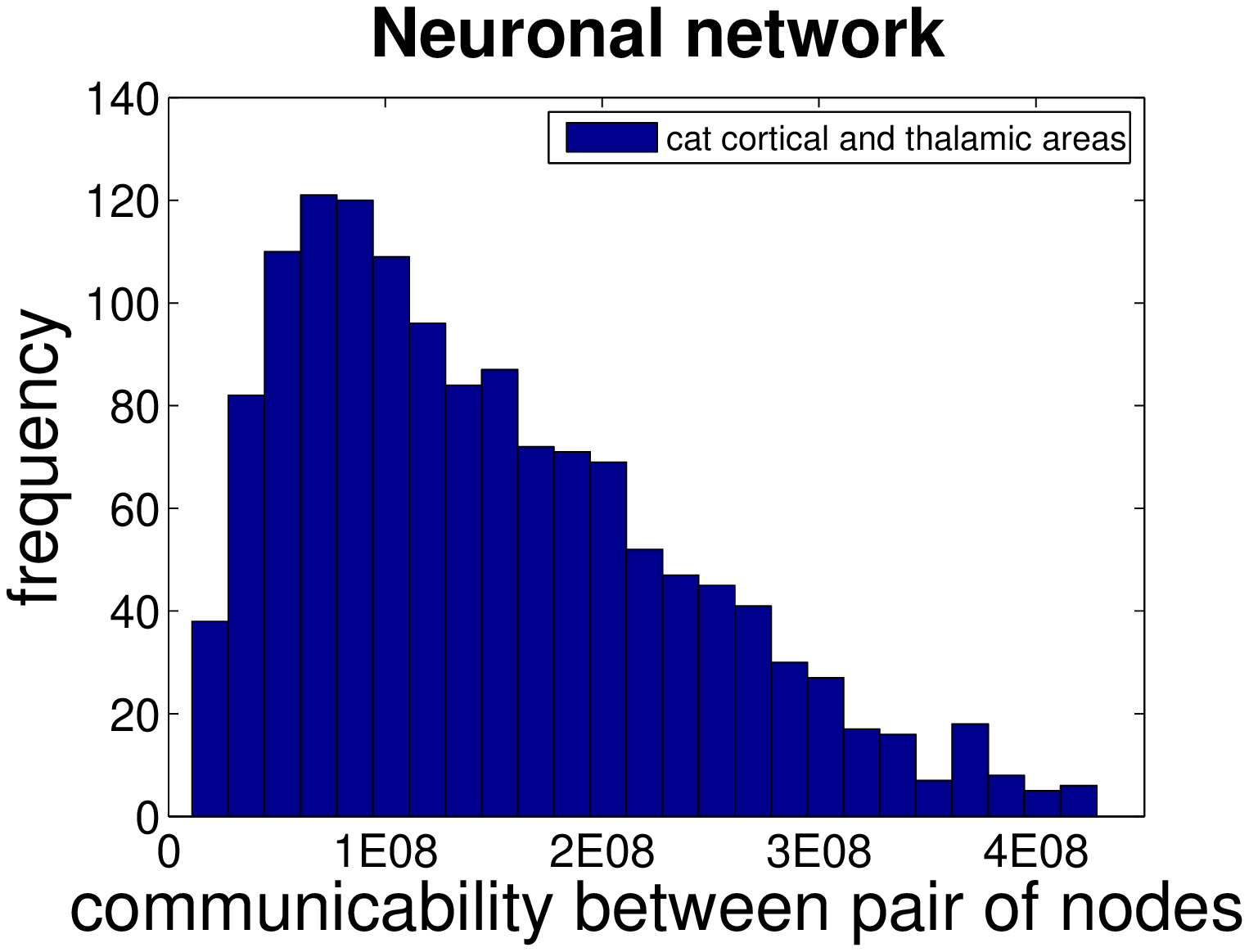}\includegraphics[scale=.33]{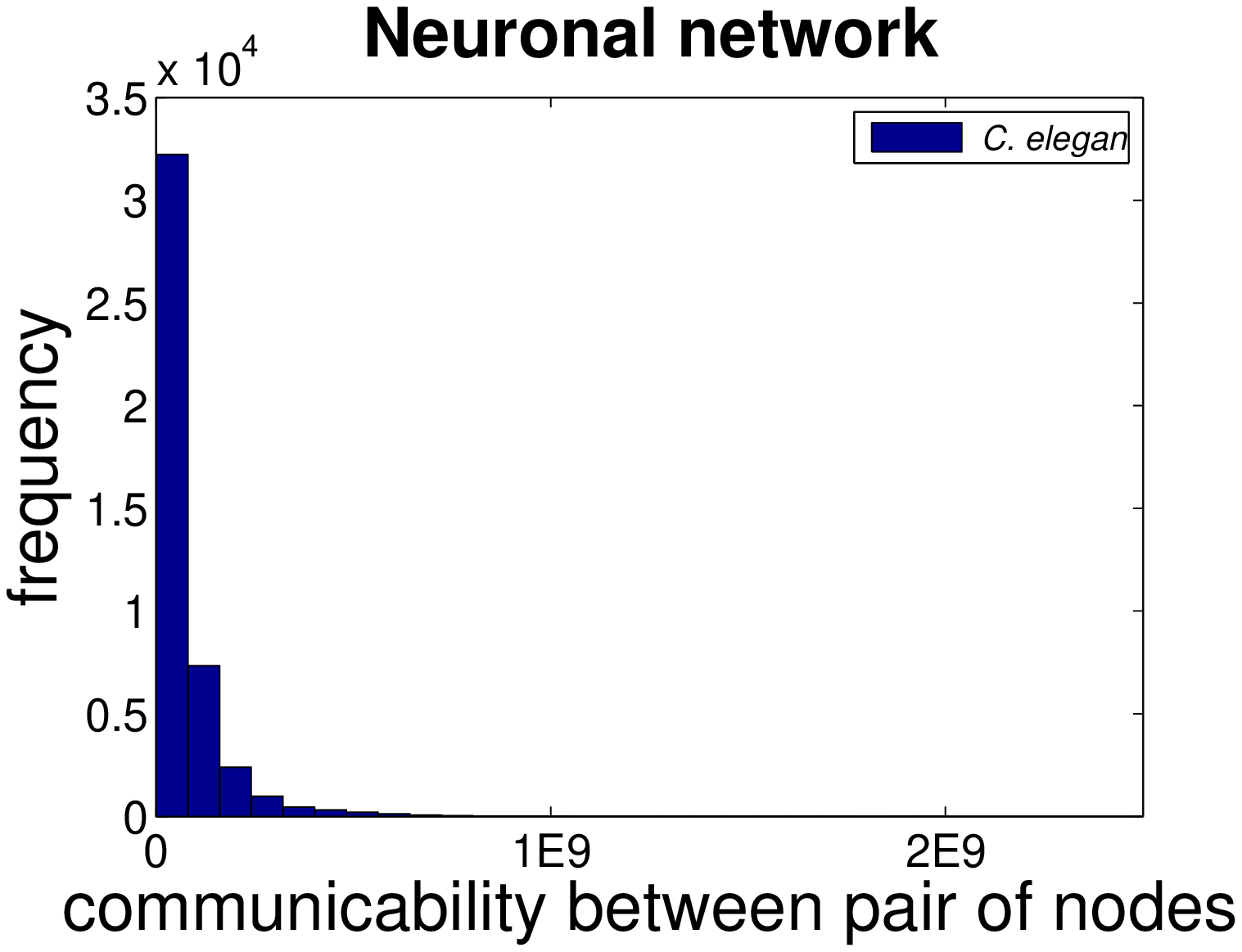}\\
\vspace{-3.2cm}
\hspace{2cm}(d)\hspace{4.4cm}(e)\hspace{4cm}(f)\\
\vspace{2.8cm}
\includegraphics[scale=.33]{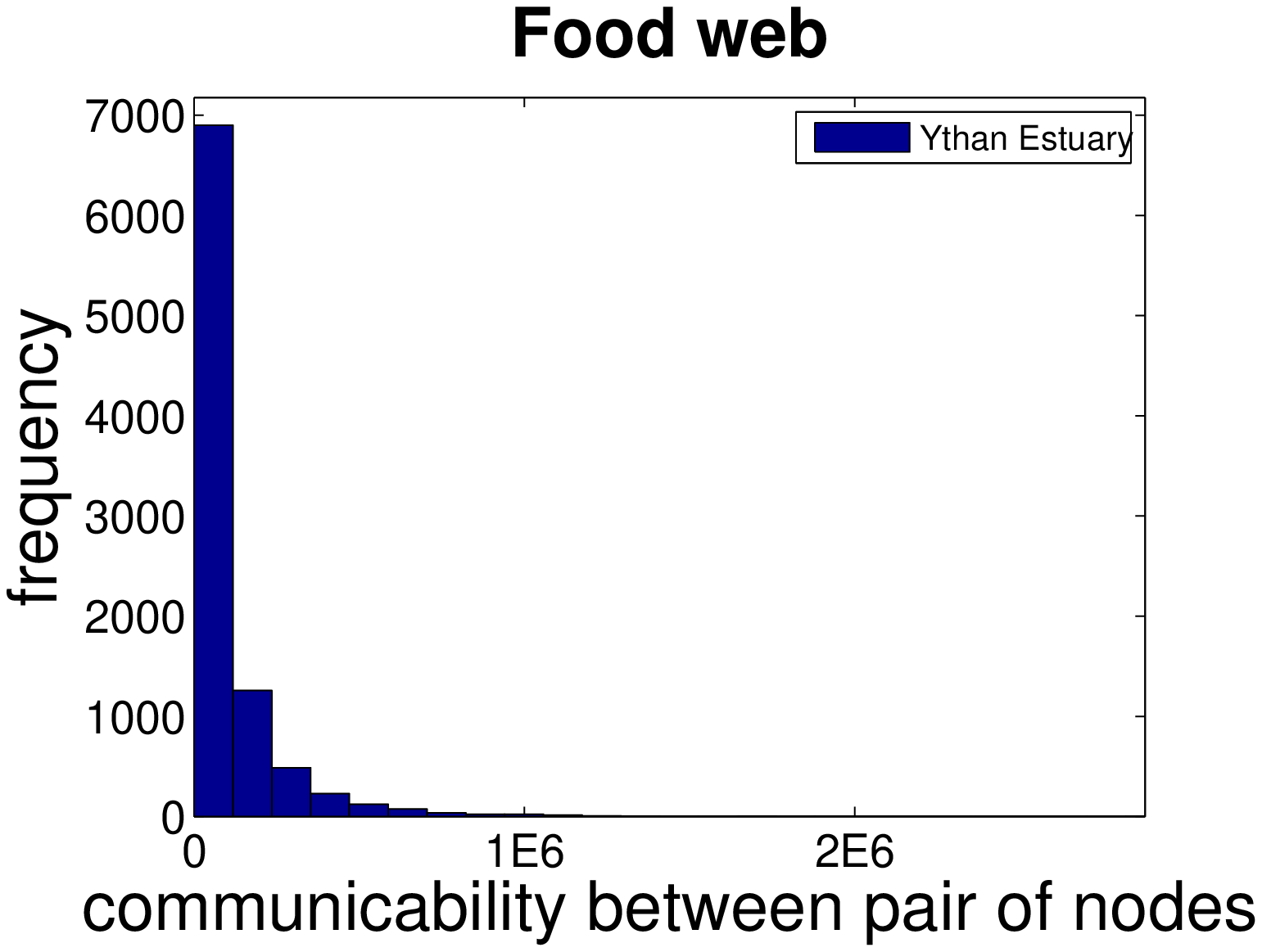}\includegraphics[scale=.33]{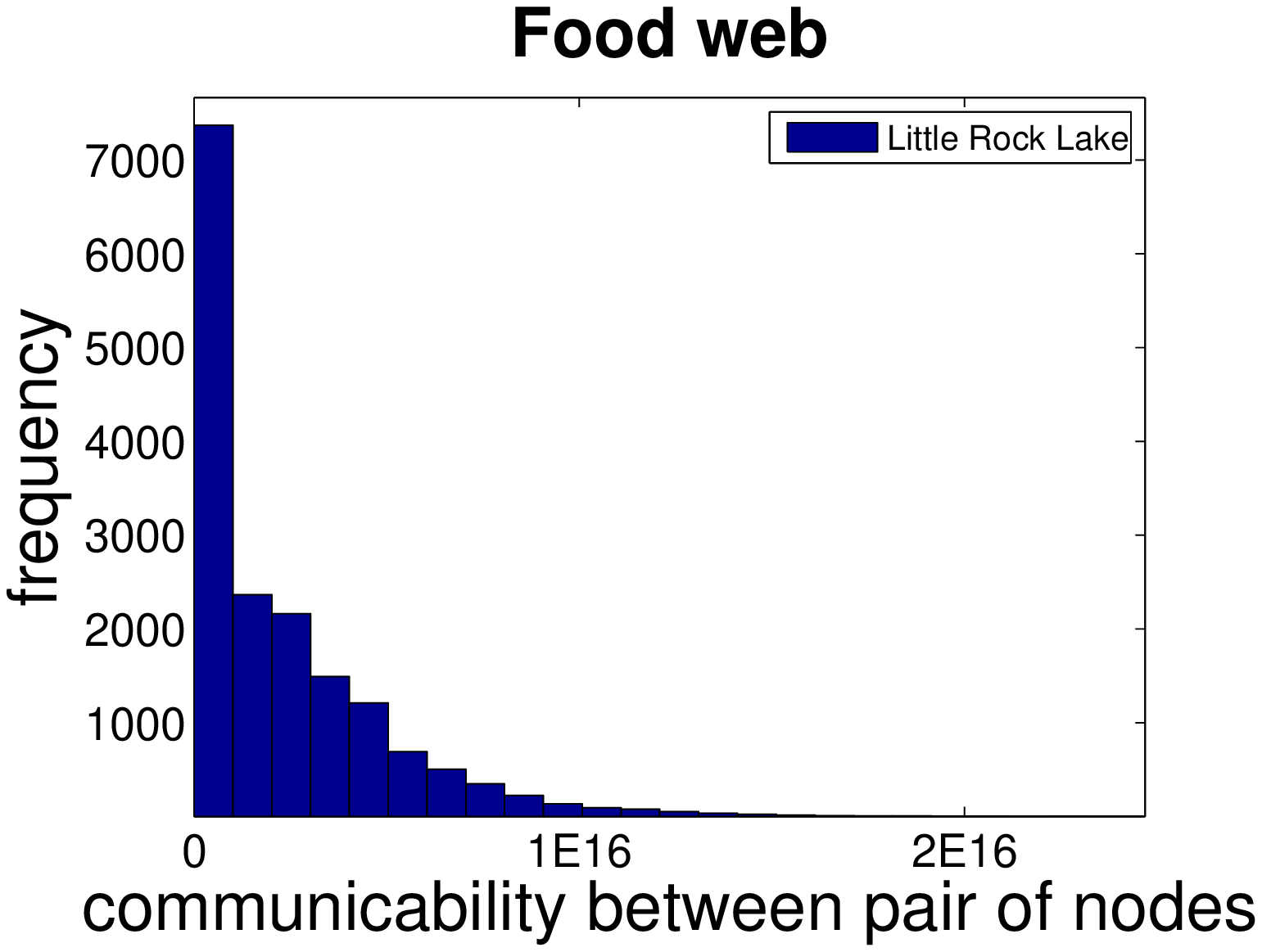}\includegraphics[scale=.33]{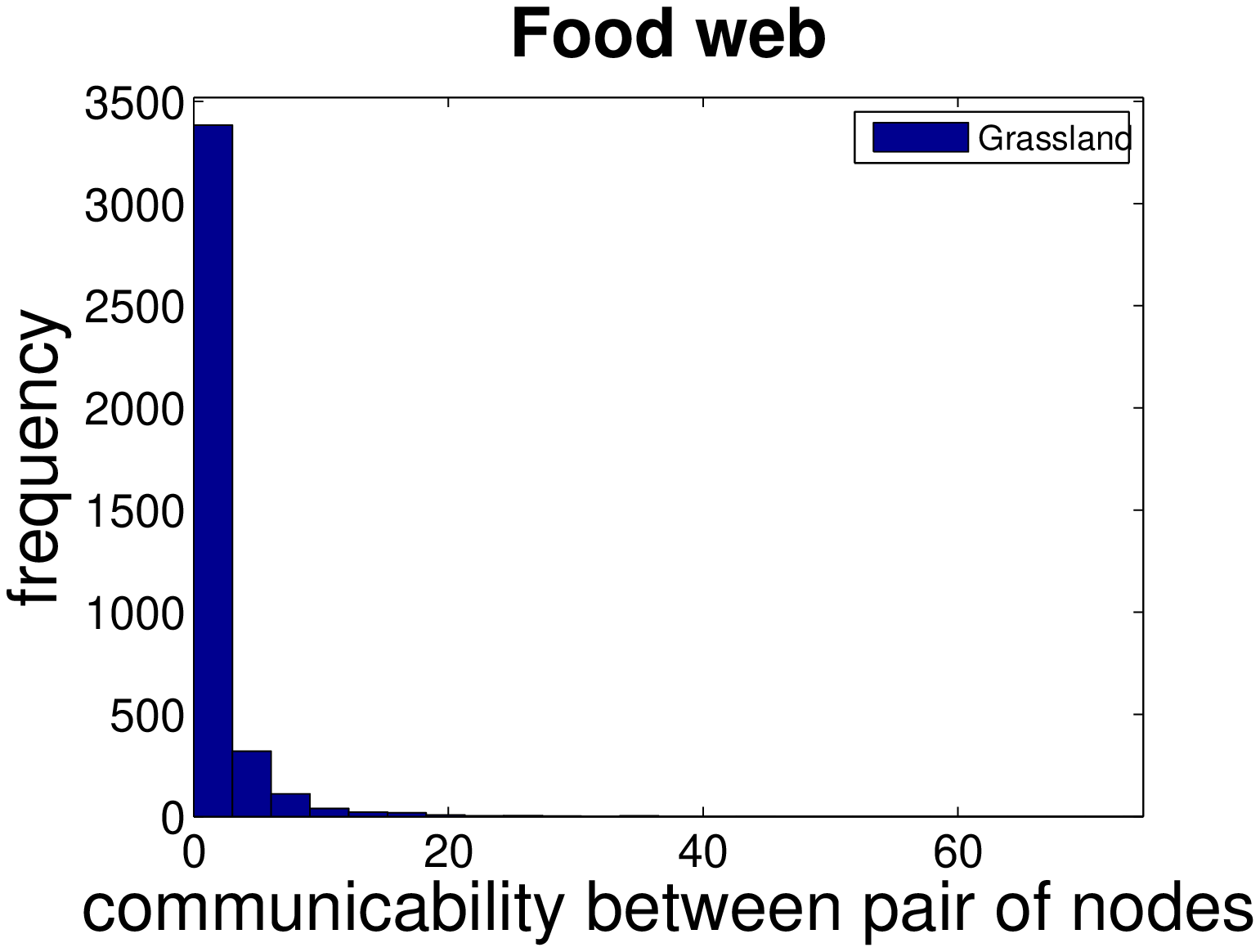}\\
\vspace{-3.2cm}
\hspace{2cm}(g)\hspace{4.4cm}(h)\hspace{4cm}(i)\\
\vspace{2.65cm}
\includegraphics[scale=.33]{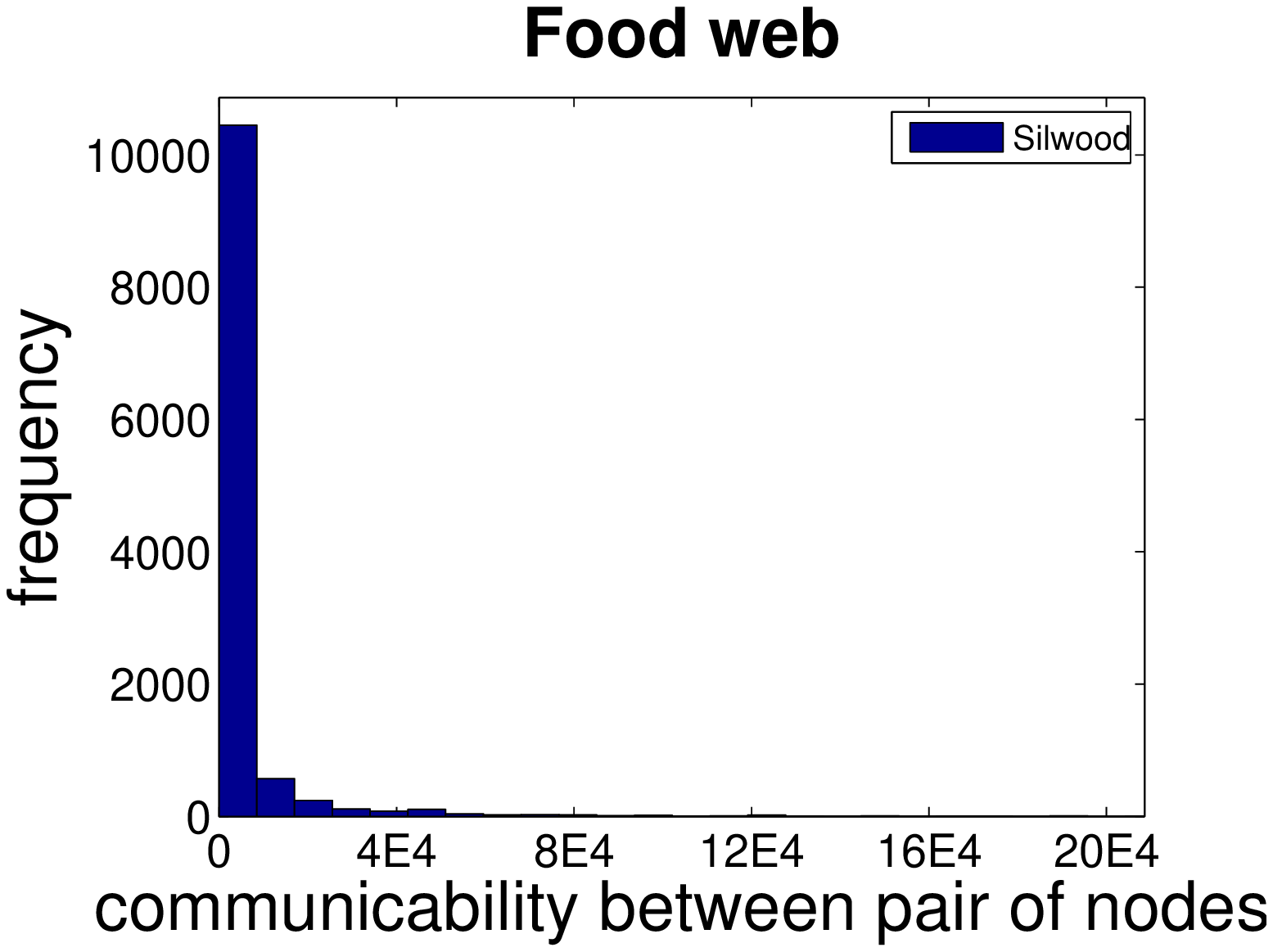}\includegraphics[scale=.33]{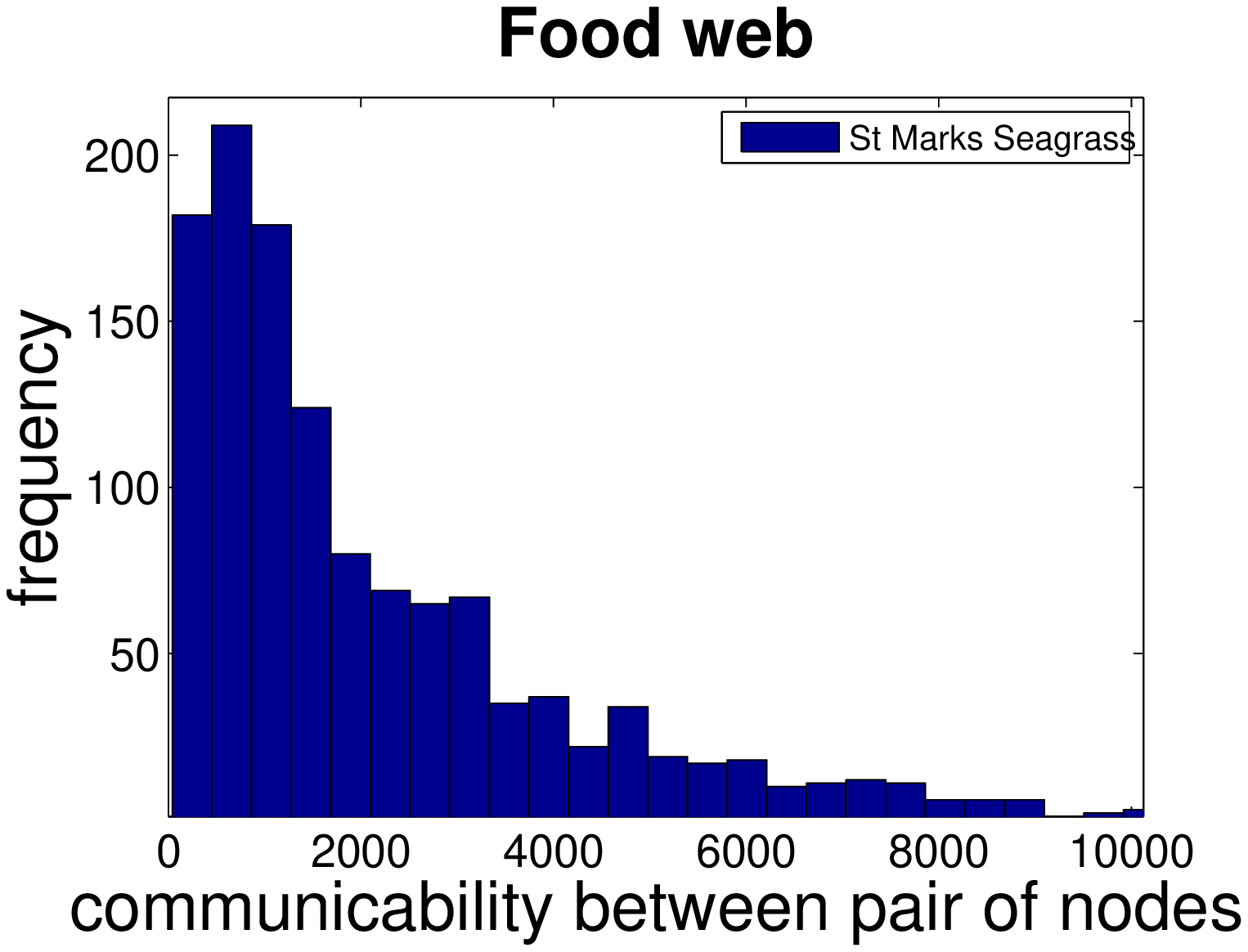}\includegraphics[scale=.33]{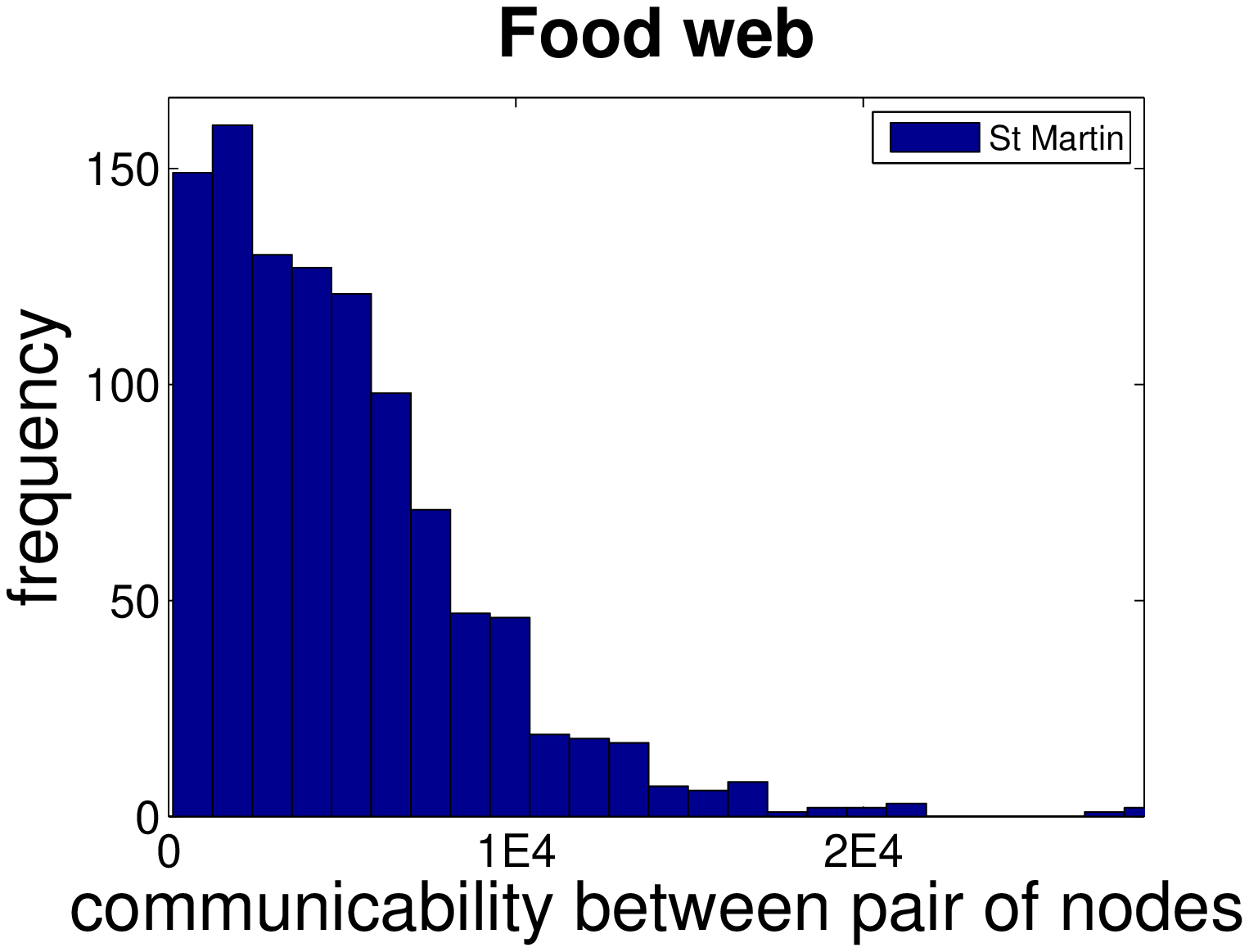}\\
\vspace{-3.6cm}
\hspace{2cm}(j)\hspace{4.4cm}(k)\hspace{4.5cm}(l)\\
\vspace{2.5cm}
\end{center}
\caption{{\bf Histogram of communicability}. Here the figures are the histogram of communicability (see equ~\ref{equ1}) of neuronal networks and food webs.
X-axis represent communicability between pair of nodes in a network and Y-axis represents frequency.
{\bf Neuronal networks}: (a) macaque visual cortex area,
(b) macaque large-scale visual and sensorimotor area corticocortical connectivity, (c)  macaque cortical connectivity, 
(d) cat cortical area, (e) cat cortical and thalamic areas, (f) {\it C. elegans}. {\bf Food webs}: (g) Ythan Estuary, (h) Little Rock Lake, (i) Grassland,
(j) Silwood, (k) St Marks Seagrass, (l) St Martin.}
\label{commu_1}
\end{figure}

\begin{figure}[!] 
\begin{center}
\includegraphics[scale=.33]{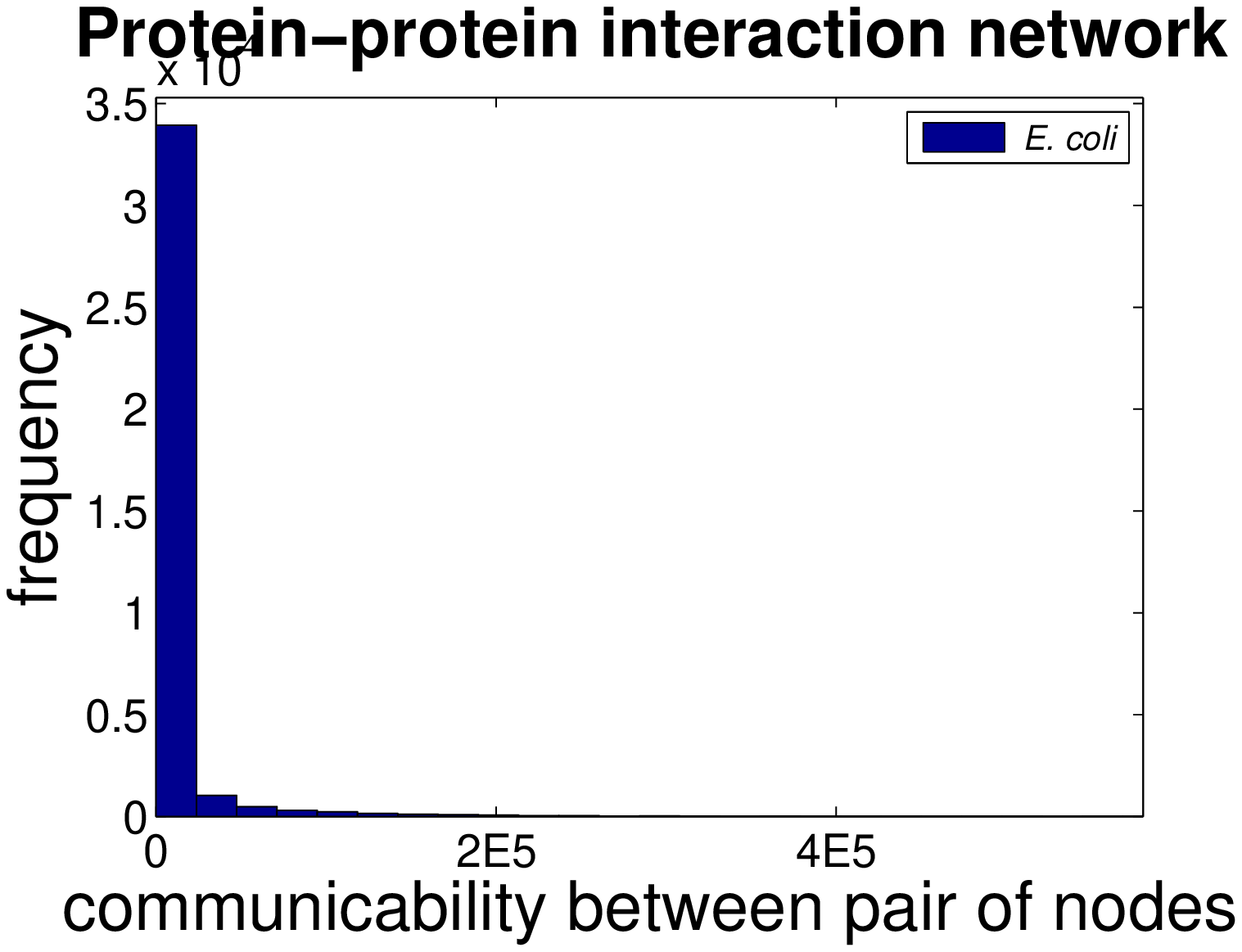}\includegraphics[scale=.33]{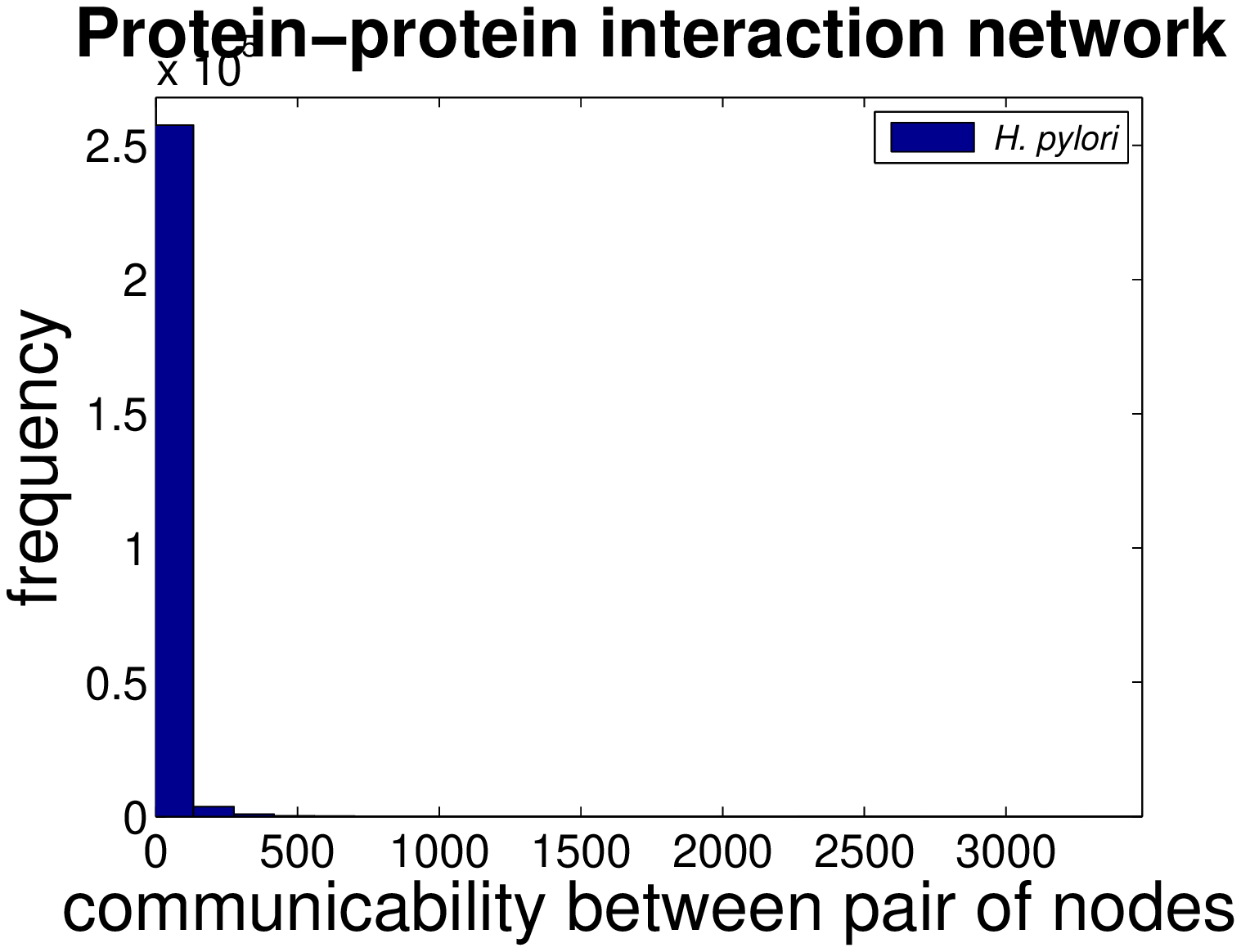}\includegraphics[scale=.33]{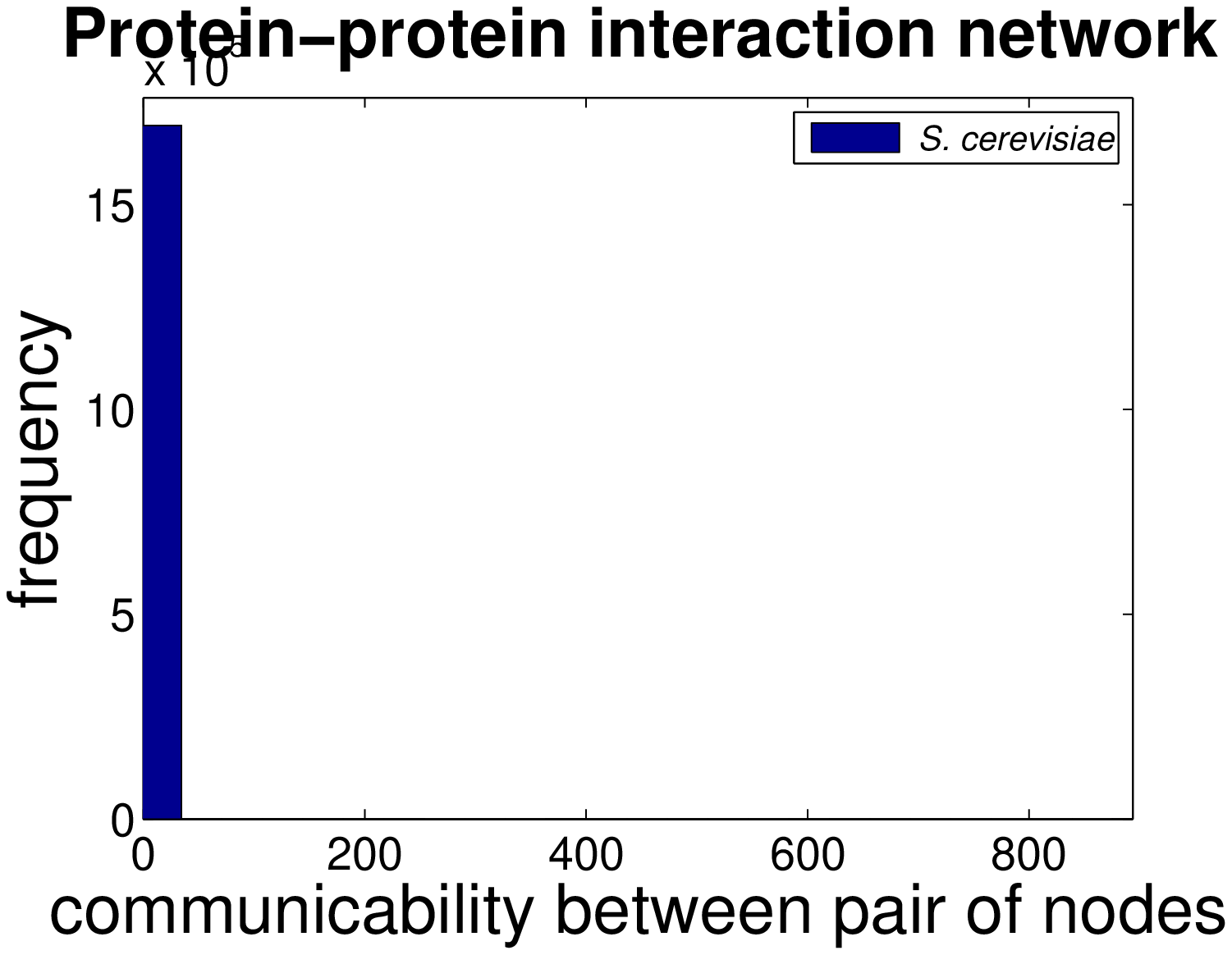}\\
\vspace{-3.2cm}
\hspace{1cm}(a)\hspace{4.8cm}(b)\hspace{4.8cm}(c)\\
\vspace{2.8cm}
\includegraphics[scale=.33]{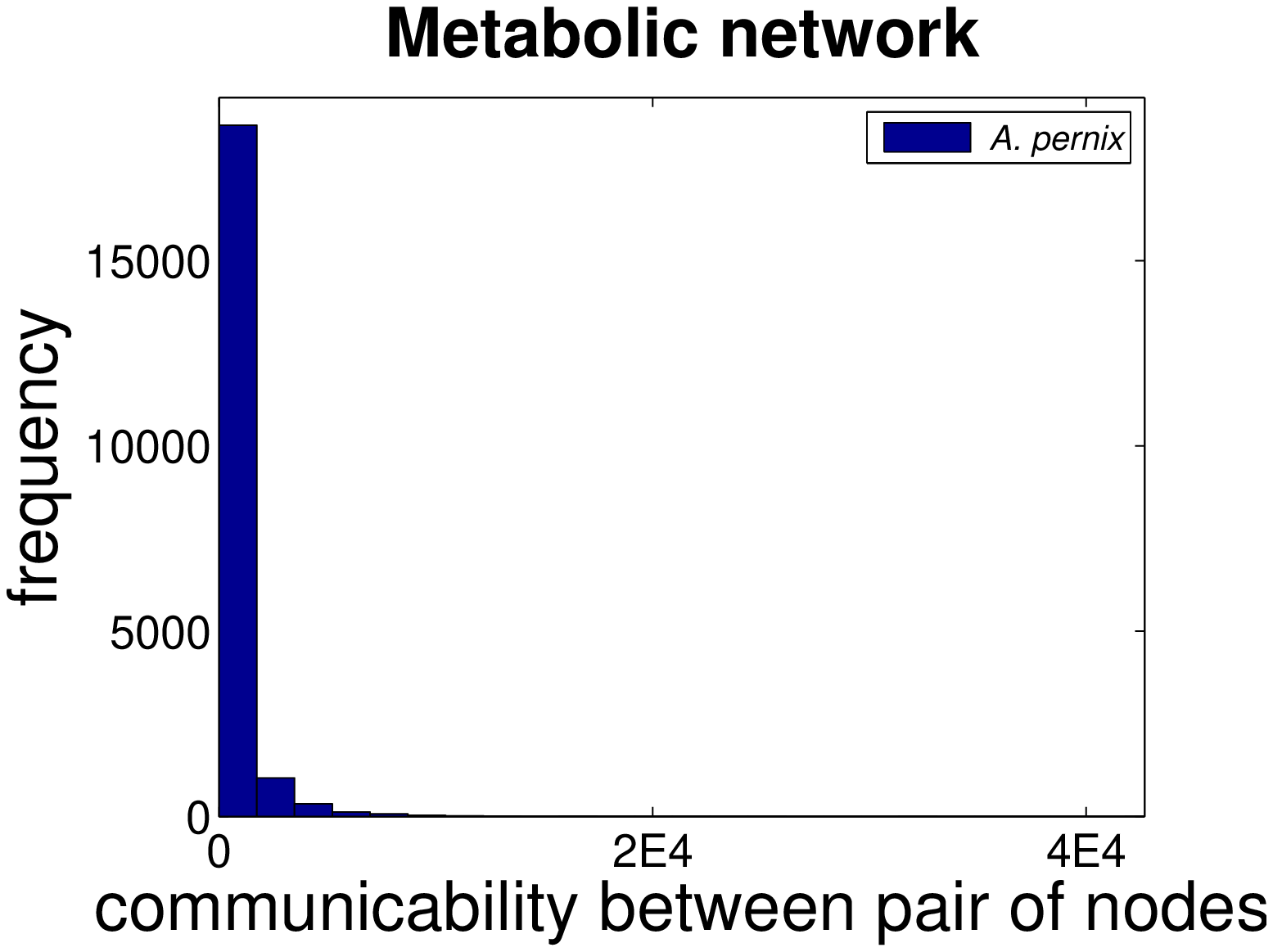}\includegraphics[scale=.33]{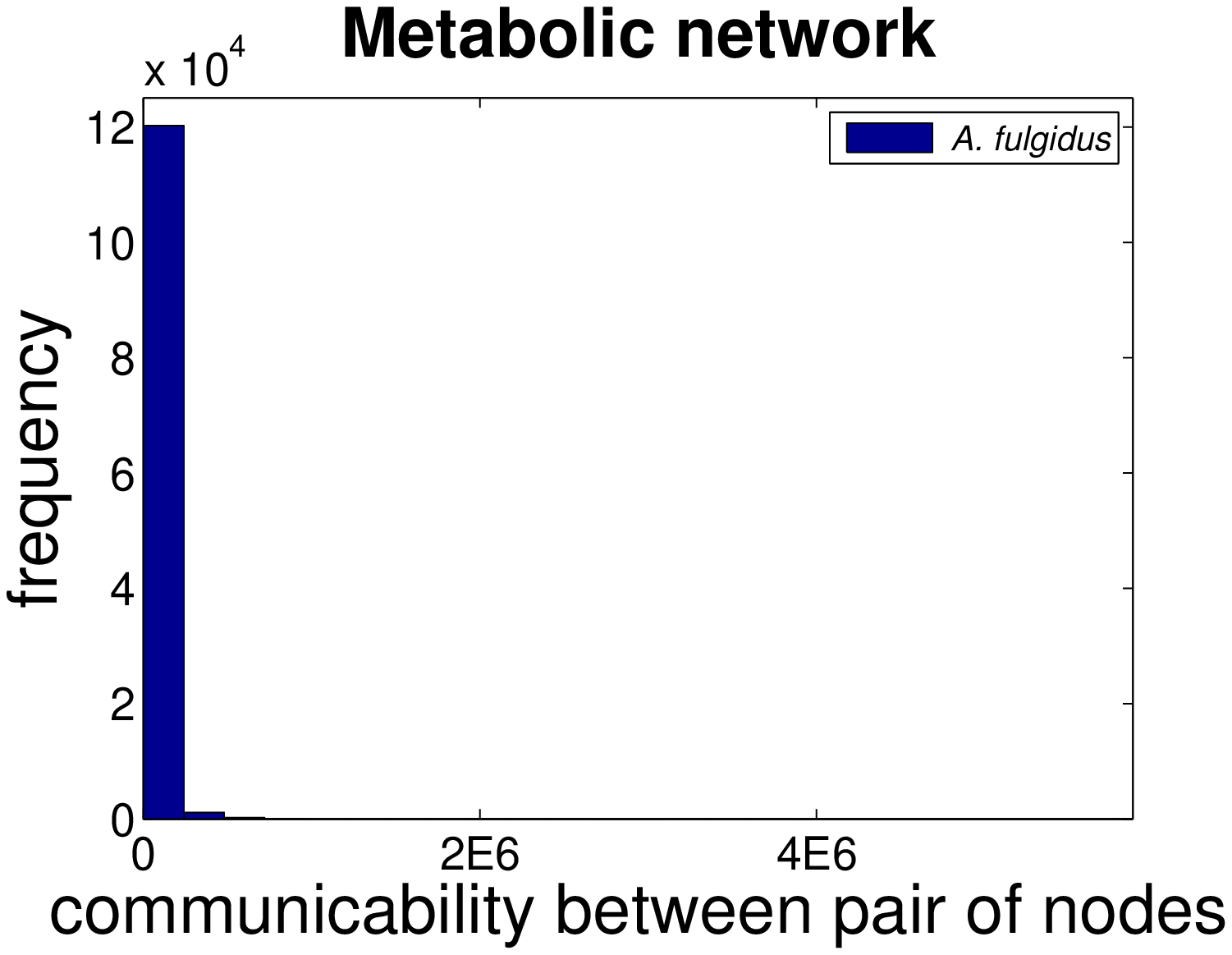}\includegraphics[scale=.33]{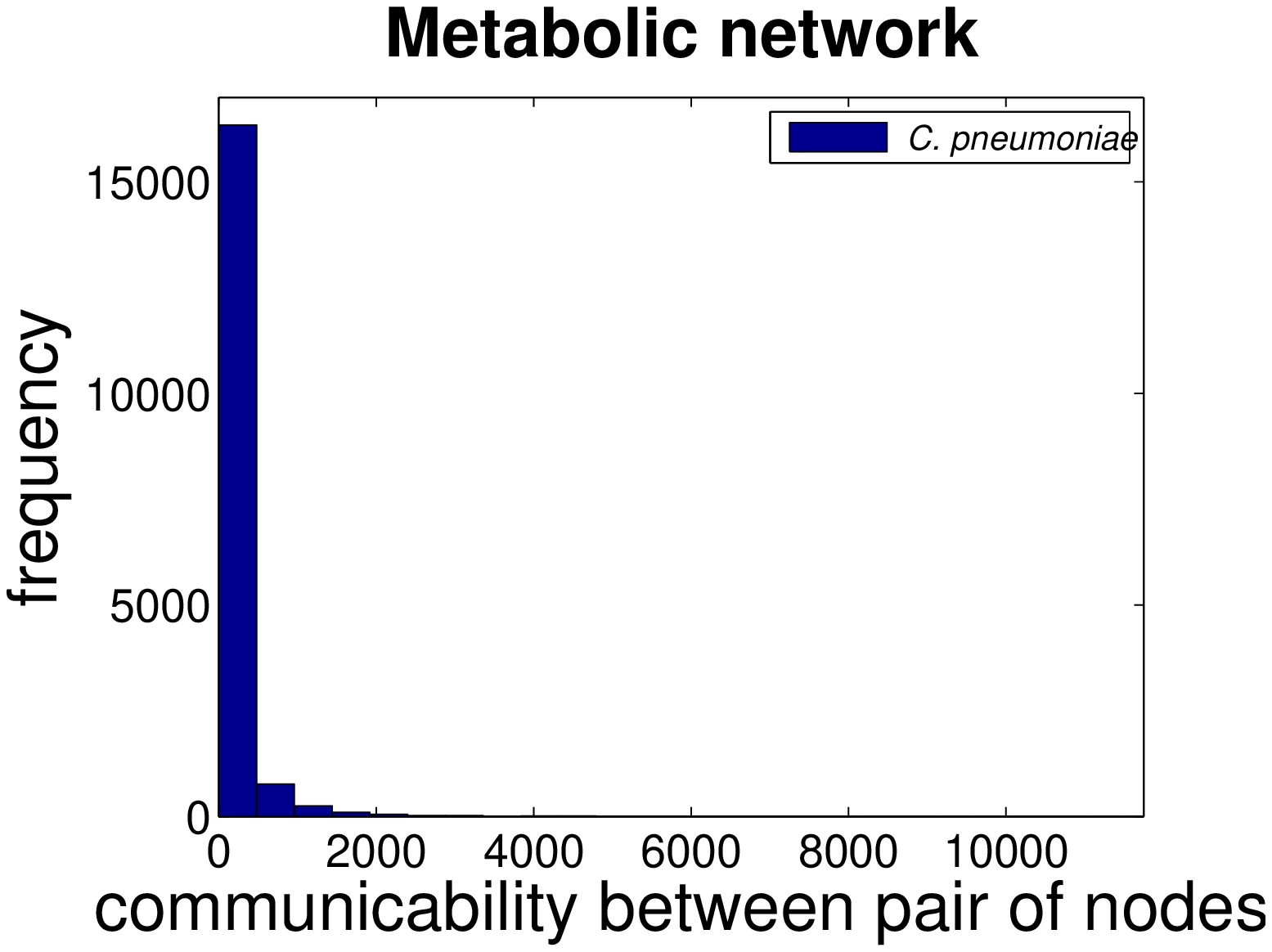}\\
\vspace{-3.2cm}
\hspace{1cm}(d)\hspace{4.8cm}(e)\hspace{4.8cm}(f)\\
\vspace{2.65cm}
\includegraphics[scale=.33]{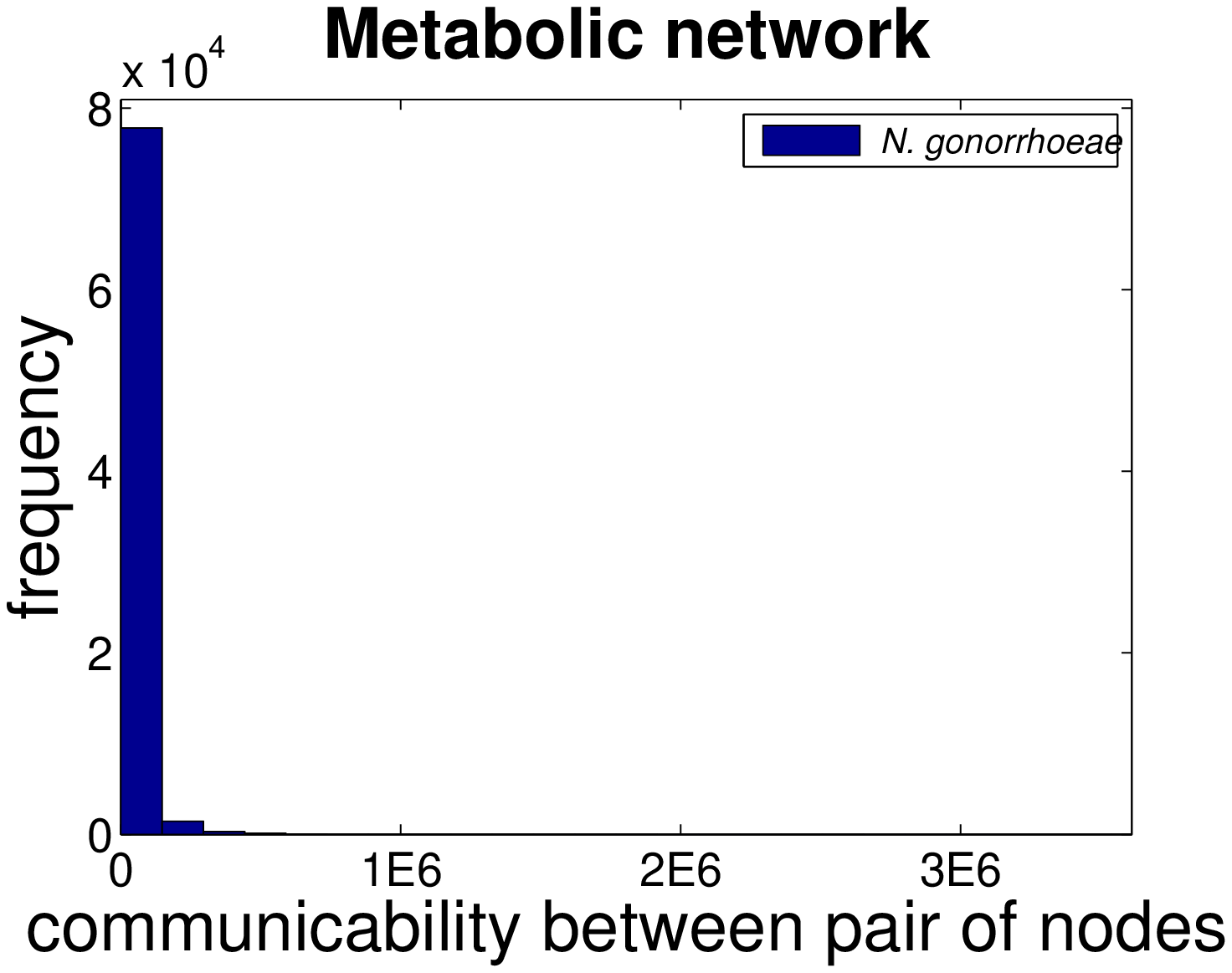}\includegraphics[scale=.33]{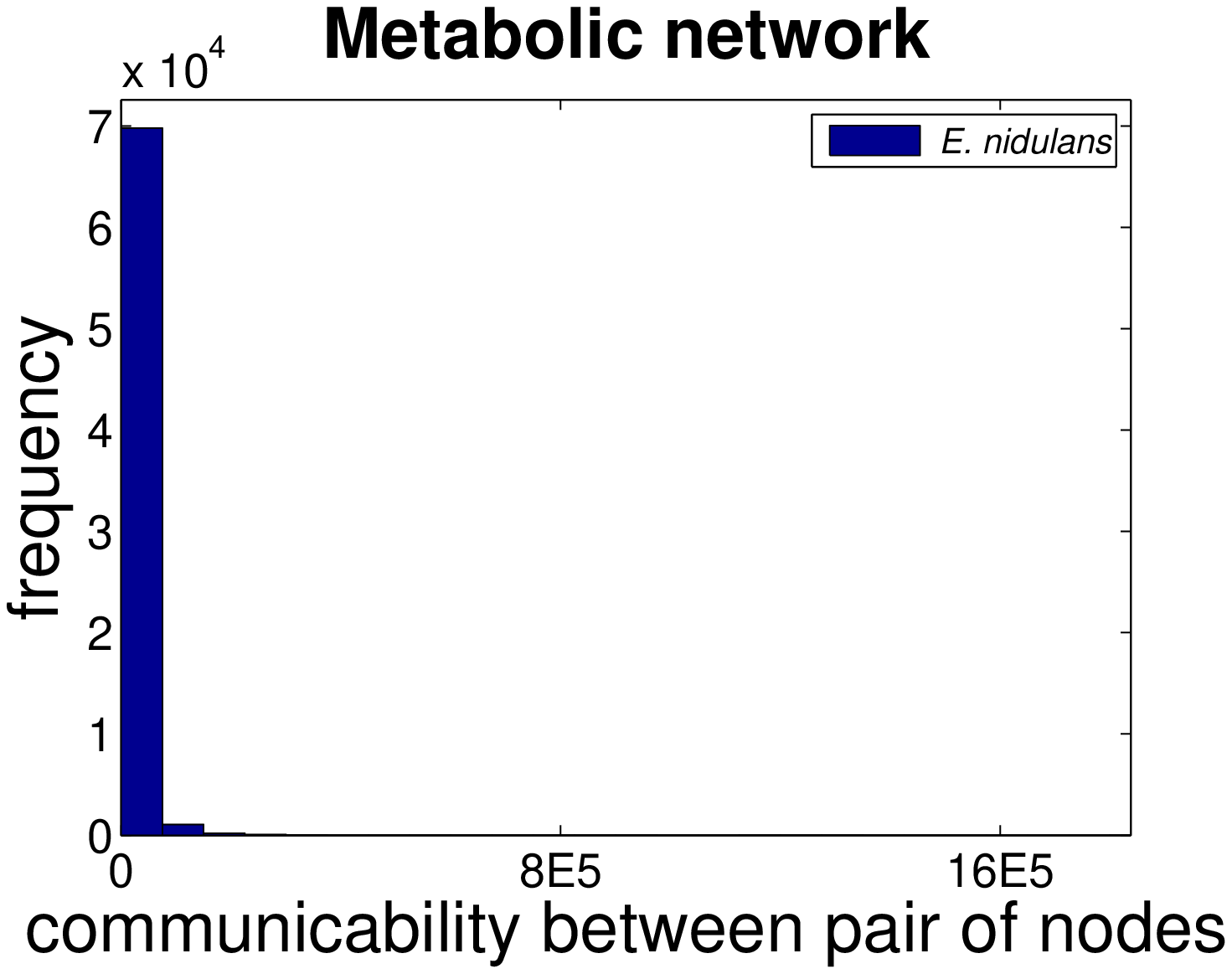}\includegraphics[scale=.33]{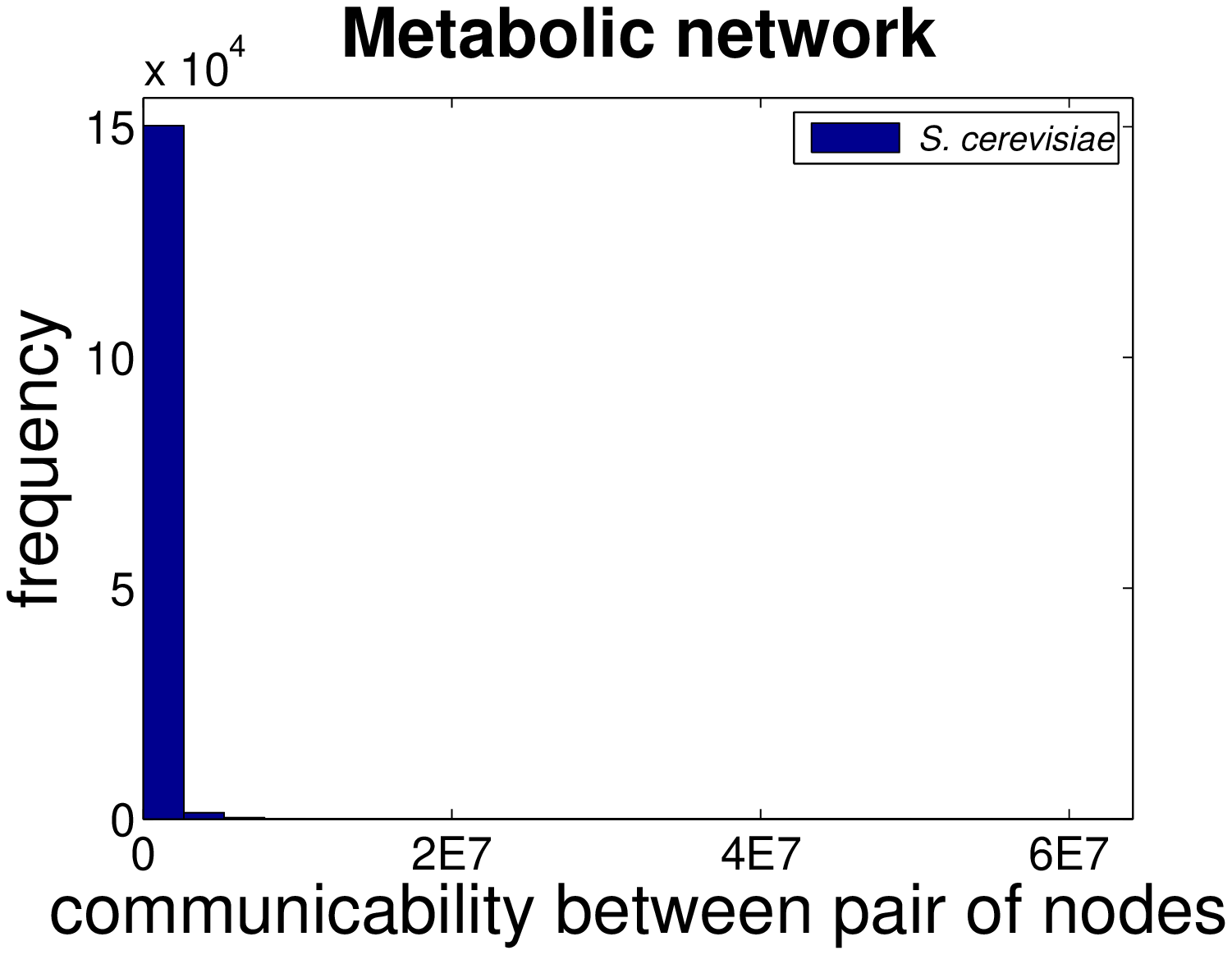}\\
\vspace{-3.2cm}
\hspace{1cm}(g)\hspace{4.8cm}(h)\hspace{4.8cm}(i)\\
\vspace{2.8cm}
\includegraphics[scale=.33]{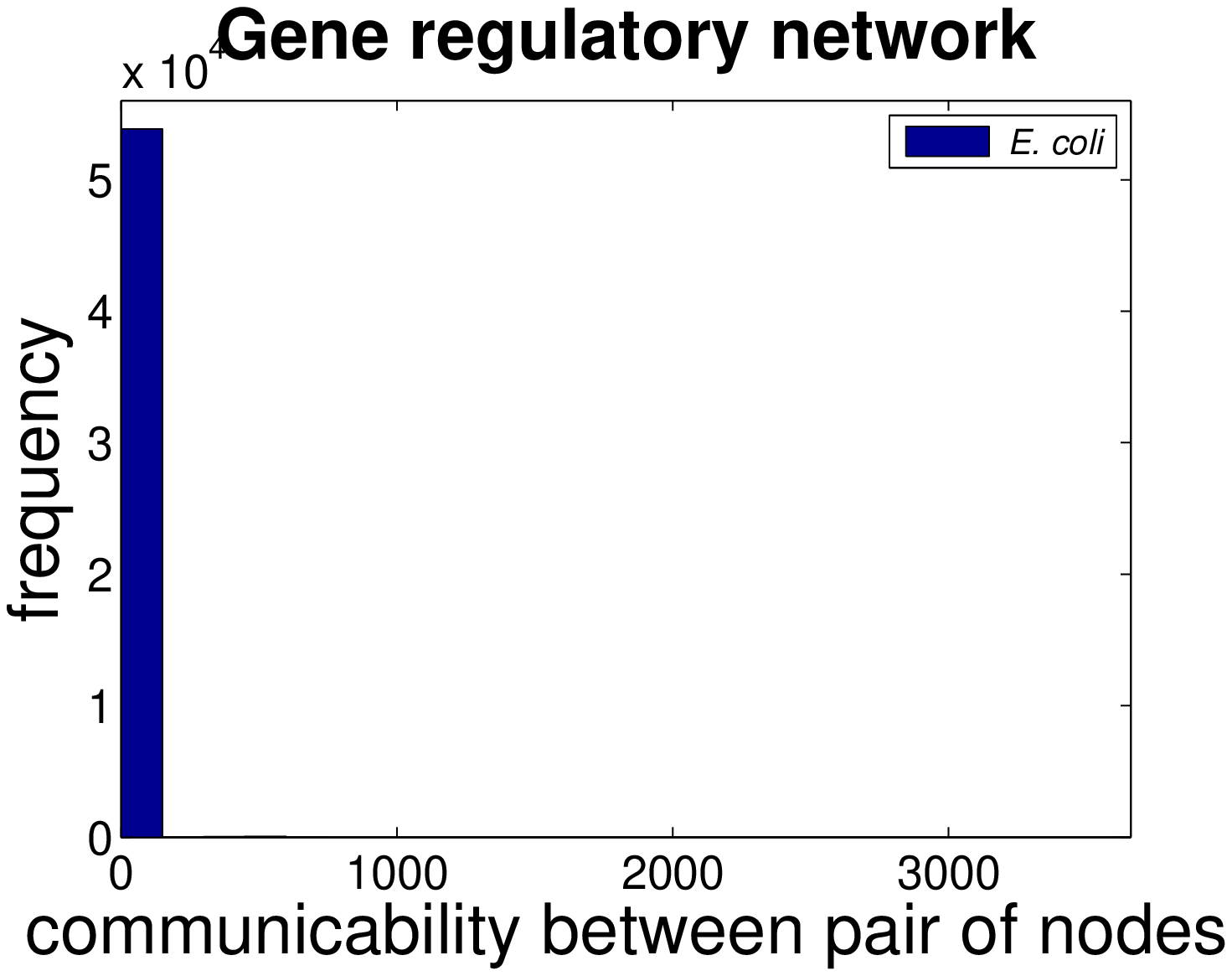}\includegraphics[scale=.33]{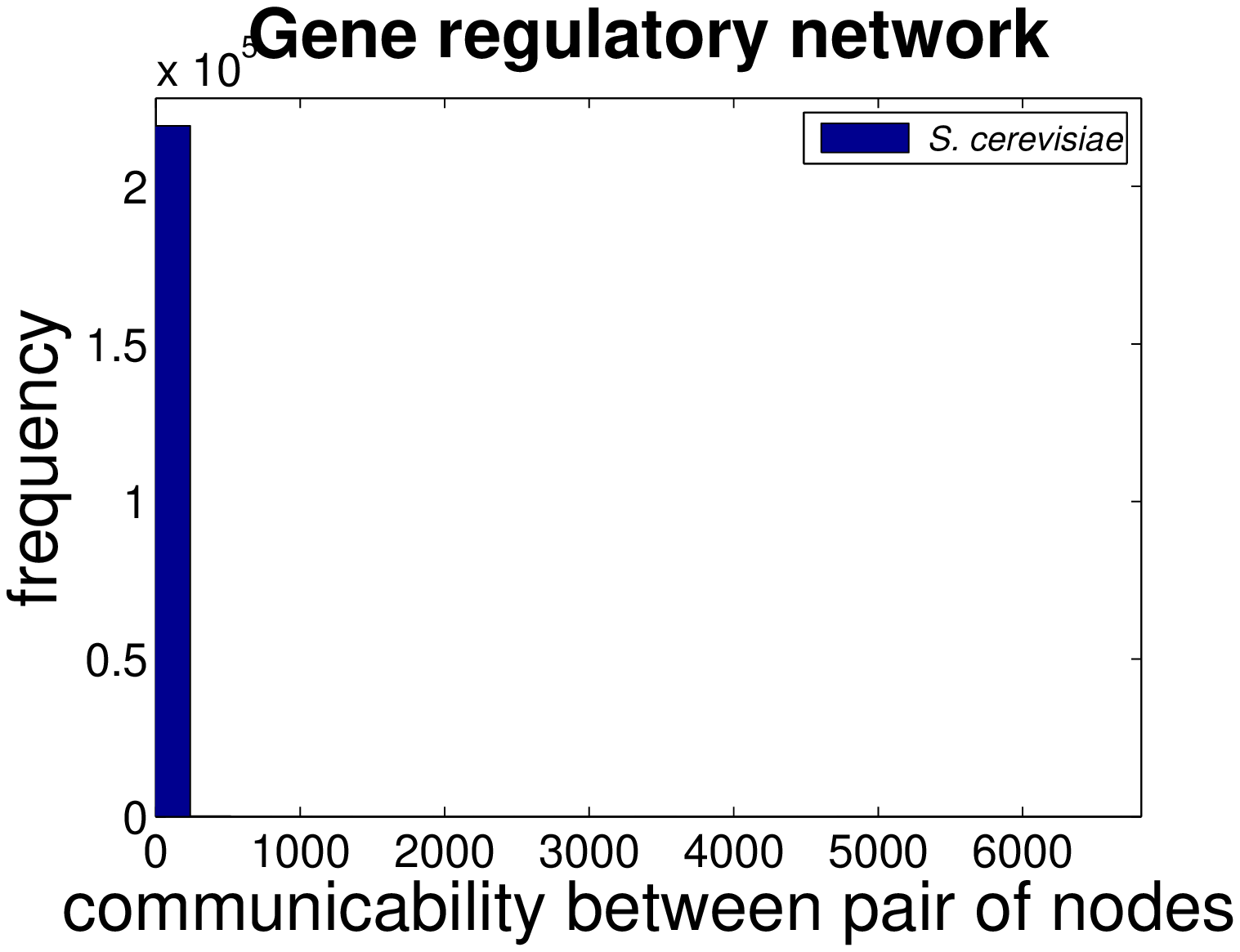}\\
\vspace{-3.2cm}
\hspace{-2cm}(j)\hspace{4.5cm}(k)\\
\vspace{2.2cm}
\end{center}
\caption{{\bf Histogram of communicability}. Here the figures are the histogram of communicability (see equ~\ref{equ1}) of protein-protein interaction networks,
metabolic networks and gene regulatory networks. X-axis represent communicability between pair of nodes in a network and Y-axis represents frequency.
 {\bf Protein-protein interaction networks}: (a) {\it E. Coli}, 
(b)  {\it S. cerevisiae}, (c) {\it H. Pylori}, {\bf Metabolic networks}: (d) {\it A. pernix}, (e)  {\it A. fulgidus}, 
(f) {\it C. pneumoniae}, (g) {\it N. gonorrhoeae}, (h) {\it E. nidulans}, (i) {\it S. cerevisiae}, {\bf Gene regulatory networks}: (j) {\it E. coli}, (k) {\it S. cerevisiae}.}
\label{commu_2}
\end{figure}

The distribution of the distances between every pair of nodes of each network (see Figure~\ref{sdneuronal}) follows a  
Gaussian like pattern, whereas, the distributions for the communicabilities for the same are different (see Figure~\ref{commu_1},~\ref{commu_2}). 
They clearly show that the data (i.e., communicability 
between pairs of nodes) are positively skewed for most of the networks and the relative frequency is highly concentrated in 
a small interval of whole range, i.e. the relative frequencies for almost every 
interval is near zero except for a few intervals. Thus most networks have a small number of 
pairs of nodes that show high communicability. Remarkably the distribution pattern for the most of
the neuronal networks and a few food webs are positively skewed and the data are spread out over the whole range
in the sense that the relative frequency of almost each interval is significant (Figure~\ref{commu_1}). 
It reflects that
the underlying undirected structure of most of the neuronal networks and a few food webs show high communicability between a relatively 
higher number of pairs of nodes within the network.

\begin{figure}[!]
\begin{center}
\includegraphics[scale=.33]{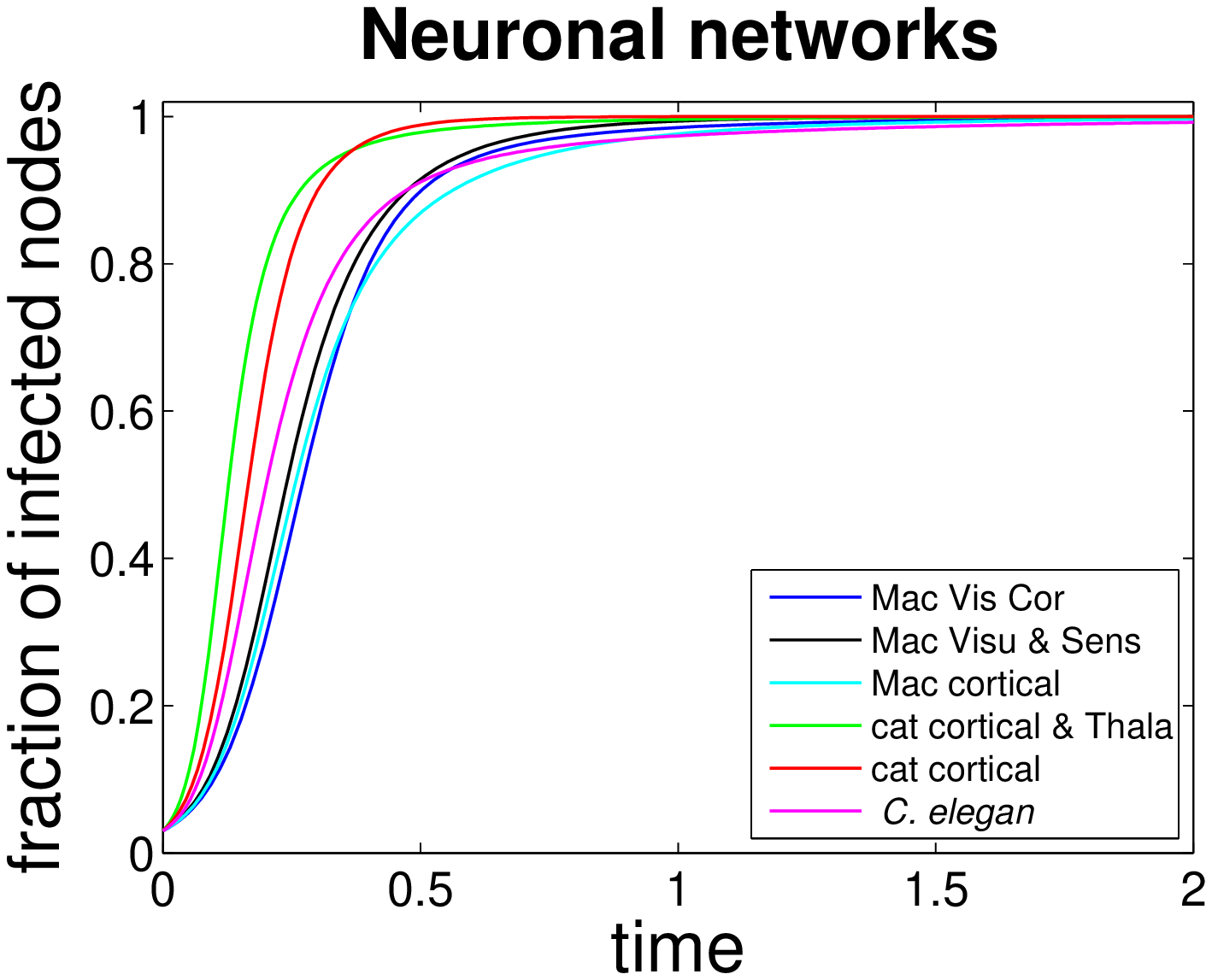}\includegraphics[scale=.33]{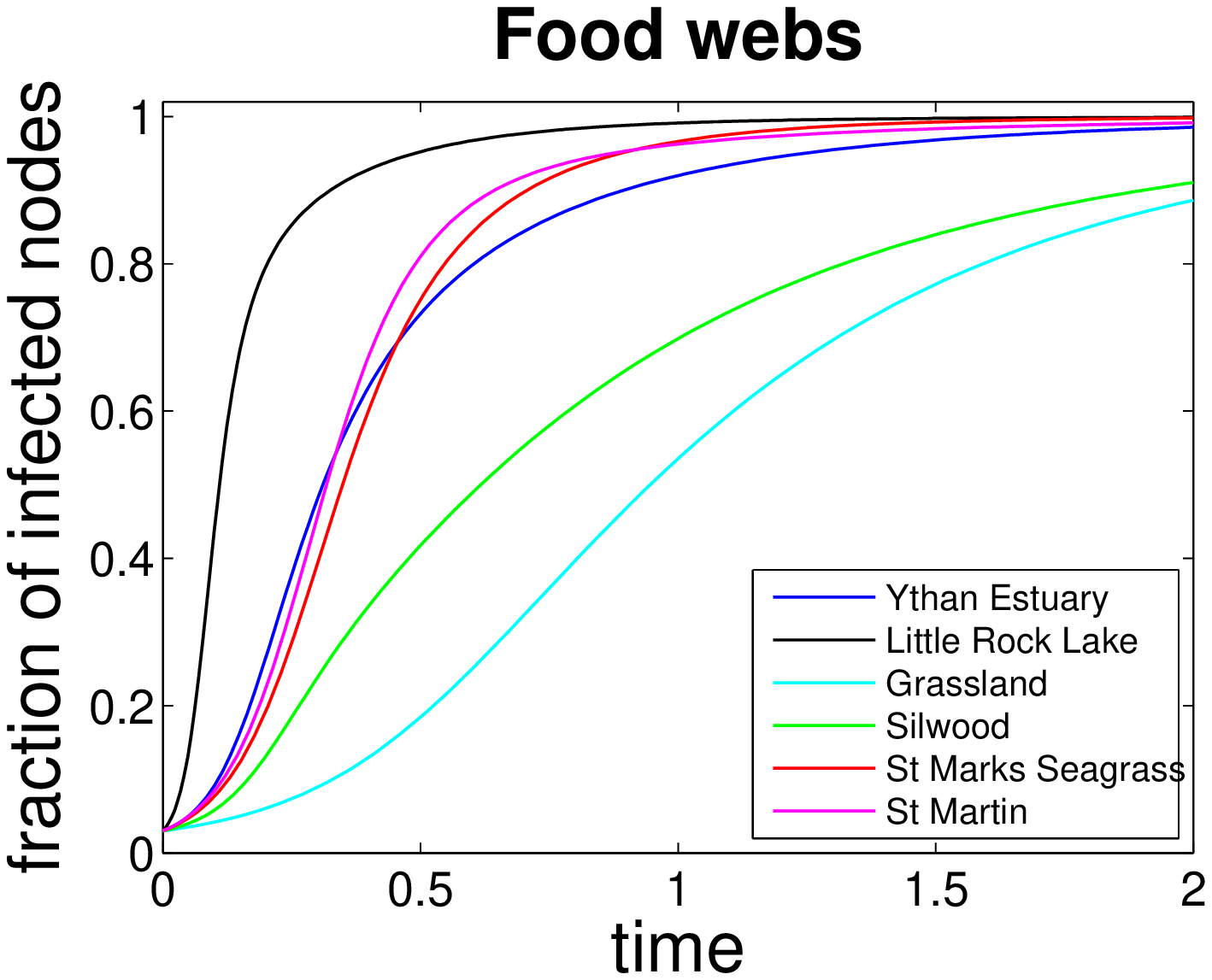}\includegraphics[scale=.33]{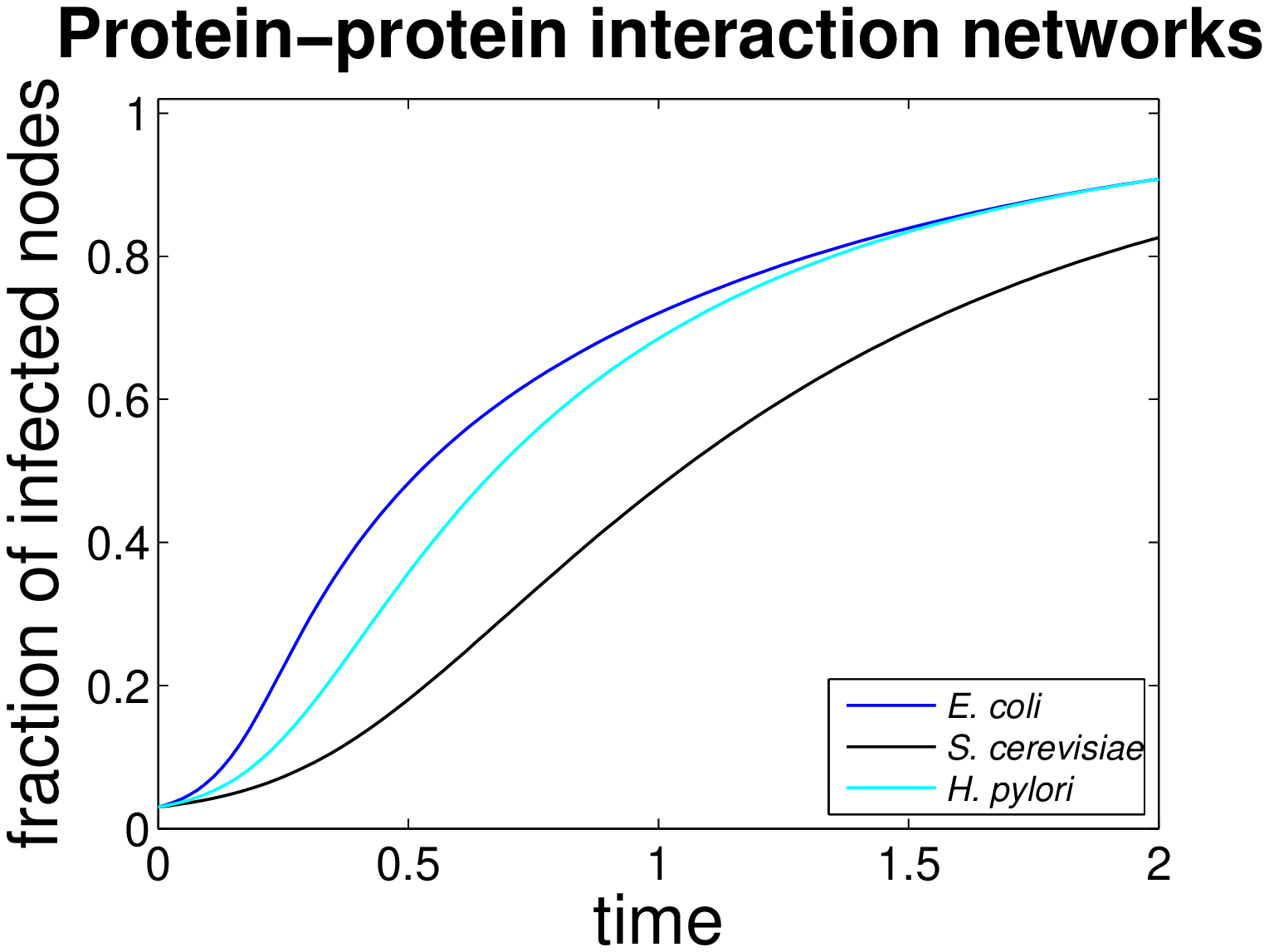}\\
\vspace{-3.2cm}
\hspace{1.5cm}(a)\hspace{5.2cm}(b)\hspace{4cm}(c)\\
\vspace{3.00cm}
\includegraphics[scale=.33]{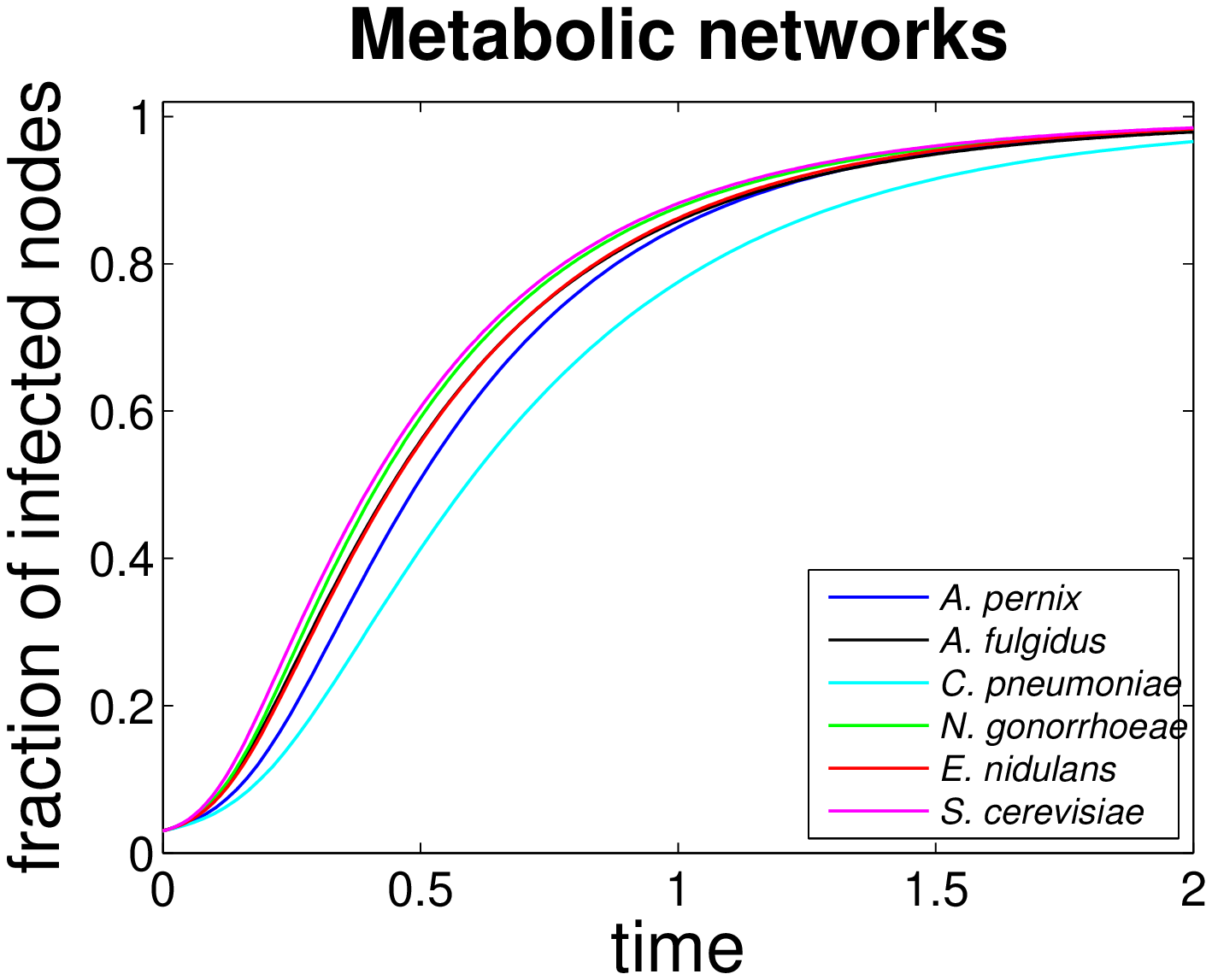}\includegraphics[scale=.33]{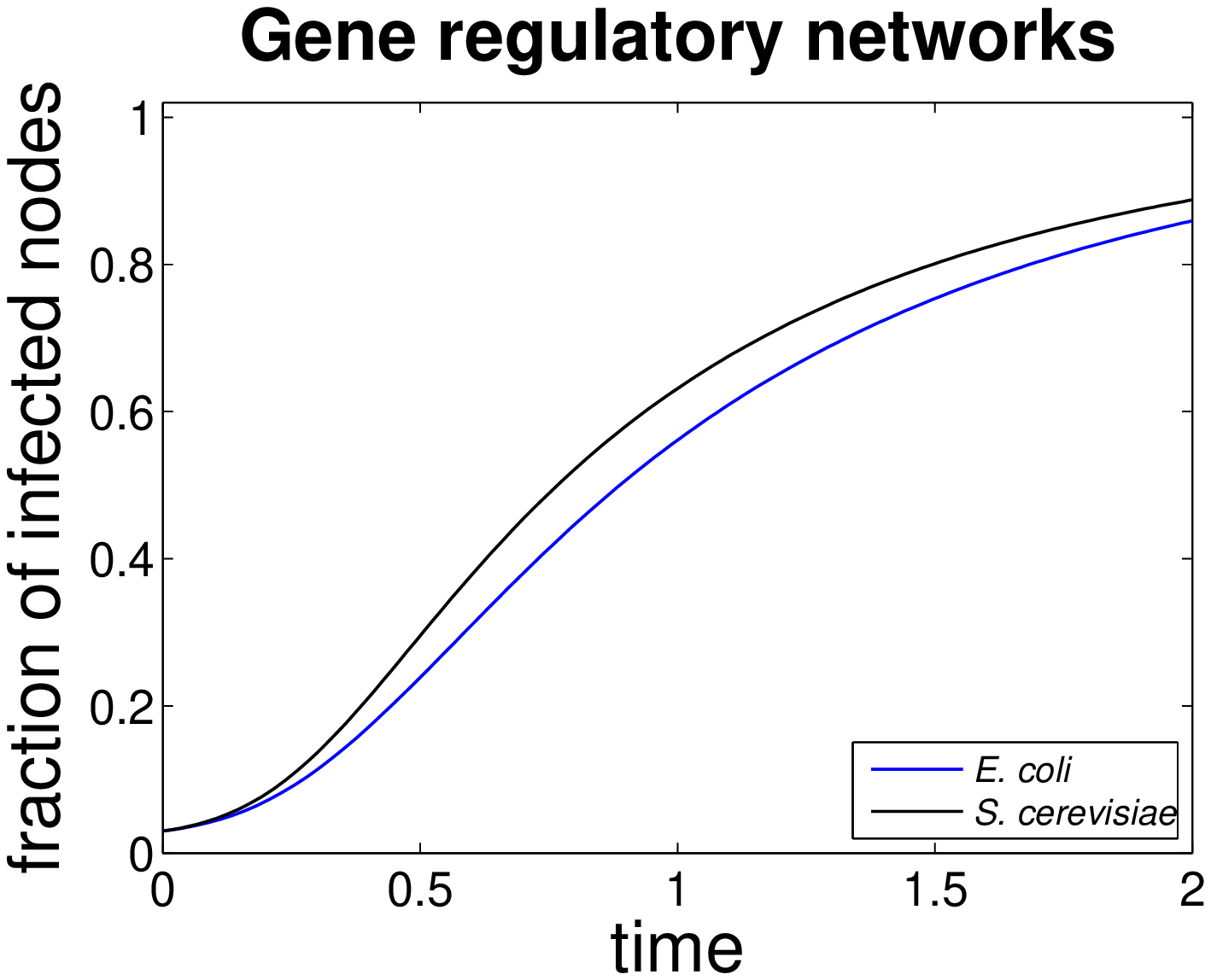}\\
\vspace{-3.4cm}
\hspace{-2.4cm}(d)\hspace{4.8cm}(e)\\
\vspace{2.4cm}
\end{center}
\caption{{\bf Dynamics of epidemics in SI model}. X-axis represents time and 
Y-axis represents fraction of infected nodes in a network. (a) {\bf Neuronal networks}: (macaque visual cortex area, 
macaque large-scale visual and sensorimotor area corticocortical connectivity, macaque cortical connectivity, 
cat cortical area, cat cortical and thalamic areas, {\it C. elegans}), (b) {\bf Food webs}: (Ythan Estuary, 
 Little Rock Lake, Grassland, Silwood, St Marks Seagrass, St Martin),
 (c) {\bf Protein-protein interaction networks}: ({\it E. coli}, {\it S. cerevisiae}, {\it H. pylori}), (d) {\bf Metabolic networks}:
         archaea ({\it A. pernix}, {\it A. fulgidus}), eukaryota ({\it E. nidulans}, {\it S. cerevisiae}), bacteria ({\it C. pneumoniae}, {\it N. gonorrhoeae}), 
(g) {\bf Gene regulatory networks}: ({\it E. coli}, {\it S. cerevisiae}).}
\label{sineuronal}
\end{figure}

\begin{figure}[!]
\begin{center}
\includegraphics[scale=.33]{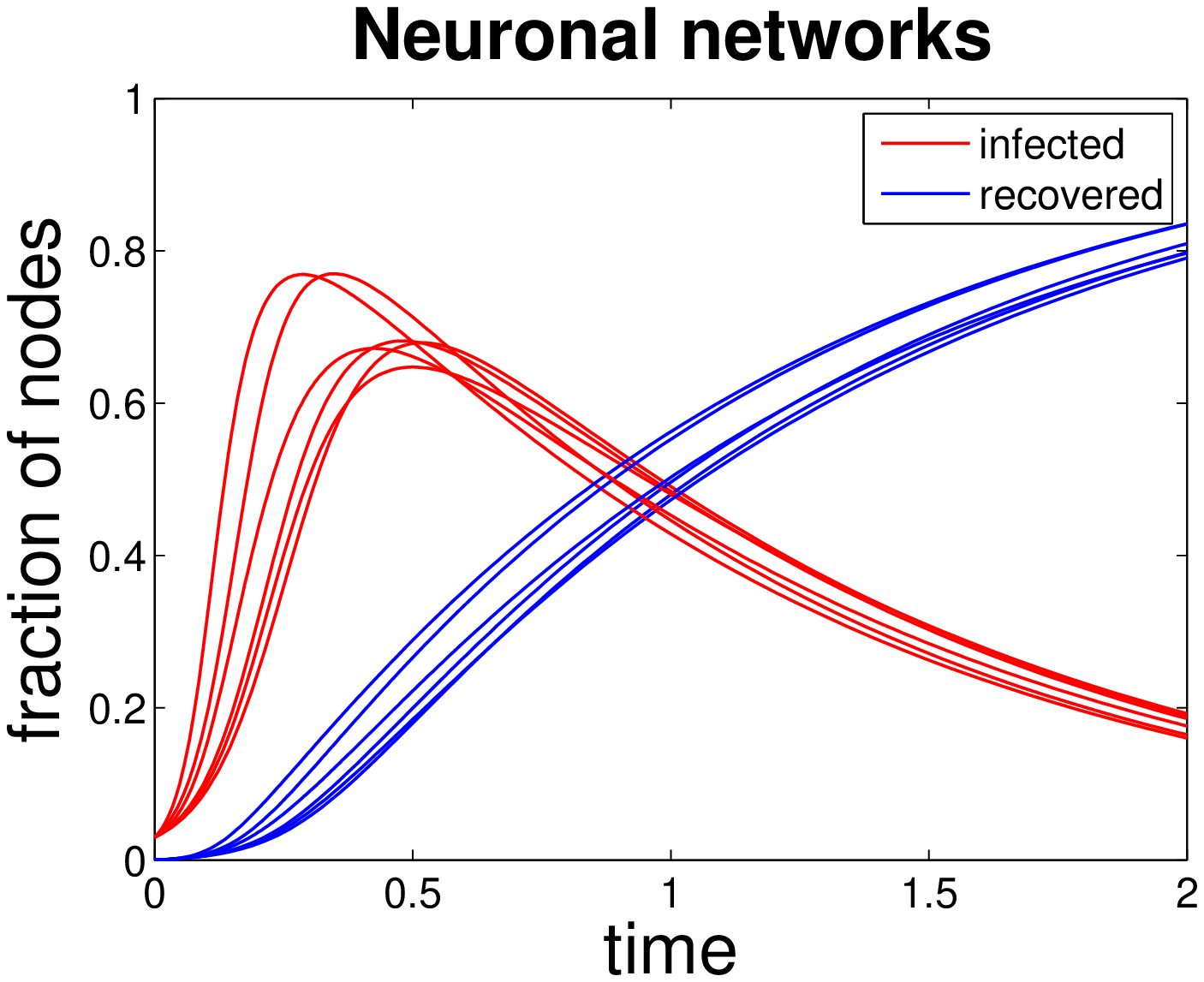}\includegraphics[scale=.33]{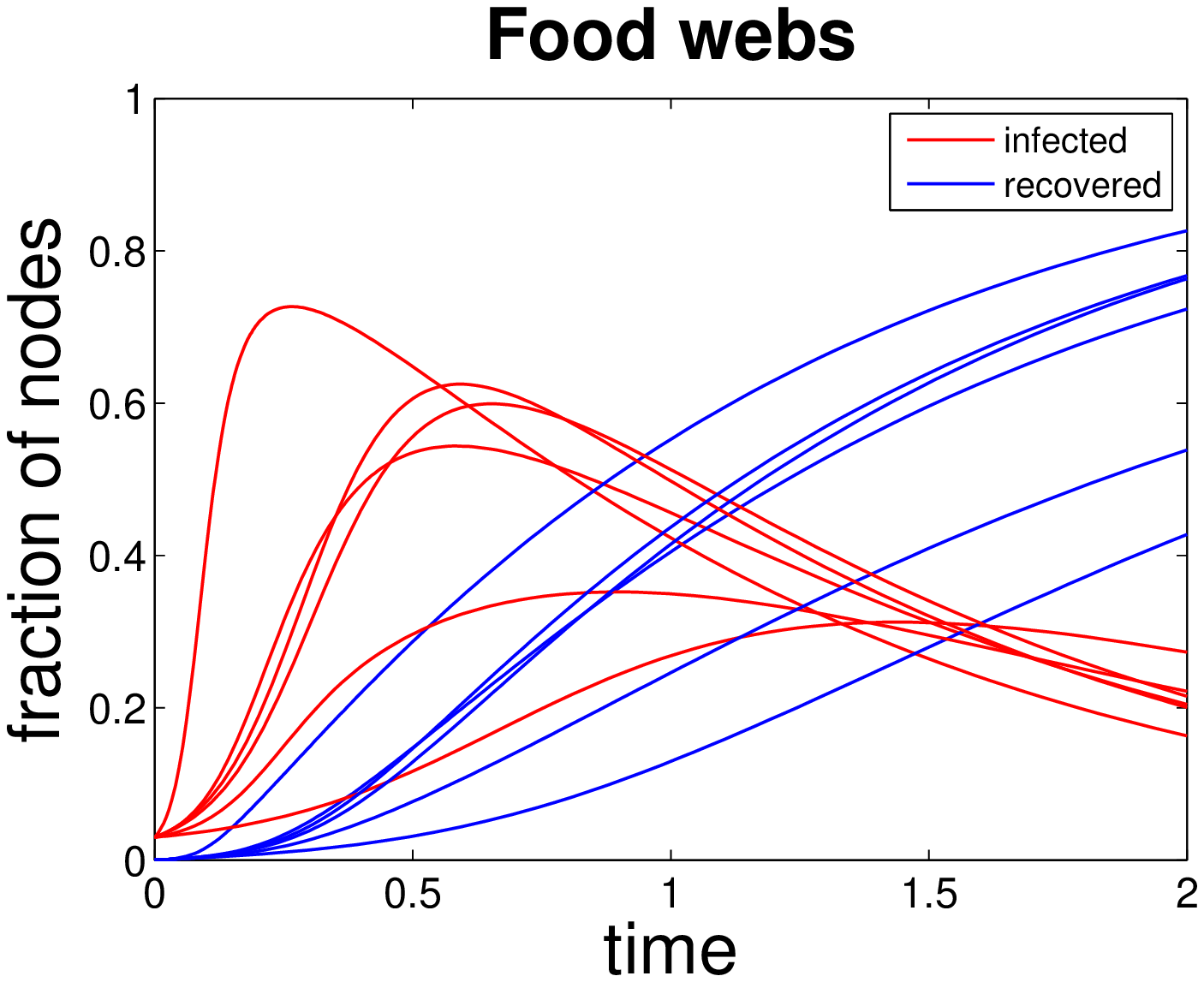}\includegraphics[scale=.33]{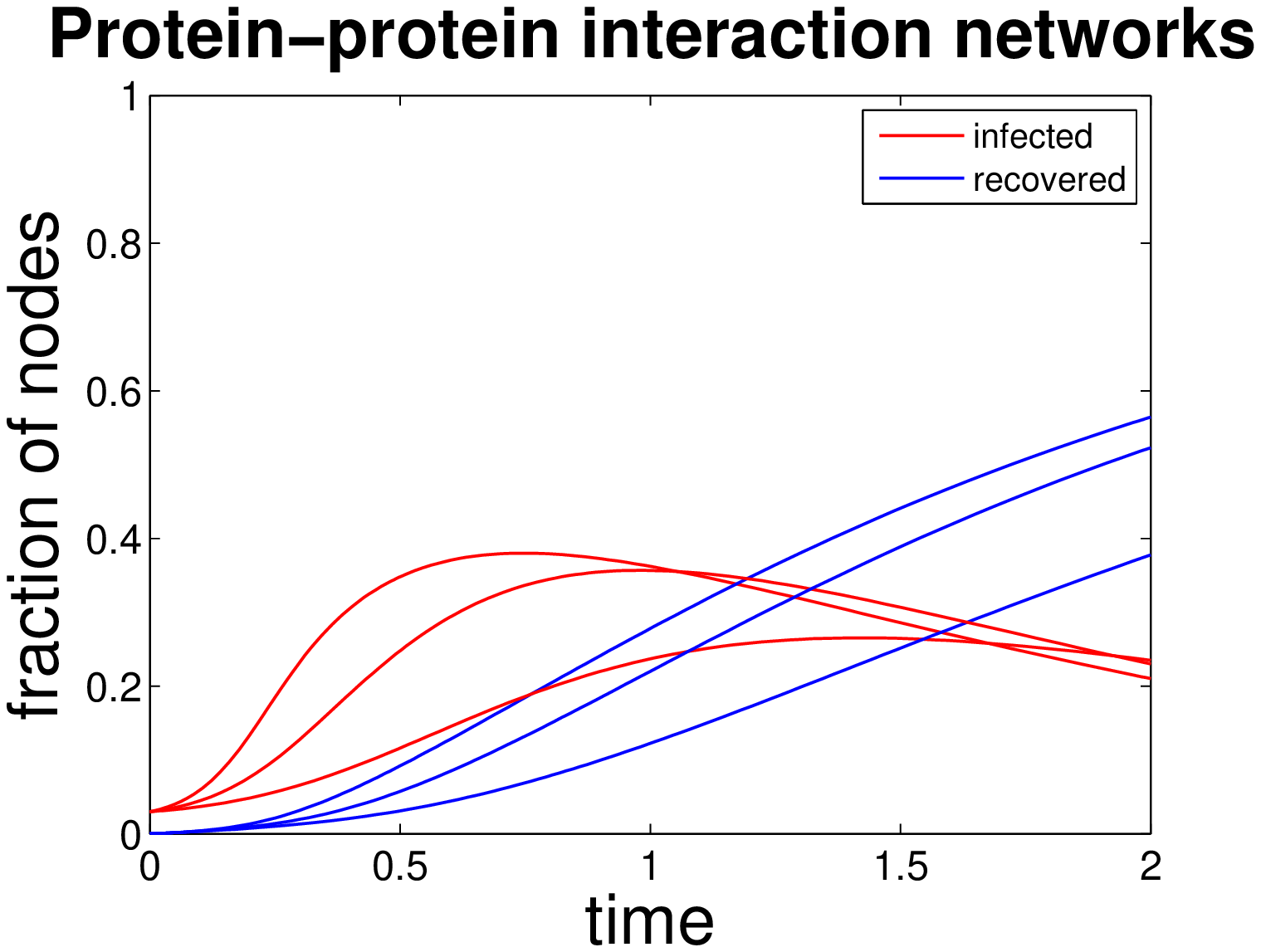}\\
\vspace{-3.45cm}
\hspace{.8cm}(a)\hspace{4.8cm}(b)\hspace{4.6cm}(c)\\
\vspace{3.2cm}
\includegraphics[scale=.33]{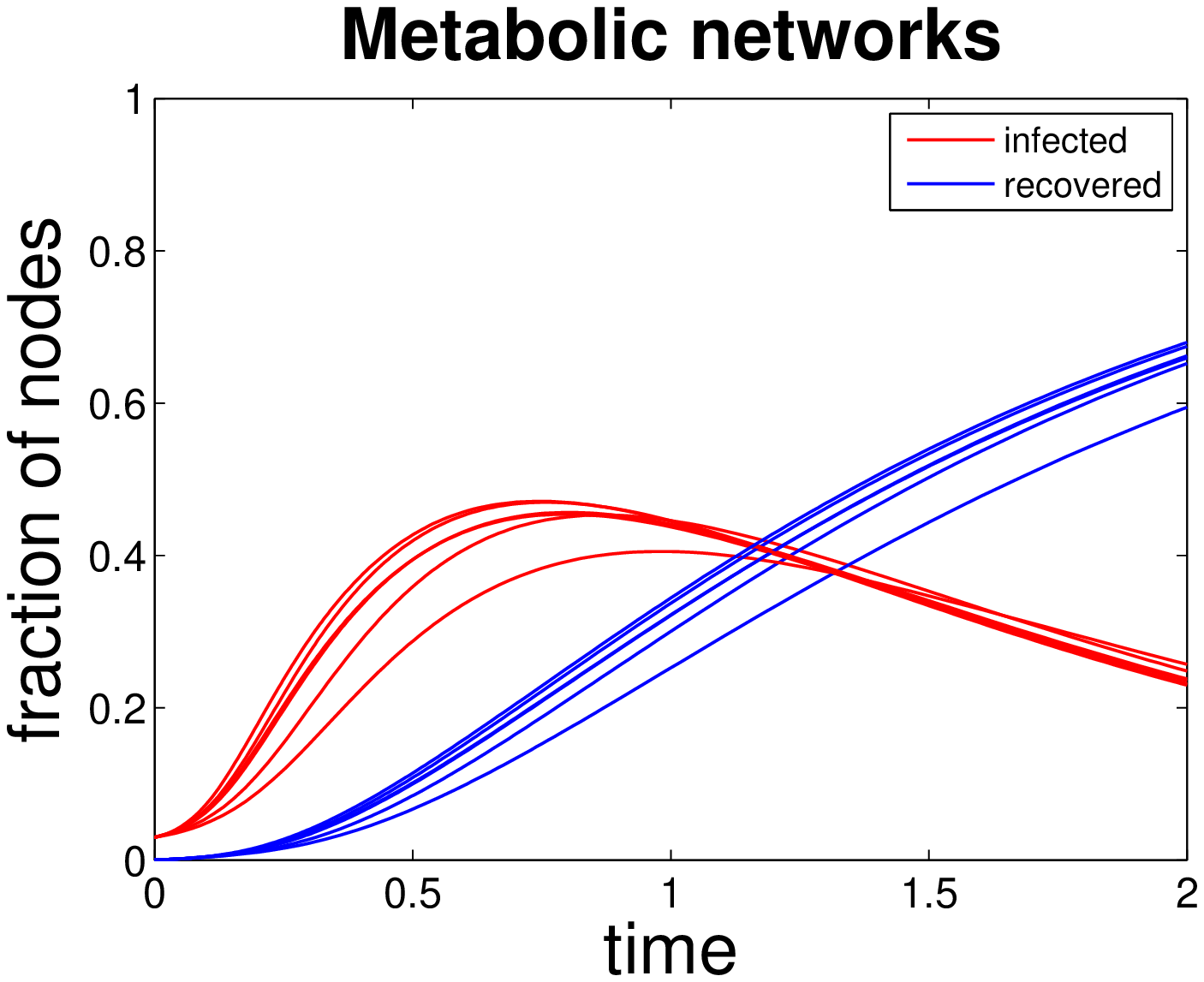}\includegraphics[scale=.33]{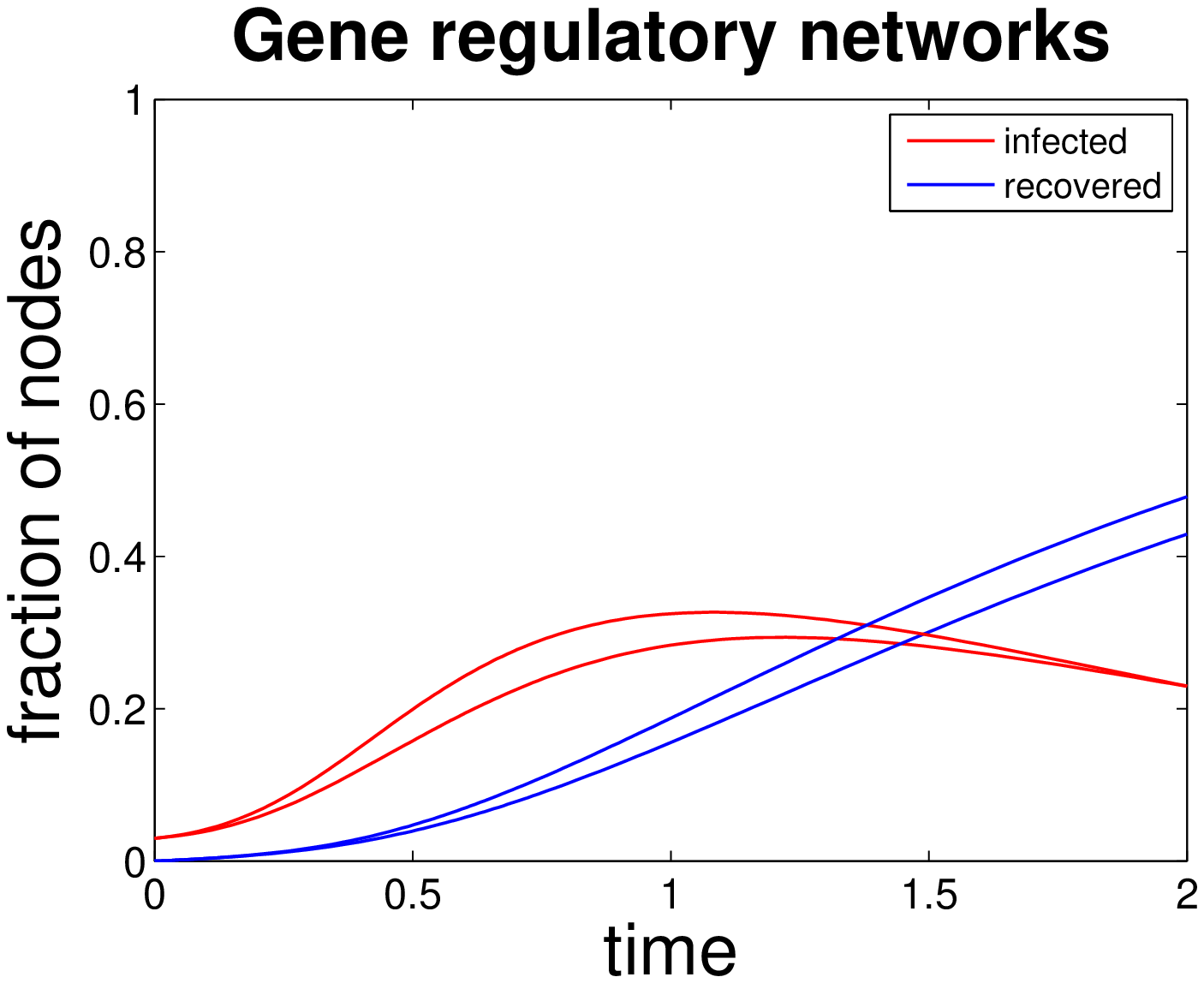}\\
\vspace{-3.4cm}
\hspace{-2.4cm}(d)\hspace{4.8cm}(e)\\
\vspace{2.4cm}
\end{center}
\caption{{\bf Dynamics of epidemics in SIR model}.
X-axis represents time and Y-axis represents fraction of nodes that are infected by red line and recovered by blue line. (a) {\bf Neuronal networks}: (macaque visual cortex area, 
macaque large-scale visual and sensorimotor area corticocortical connectivity, macaque cortical connectivity, 
cat cortical area, cat cortical and thalamic areas, {\it C. elegans}), (b) {\bf Food webs}: (Ythan Estuary, 
 Little Rock Lake, Grassland, Silwood, St Marks Seagrass, St Martin),
 (c) {\bf Protein-protein interaction networks}: ({\it E. coli}, {\it S. cerevisiae}, {\it H. pylori}), (d) {\bf Metabolic networks}:
         archaea ({\it A. pernix}, {\it A. fulgidus}), eukaryota ({\it E. nidulans}, {\it S. cerevisiae}), bacteria ({\it C. pneumoniae}, {\it N. gonorrhoeae}), 
(g) {\bf Gene regulatory networks}: ({\it E. coli}, {\it S. cerevisiae}).}
\label{sirneuronal}
\end{figure}

\begin{figure}[!]
\begin{center}
\includegraphics[scale=.33]{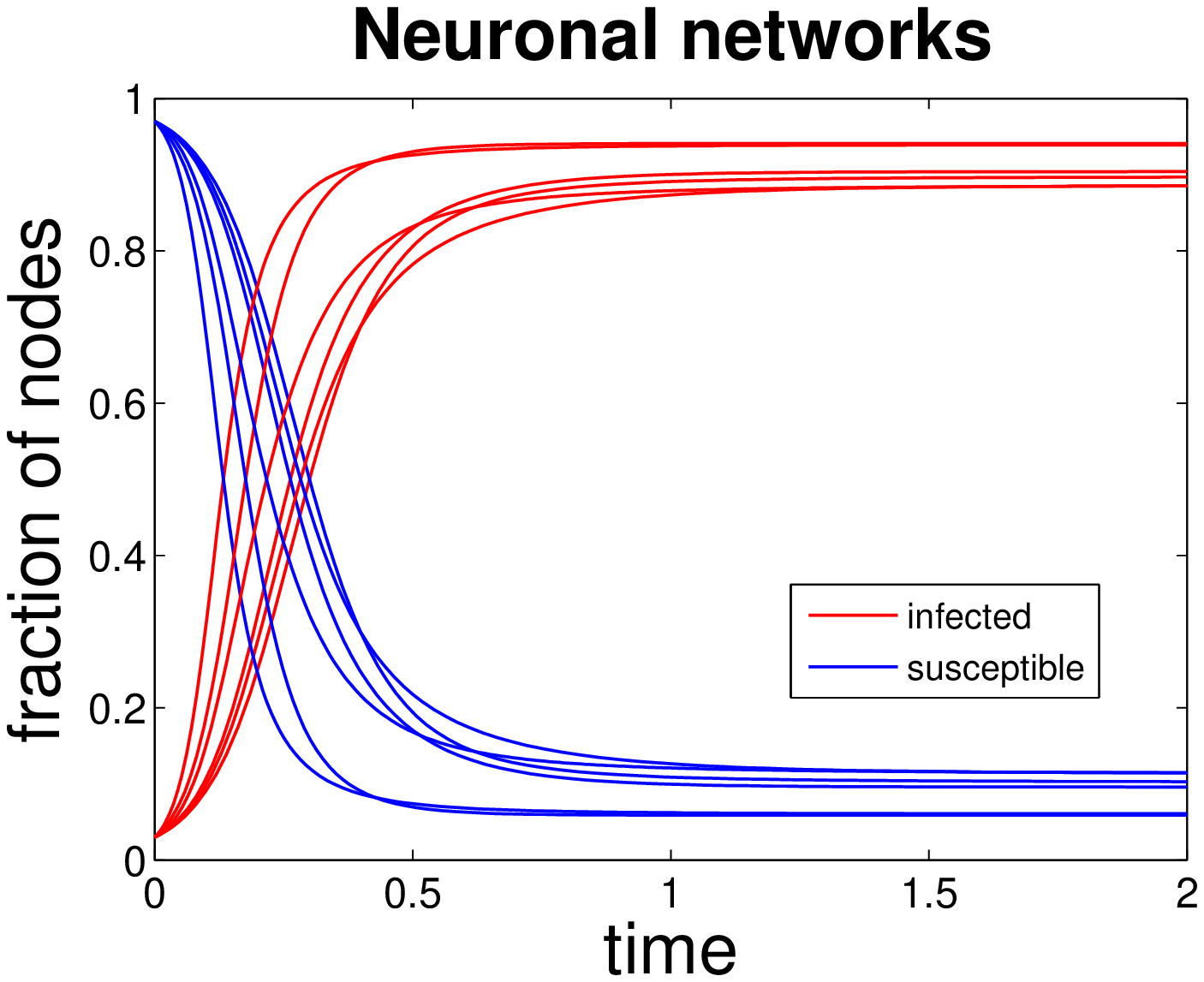}\includegraphics[scale=.33]{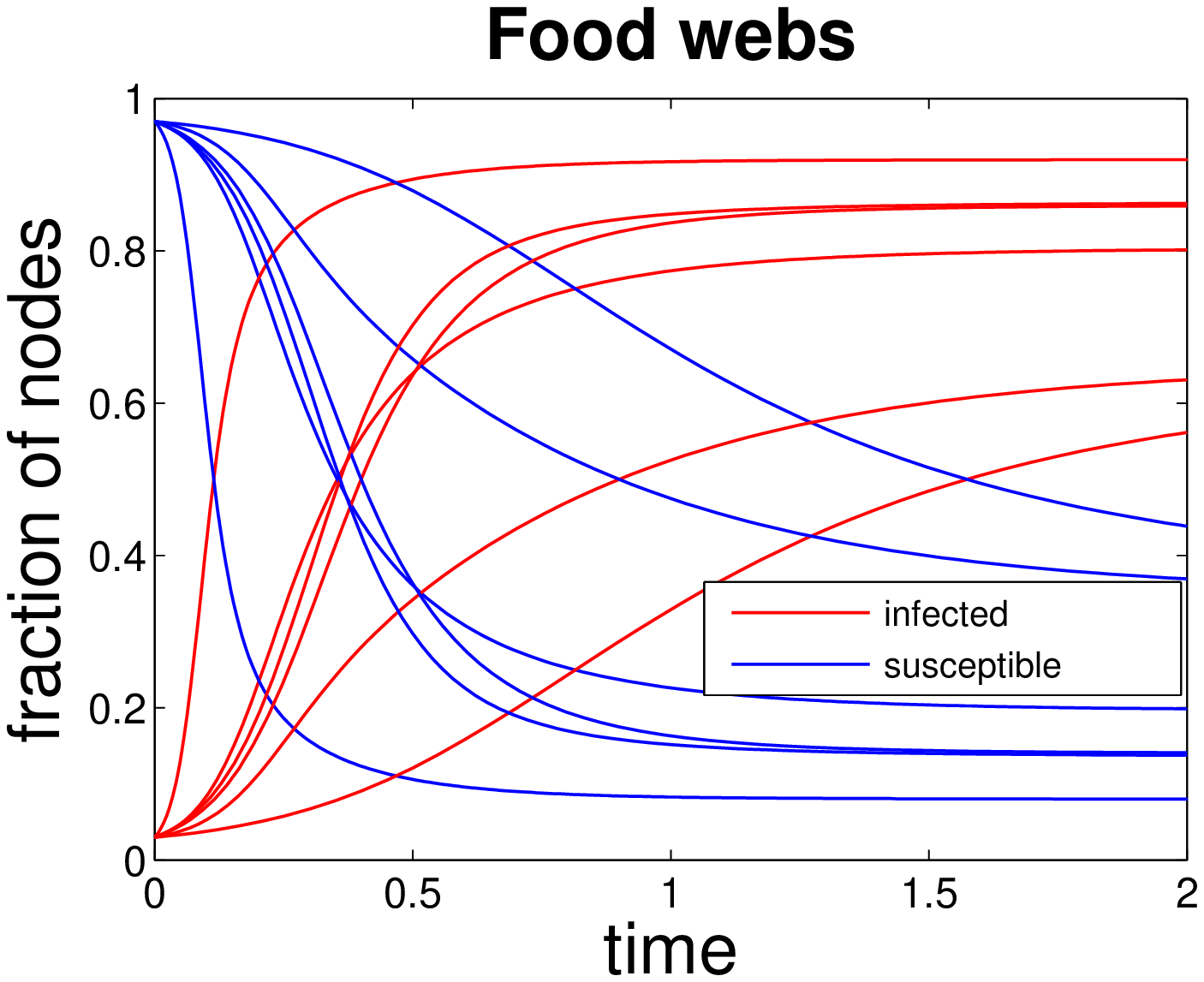}\includegraphics[scale=.33]{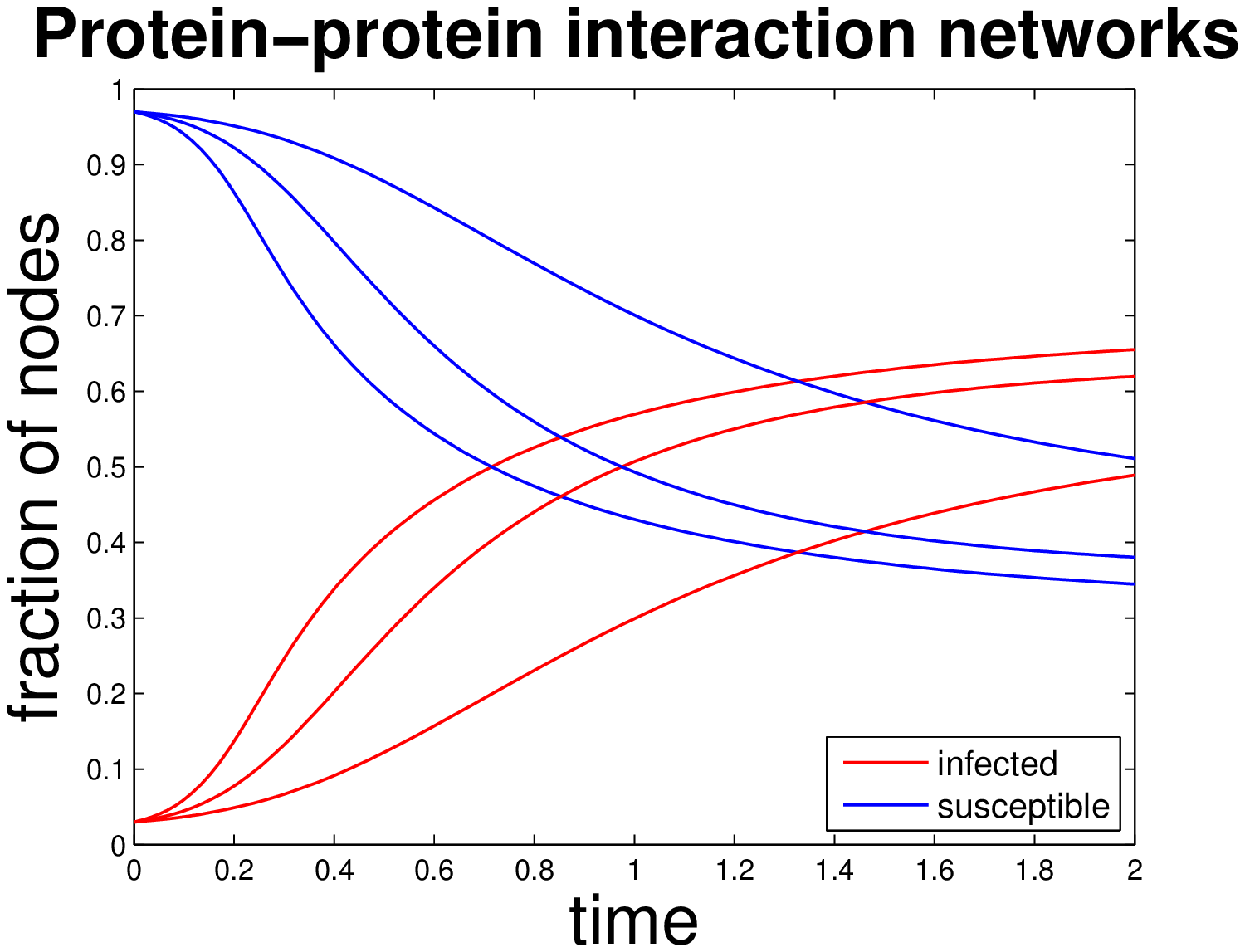}\\
\vspace{-3cm}
\hspace{2cm}(a)\hspace{5.5cm}(b)\hspace{2.5cm}(c)\\
\vspace{2.8cm}
\includegraphics[scale=.33]{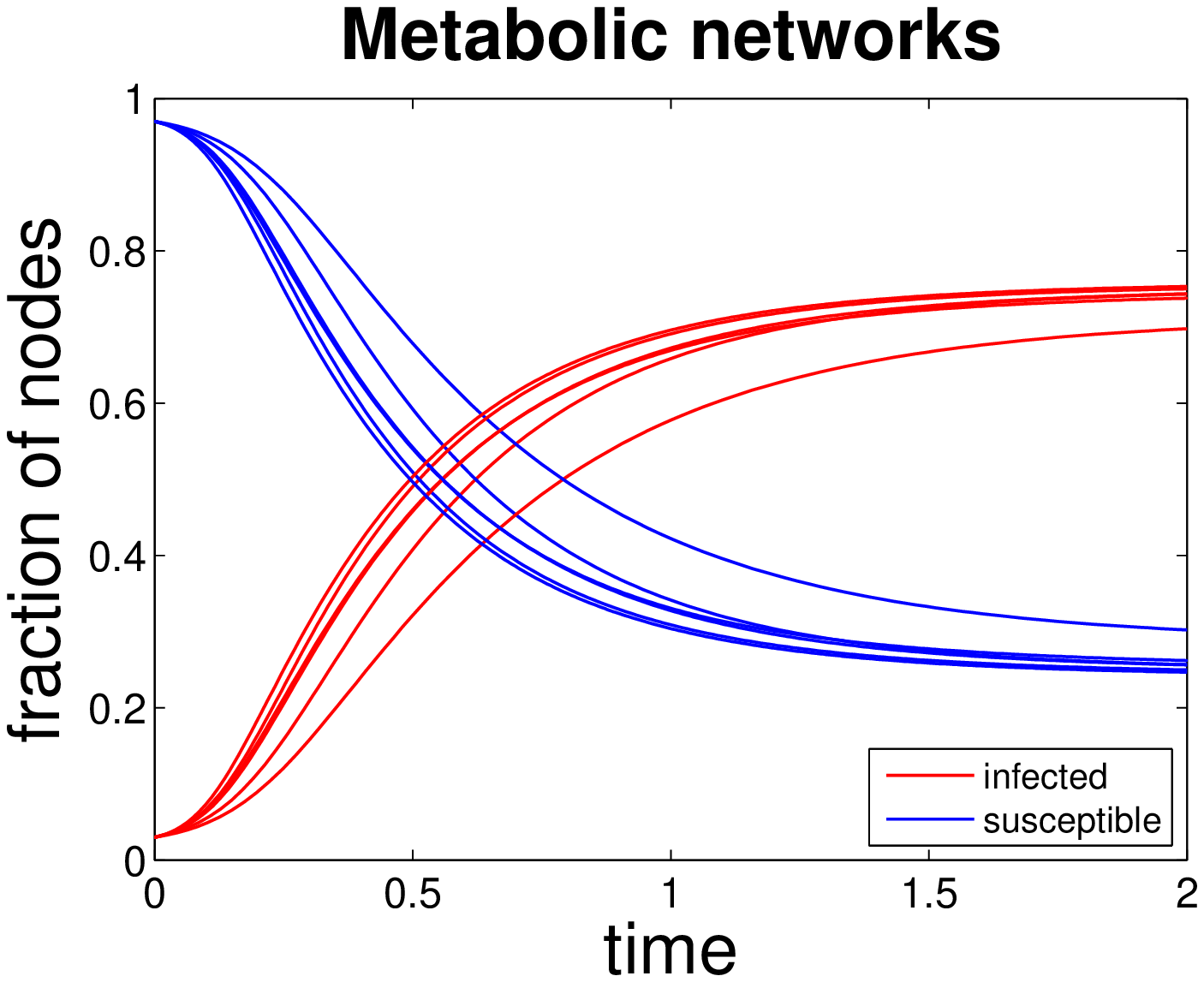}\includegraphics[scale=.33]{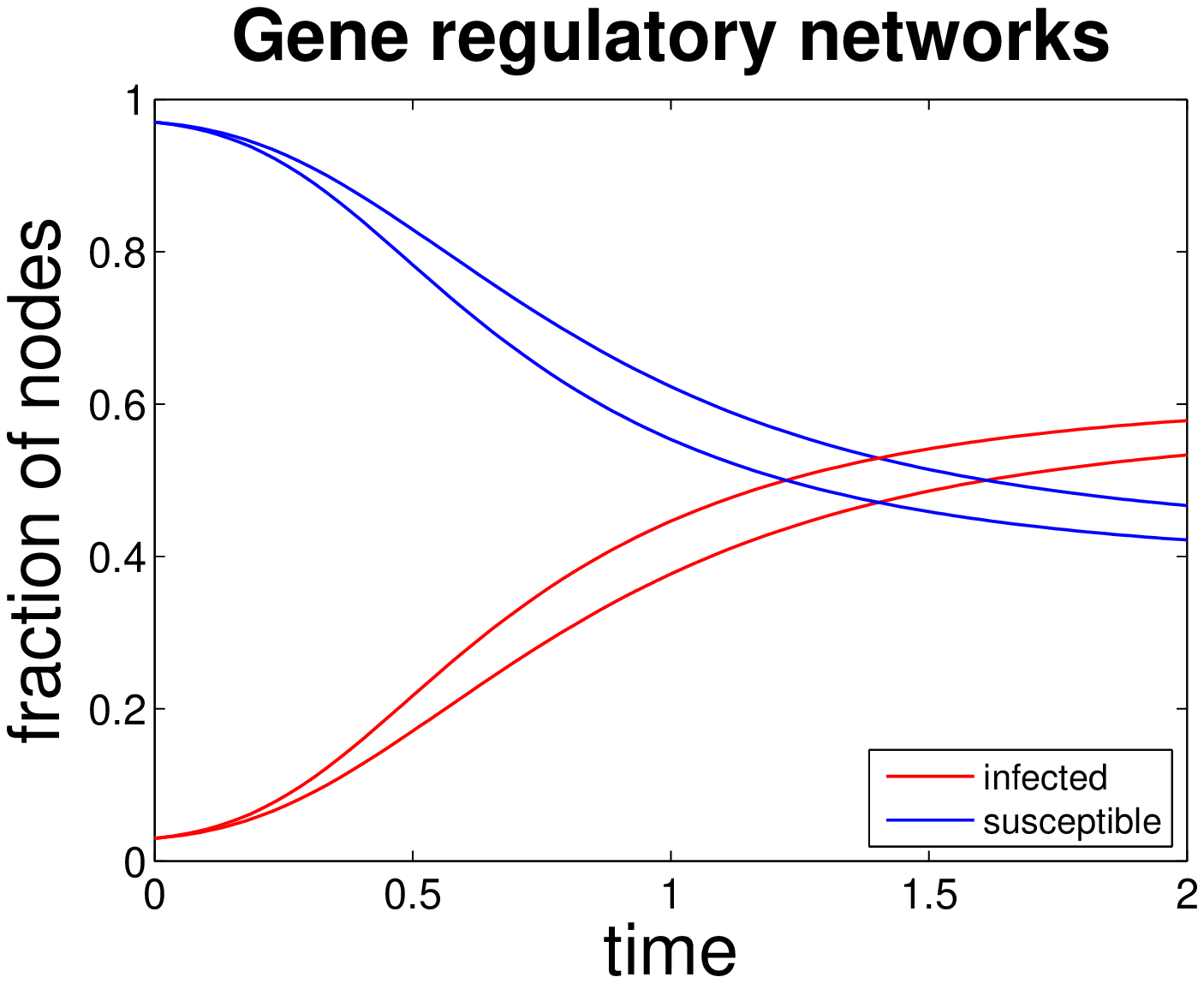}\\
\vspace{-3.2cm}
\hspace{-2cm}(d)\hspace{4cm}(e)\\
\vspace{2.2cm}
\end{center}
\caption{{\bf Dynamics of epidemics in SIS model}. X-axis represents time and Y-axis represents fraction of nodes that are infected by red line and susceptible by blue line.
(a) {\bf Neuronal networks}: (macaque visual cortex area, 
macaque large-scale visual and sensorimotor area corticocortical connectivity, macaque cortical connectivity, 
cat cortical area, cat cortical and thalamic areas, {\it C. elegans}), (b) {\bf Food webs}: (Ythan Estuary, 
 Little Rock Lake, Grassland, Silwood, St Marks Seagrass, St Martin),
 (c) {\bf Protein-protein interaction networks}: ({\it E. coli}, {\it S. cerevisiae}, {\it H. pylori}), (d) {\bf Metabolic networks}:
         archaea ({\it A. pernix}, {\it A. fulgidus}), eukaryota ({\it E. nidulans}, {\it S. cerevisiae}), bacteria ({\it C. pneumoniae}, {\it N. gonorrhoeae}), 
(g) {\bf Gene regulatory networks}: ({\it E. coli}, {\it S. cerevisiae}).}
\label{sisneuronal}
\end{figure}

While studying the three epidemic models, we observe that in the SI model, the infection spreads faster on the underlying undirected
structure of all neuronal networks than that of the other biological networks (see Figure~\ref{sineuronal}).
In the SIR model, the results show that the entire underlying undirected structure of all neuronal networks get infected, 
and also recover more rapidly compared to that of the other biological networks (see Figure~\ref{sirneuronal}).  
Similar results also hold in the SIS model for neuronal networks compared to the rest of the biological networks. 
In neuronal networks, states change quickly from susceptible to infected, and back again to susceptible, compared to the
other biological networks studied here (see Figure~\ref{sisneuronal}).

\subsection*{Structural basis of information transfer:}
We see that the underlying undirected structure of neuronal networks show high communicability and possess a good information
spreading characteristic which is derived from the spectral gap. An epidemic can also spread faster on neuronal networks 
than the other biological networks studied here. Thus the underlying undirected architecture of a neuronal network possesses  
certain conformation which is favourable for  spreading of  different entities or information. 

Now, to explore what are the topological characteristics that make the underlying undirected structure of a neuronal 
network very supportive for faster spreading of information, we investigate the small-world property and the small-world-ness
of all the networks.  We also study the same by randomizing the network, while conserving the degree sequence, to
understand how small-world-ness relatively varies  across a family of networks with the same degree sequence as
in the given  (undirected) network structure. To further investigate the architecture across  various biological networks,
we decompose the underlying undirected structure of a network into cores or shells.

\subsubsection*{The small-world property  and small-world-ness:}

In a small-world network, two nodes may not be directly
connected, but, one can be reached from the other by a finite number of steps.
Usually we see that small-world networks have a low average shortest path length and a high clustering coefficient (or transitivity) (\cite{sw98}). 
  A measure is defined on small-world property, called small-world-ness \cite{humphries08}, as
$$SW_{G}=\frac{T_{G}/ L_{G}}{ T_{ER}/ L_{ER}},$$
where $T_{G}, L_{G}$ are the transitivity and average 
shortest path length of the network $G$ respectively. $T_{ER}$, $L_{ER}$ denote the same quantities for an Erd\"{o}s-R\'{e}nyi's 
random graph  with the same number of vertices and edges as  $G$ (see \cite{erdos59,erdos60}). 
It is considered that  the network $G$ has the small-world
property  if,
		$$SW_{G}>1$$.

Obviously, if a network $G$ has the small-world property, the ratio ($T_{G}/L_{G}$) is
 strictly higher than ($T_{ER}/L_{ER}$).

The underlying undirected structure of all the biological networks, studied here, have the small-world property. Now, we 
perform a relative study between small-world-ness of a network $G$ with its family $F_G$
for measuring the quality of the small-world property in $G$. A family $F_G$ of a network $G$, is 
a group of randomly generated networks which not only have the same number of vertices and
edges but also have the same degree sequence as that of $G$.
We  see that not only the underlying undirected structure of all biological networks have the 
small-world property, but also, the families of all those networks  possess the same property.
For a qualitative study we define the z-score of small-world-ness of a network $G$ as

                   $$Z_{G}=\frac{SW_{G}-<SW_{F_G}>}{std(SW_{F_G})},$$
                   
where $SW_{G}$ is the small-world-ness of the network $G$, $<SW_{F_G}>$
is the mean of small-world-ness of the family $F_G$
and $std(SW_{F_G})$ is the standard deviation of the family $F_G$. \\

\begin{figure}[!]
\begin{center}
\includegraphics[scale=.33]{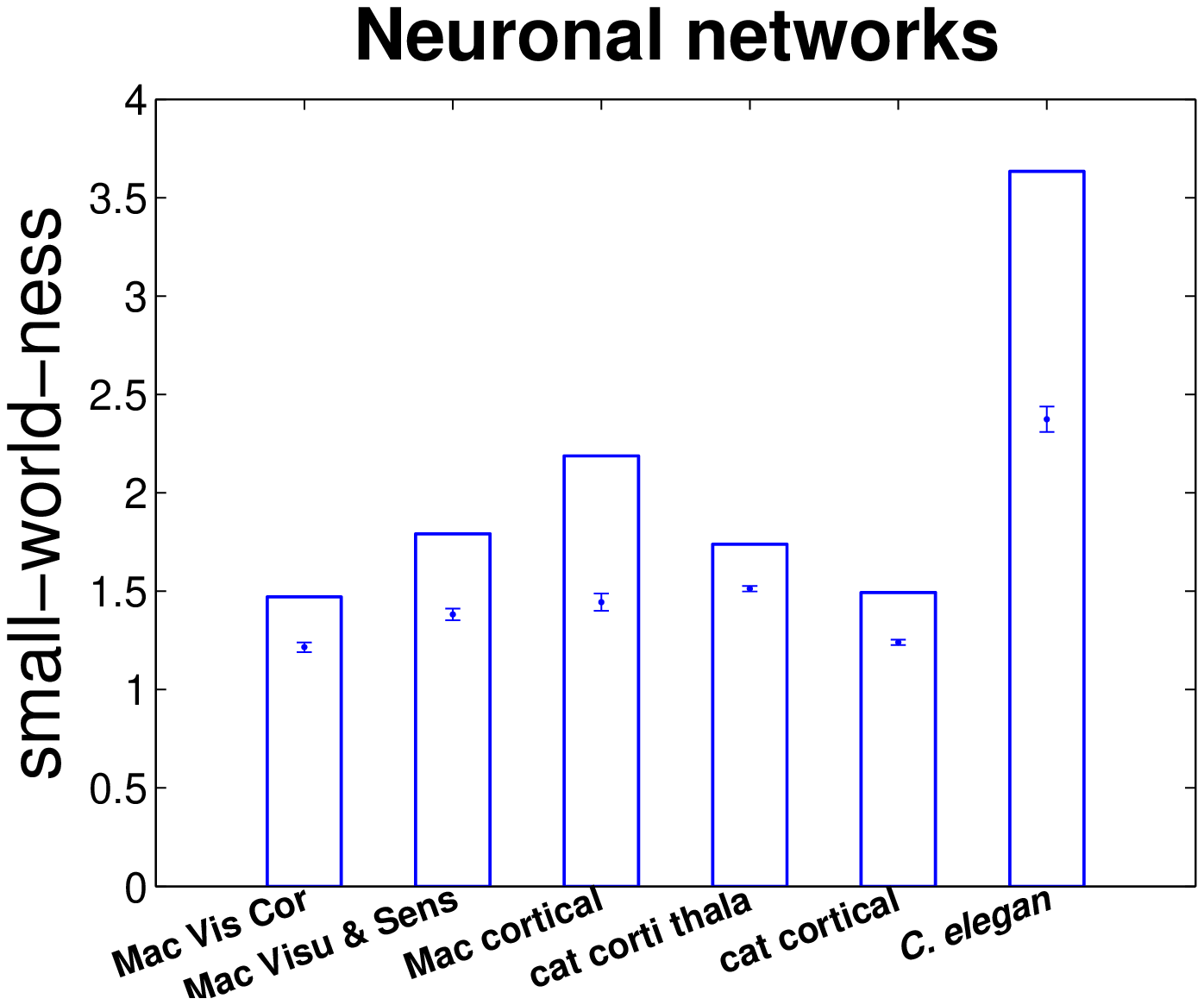}\includegraphics[scale=.33]{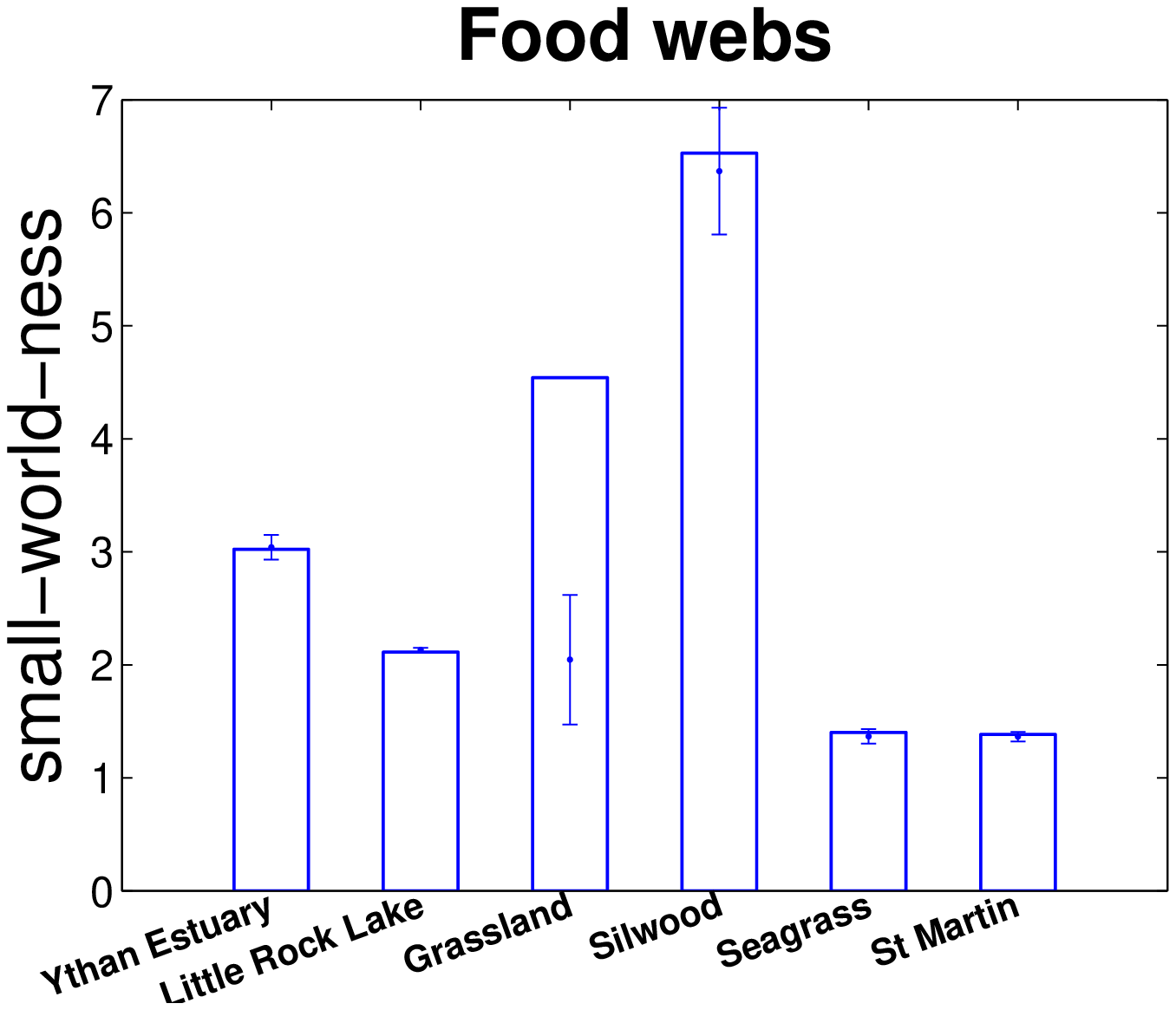}\includegraphics[scale=.33]{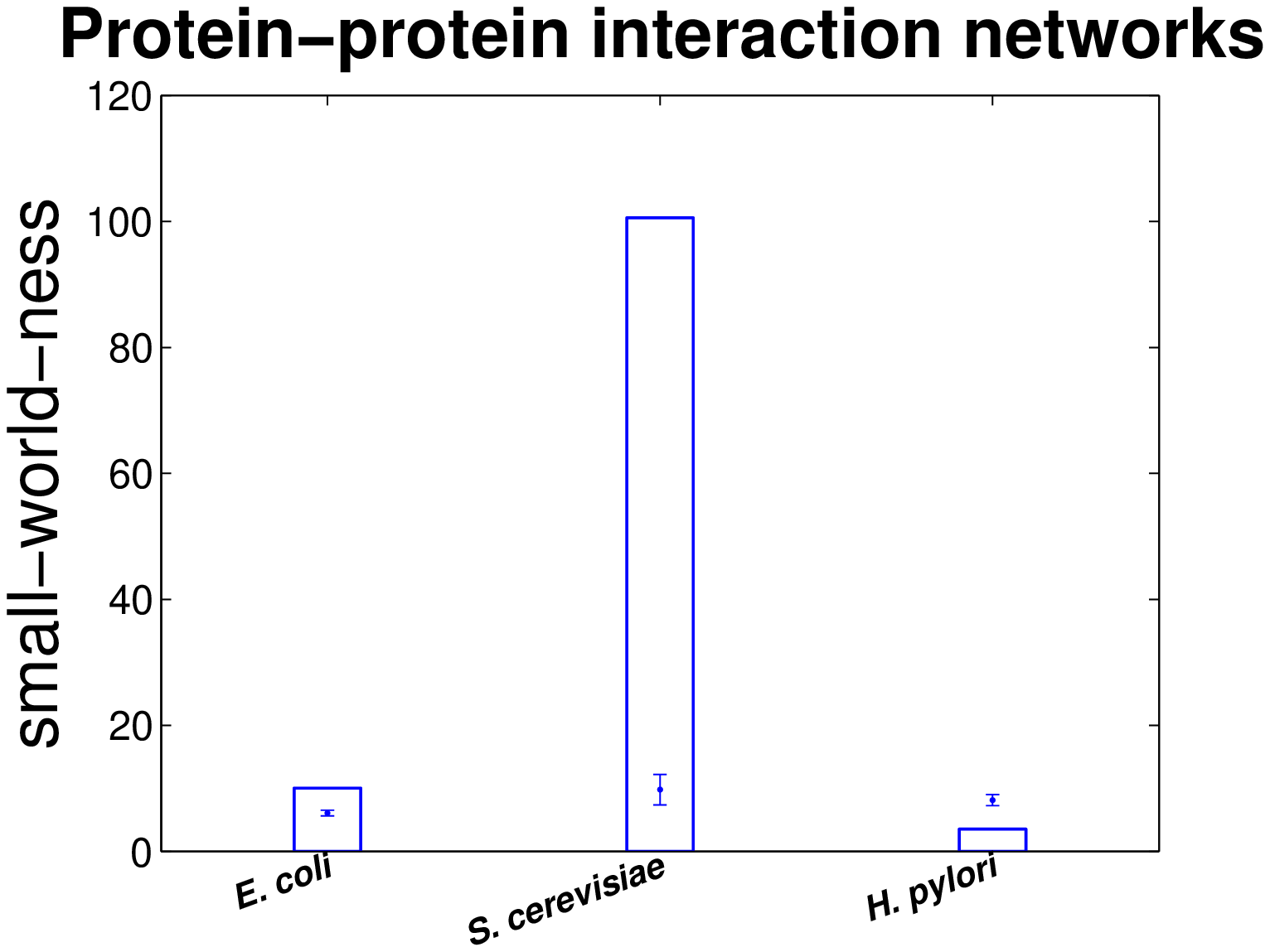}\\
\vspace{-3.4cm}
\hspace{.8cm}(a)\hspace{4.8cm}(b)\hspace{4.6cm}(c)\\
\vspace{3.2cm}
\includegraphics[scale=.33]{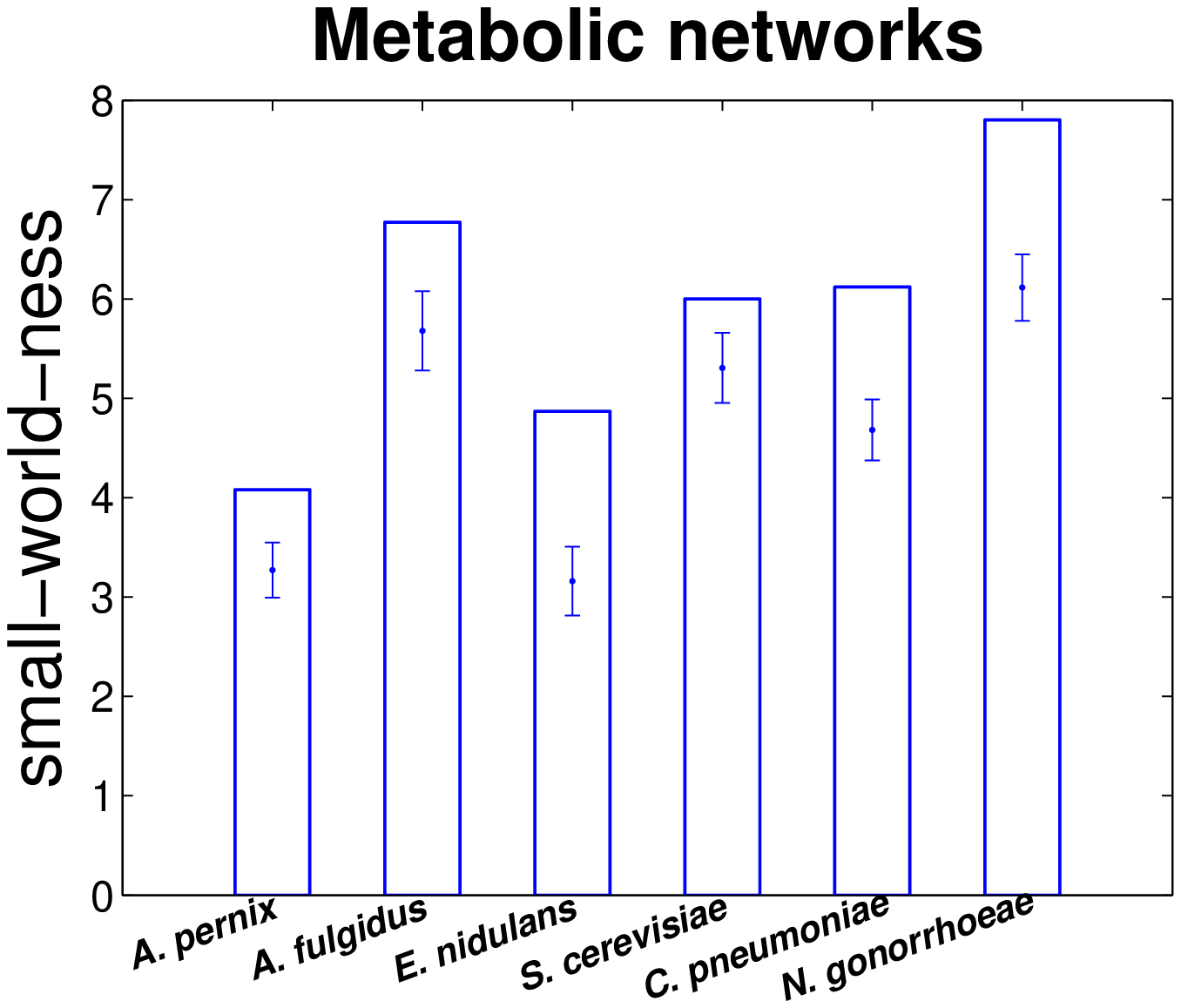}\includegraphics[scale=.33]{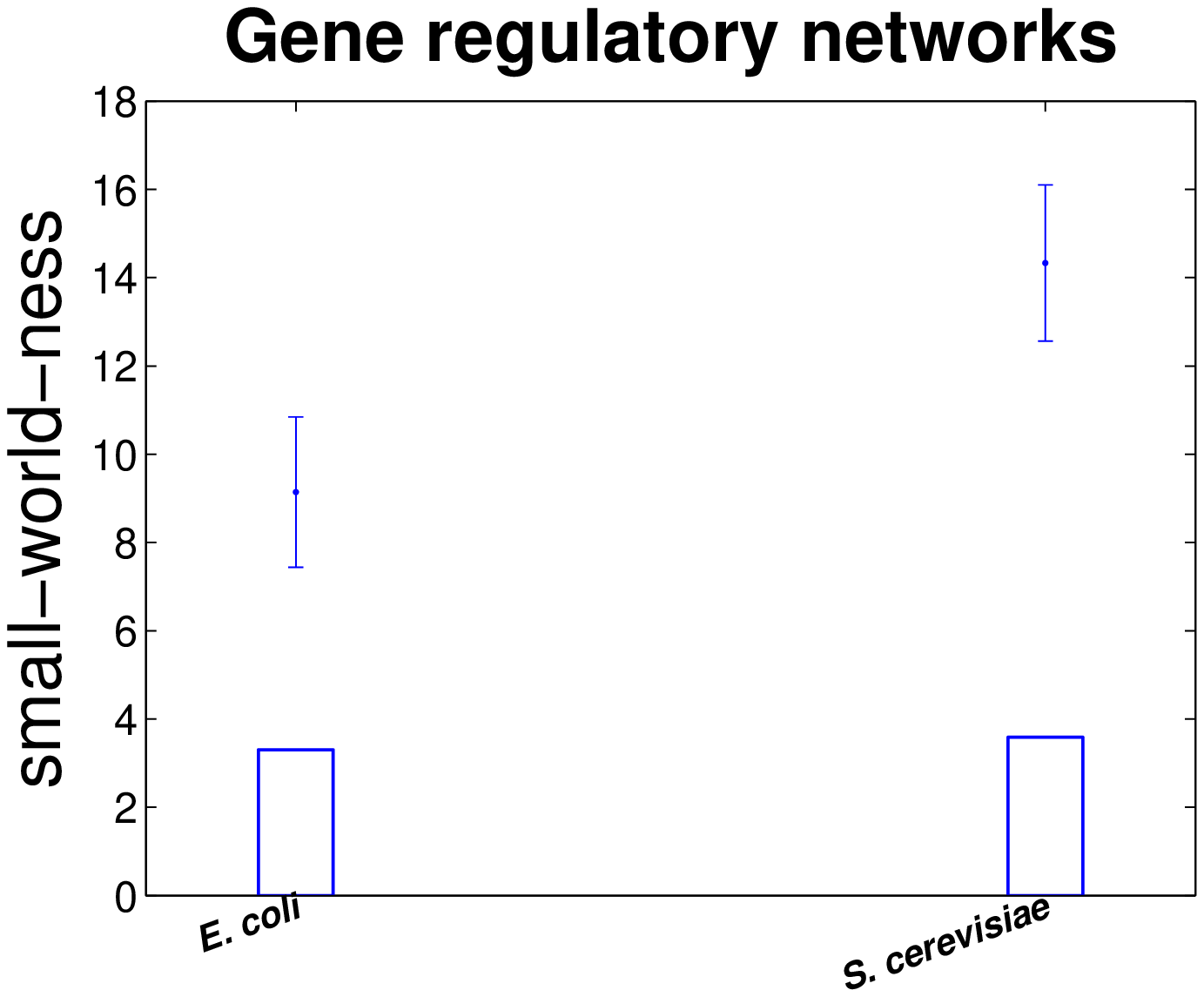}\\
\vspace{-3.4cm}
\hspace{-2.6cm}(d)\hspace{4.8cm}(e)\\

\vspace{2.4cm}
\end{center}
\caption{{\bf Histogram of small-world-ness of all biological networks}. Here, a bin represents the small-world-ness of a network $G$ and errorbar
represents the variability of small-world-ness of the family $F_G$ of $G$. $F_G$ is a set of networks have the same number of nodes and degree sequence as $G$ has.
X-axis represents different networks of same class and y-axis 
represents small-world-ness of the corresponding networks. (a) {\bf Neuronal networks}: (macaque visual cortex area, 
macaque large-scale visual and sensorimotor area corticocortical connectivity, macaque cortical connectivity,
cat cortical area, cat cortical and thalamic areas, {\it C. elegans}), (b) {\bf Food webs}: (Ythan Estuary,
Little Rock Lake, Grassland, Silwood, St Marks Seagrass, St Martin),
 (c) {\bf Protein-protein interaction networks}: ({\it E. coli},
 {\it S. cerevisiae}, {\it H. pylori}), (d) {\bf Metabolic networks}:
         archaea ({\it A. pernix}, {\it A. fulgidus}), eukaryota ({\it E. nidulans}, 
 {\it S. cerevisiae}), bacteria ({\it C. pneumoniae}, {\it N. gonorrhoeae}), 
(g) {\bf Gene regulatory networks}: ({\it E. coli}, {\it S. cerevisiae})}.
\label{swness}
\end{figure}

We  observe that all the neuronal networks have a  positive  and very high
z-score (see  Table~\ref{swness_table}, Figure~\ref{swness}(a)).  Four among six food webs
(Grassland, Silwood, St Marks Seagrass and St Martin) have positive, but low, z-scores and the rest possess negative z-scores (see Table ~\ref{swness_table},
Figure ~\ref{swness}(b)). 
Among all protein-protein interaction networks, 
{\it E. coli} and {\it S. cerevisiae} have positive z-scores unlike
{\it H. Pylori}, which has a negative z-score (see Figure~\ref{swness}(c)).
We study metabolic networks from three different domains, namely Archaea, Bacteria and Eucaryota. All of them 
have positive, but not high z-scores (see Figure~\ref{swness}(d),(e), (f)).
All the gene regulatory networks studied here, {\it E. coli} and {\it S. cerevisiae }, have negative
z-scores (see Figure~\ref{swness}(g)).

These z-scores topologically signify that the small-world-ness of each neuronal (undirected) network is higher than
that of its family of networks and also possesses a highly positive z-score. Two gene regulatory networks {\it E. coli} 
and {\it S. cerevisiae} and one protein-protein interaction network {\it S. cerevisiae} show varied characteristics,
 and their small-world-ness is drastically different from their family. So among all biological networks, studied here,
the underlying undirected structure of a neuronal network has special conformation. Not only, it has the small-world property,
but also, it is expressed remarkably to a higher degree than any randomly generated network with the same (undirected) 
degree sequence. Thus, we see that the (undirected) structure of a neuronal network is more suitable for communication 
and information transfer.

The results above do not vary much even if the same study is done by generating 100, 200, or 300 networks in a family.
Here, we show all the results over 200 realizations. 

\subsubsection*{\emph{k}-core decomposition:}
It was considered that the dynamics of spreading information is very fast on  a network having high degree nodes. 
Later on, it was shown that the vertices, which spread information efficiently, are not those with high degree or high betweenness centrality, 
but those that belong to a high \emph{k}-core in the network (\cite{kitsak10}). 

A \emph{k}-core of a graph $G$ is a maximal induced subgraph such that the degree of any vertex in that subgraph is grater than or equal to \emph{k}.
Thus a \emph{k}-core of a graph can be obtained by recursively removing all the vertices of degree less than $k$, until the degrees of all nodes become at least $k$. 
A vertex with high degree may not belong to a high core, e.g. the centre vertex in a star graph is not located in a \emph{k}-core for $k>1$.
A vertex or node  is assigned a \textit{shell index} or equivalently \textit{coreness} $k$, if it belongs to a \emph{k}-core but not a \emph{k+1}-core. 
All the vertices with shell index $k$ form a $k$-\textit{shell} $S_k$. 
( for more details on \emph{k}-core decomposition see \cite{newmanbook}).\\

\begin{figure}[!]
\begin{center}
\includegraphics[scale=.33]{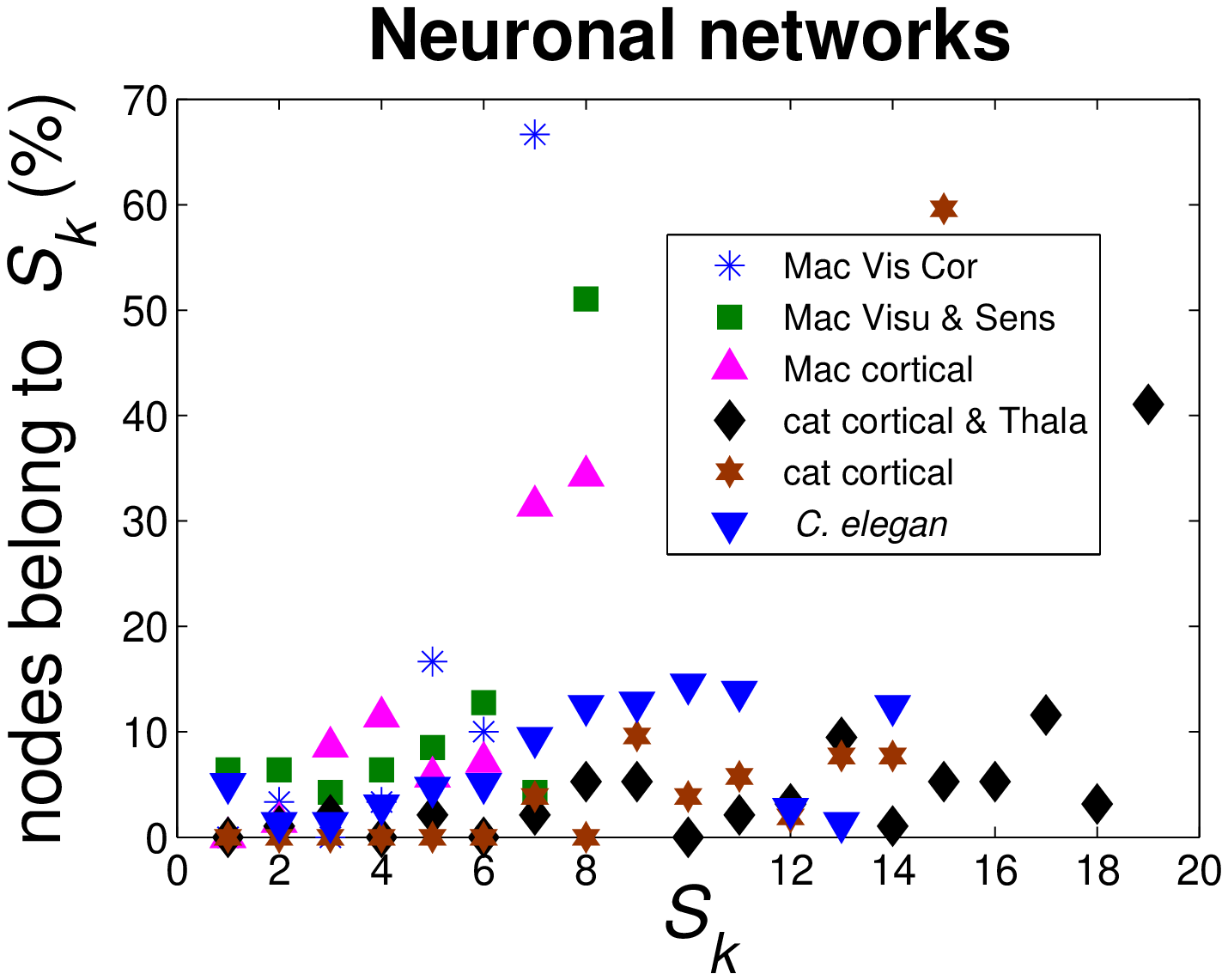}\includegraphics[scale=.33]{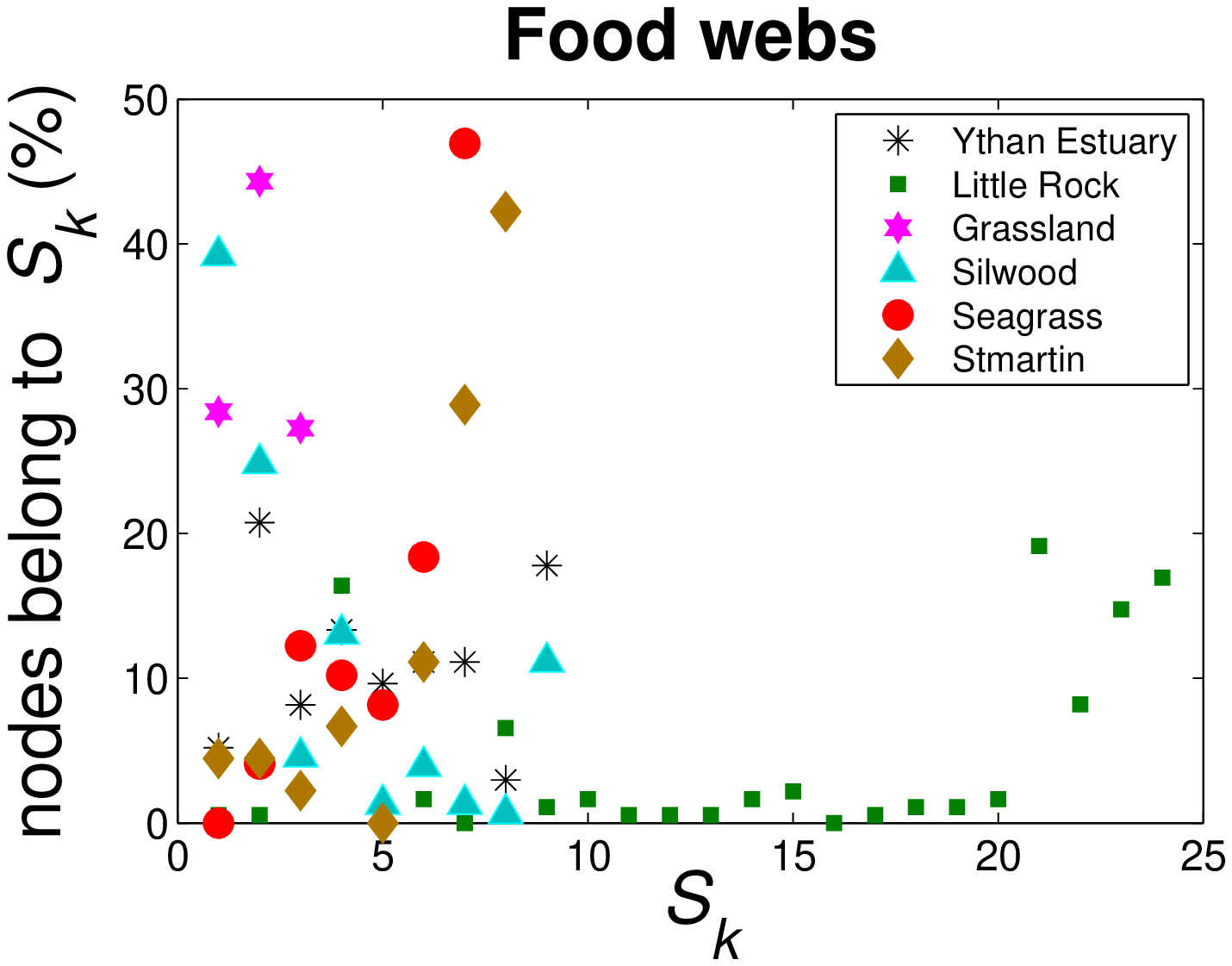}\includegraphics[scale=.33]{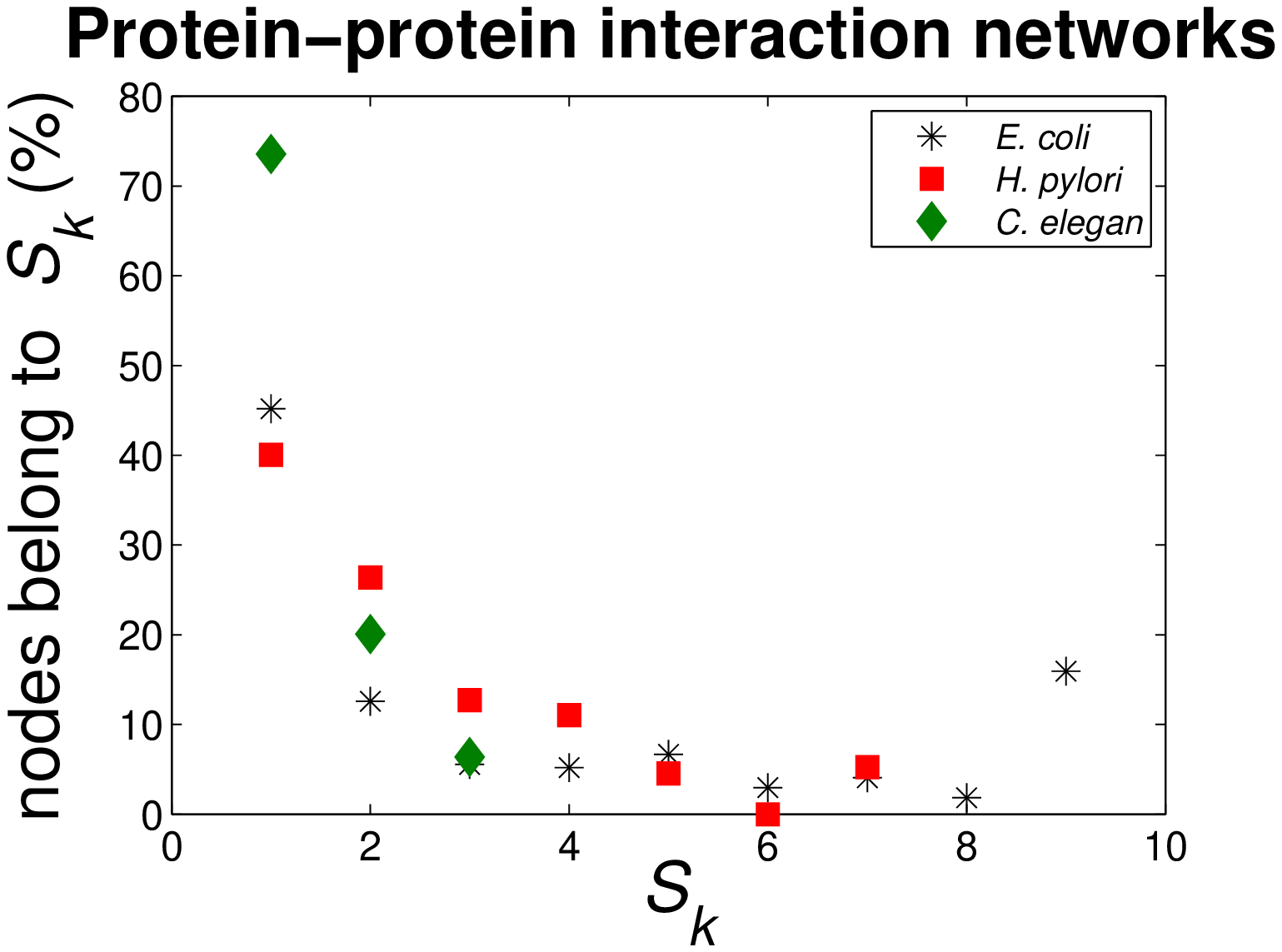}\\
\vspace{-3.4cm}
\hspace{.8cm}(a)\hspace{4.8cm}(b)\hspace{4.6cm}(c)\\
\vspace{3.2cm}
\includegraphics[scale=.33]{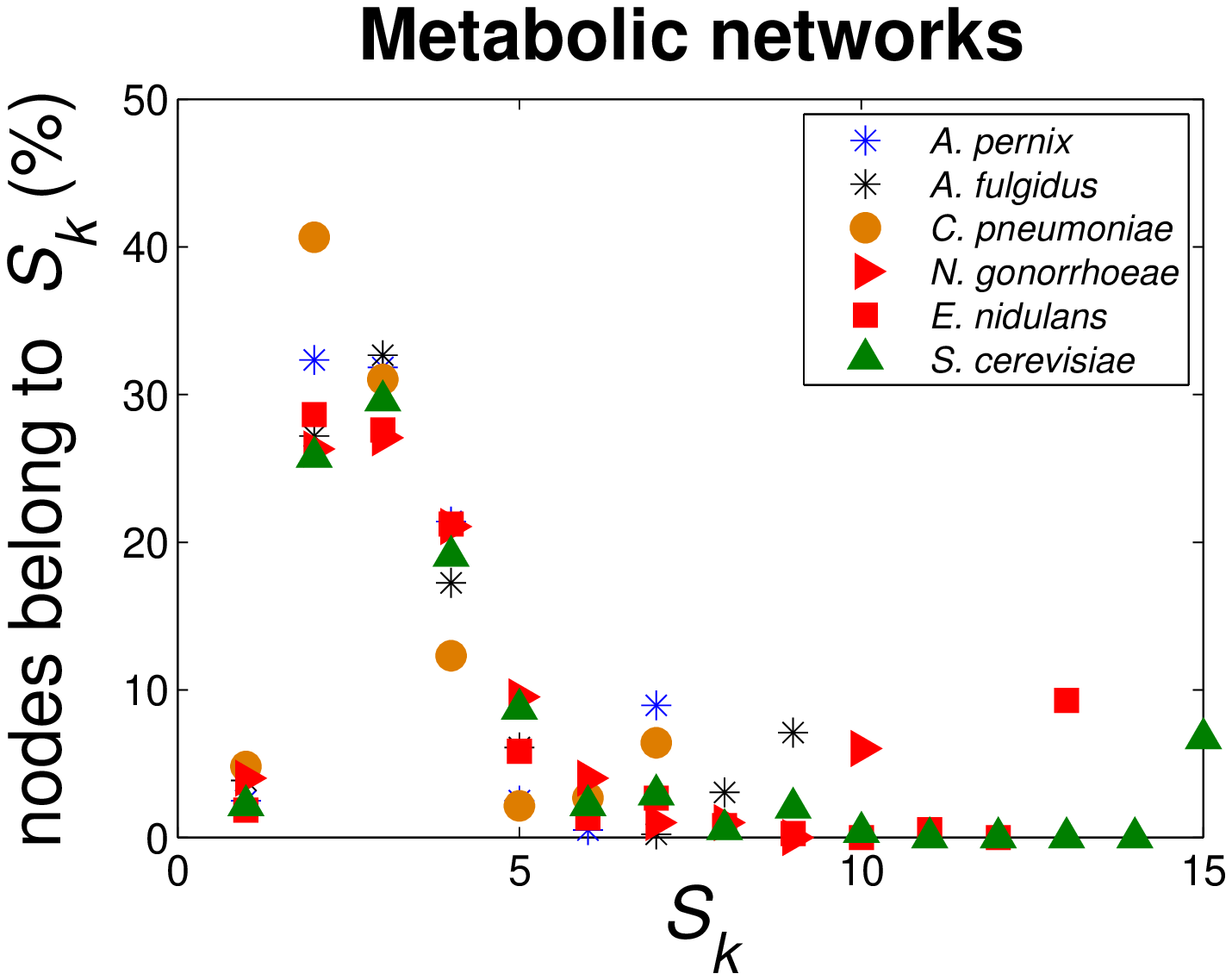}\includegraphics[scale=.33]{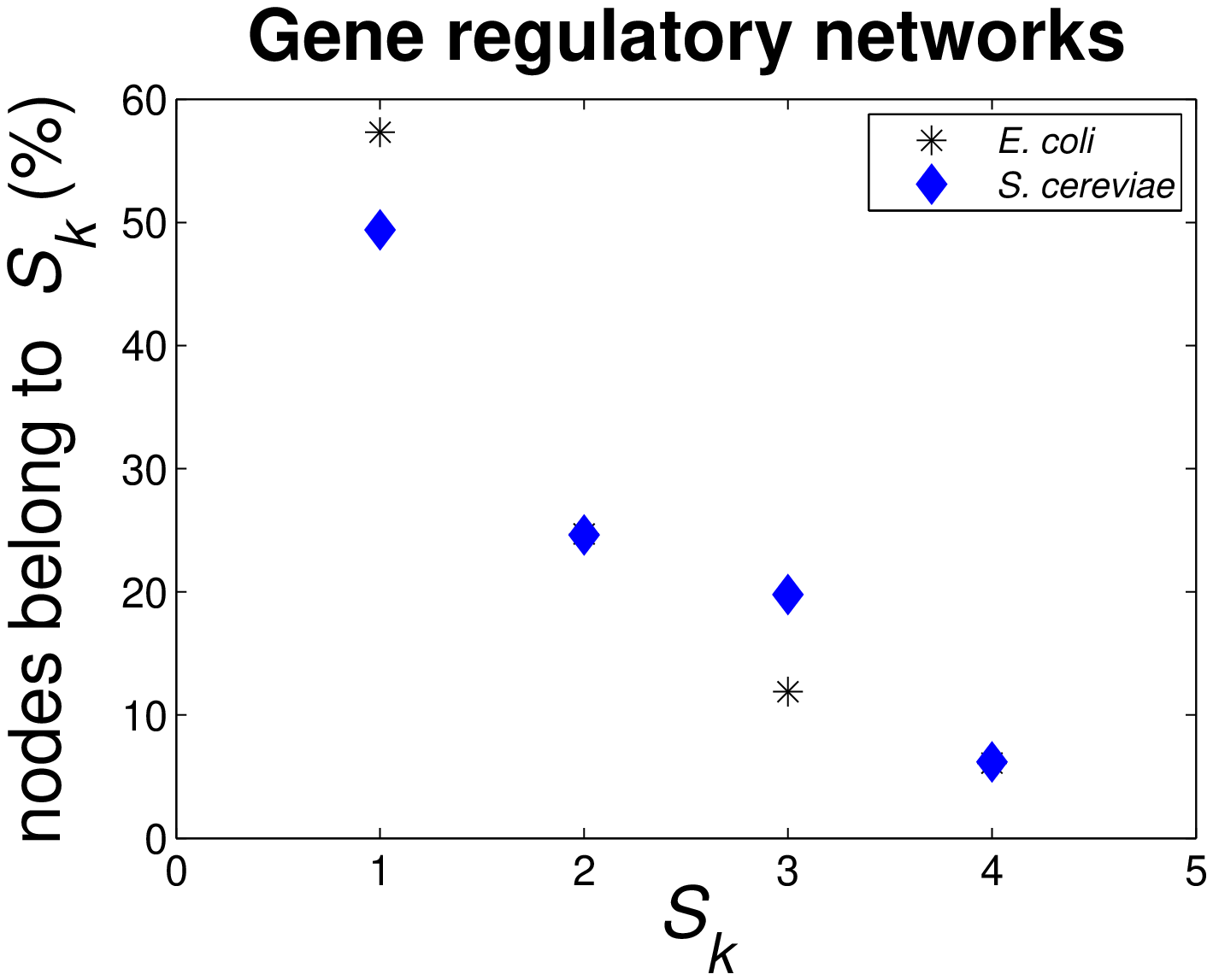}\\
\vspace{-3.4cm}
\hspace{-2.6cm}(d)\hspace{4.5cm}(e)\\
\vspace{2.4cm}
\end{center}
\caption{{\bf \emph{k}-core decomposition of all biological networks}. X-axis represents \emph{shell} index \emph{k} of \emph{k-shell} $S_\emph{k}$ and Y-axis represents  
percentage of nodes belong to $S_\emph{k}$. (a) {\bf Neuronal networks}: (macaque visual cortex area, 
macaque large-scale visual and sensorimotor area corticocortical connectivity, macaque cortical connectivity, 
cat cortical area, cat cortical and thalamic areas, {\it C. elegans}), (b) {\bf Food webs}: (Ythan Estuary, 
 Little Rock Lake, Grassland, Silwood, St Marks Seagrass, St Martin),
 (c) {\bf Protein-protein interaction networks}: ({\it E. coli}, {\it S. cerevisiae}, {\it H. pylori}), (d) {\bf Metabolic networks}:
         archaea ({\it A. pernix}, {\it A. fulgidus}), eukaryota ({\it E. nidulans}, {\it S. cerevisiae}), bacteria ({\it C. pneumoniae}, {\it N. gonorrhoeae}), 
(g) {\bf Gene regulatory networks}: ({\it E. coli}, {\it S. cerevisiae}).}
\label{shell}
\end{figure}

 Using the above information we analyze the core-structure of the  underlying undirected architecture of our networks for comparing the spreading capability in them.
 Here we estimate the percentage of the nodes present  in each shell of a network.
We observe  that most of the nodes of the food webs, metabolic networks, gene regulatory networks and  protein-protein interaction networks lie
in the \emph{periphery} (i.e. in the lower shell) of the network, whereas, a relatively high percentage of nodes form the higher core 
in neuronal networks and in a few food webs (see Figure~\ref{shell}).
As a result, we see that the undirected structure of the neuronal networks is more compact than
 the other biological networks. 
 Thus, in neuronal  networks, the deletion of a node from a higher core does not affect the spreading process much, unlike in other networks.
This shows that the spreading dynamics is more robust in (undirected) neuronal networks than the others.


\section{Conclusion}
We have empirically studied the underlying undirected topology of biological networks from  five different classes, 
namely,  neuronal networks, food webs, protein-protein interaction networks, metabolic networks and gene 
regulatory networks. Here, we have investigated which structures support faster spreading (of information, etc.) and are
better in communication. In this regard, we have  analyzed the good expansion property, using the spectral gap, and communicability
between nodes. Among all the networks, studied here, the undirected structure of all neuronal networks (and a few food webs)
possess  better expansion properties and have relatively higher number of pairs of nodes that show high communicability than the other biological networks. 

The underlying topology in neuronal networks may have evolved in such a way that they inherit
a (undirected) structure which is excellent and robust in communication. The speciality in the structure of neuronal networks has 
been investigated more with small-world-ness and \emph{k}-core decomposition. Though, the undirected topology of all the biological networks, 
studied here, show the small-world property, but in contrast, all the neuronal networks possess very high small-world-ness than any randomly
generated network with the same degree sequence. This strongly demonstrates that the topology of neuronal networks is special than 
the structure of the other biological networks. Moreover, comparatively a higher number of nodes in (undirected) neuronal networks belong 
to higher shell/core, in the \emph{k}-core decomposition, than in the other biological networks. This also shows the robustness of the (undirected)
structure of neuronal networks in communication.

\section{Acknowledgements}
 Authors are thankful to Sriram Balasubramanian for helping to prepare the manuscript. Special thanks to Satyaki Mazumder
  for fruitful  discussions on statistical significance of the figures. KD gratefully acknowledges the financial support from
  {\it CSIR} (file number 09/921(0070)/2012-EMR-I), Government of India.

\newpage

\begin{table}[!]
\centering
{\scriptsize
\caption{{\bf Spectral gap of biological networks}.}

\begin{tabular}{|l|c|c|r|}

\hline
Network & $|\lambda_1|$ & $|\lambda_2|$ & Spectral Gap\\
\hline
\hline
\multicolumn{4}{|c|}{Neuronal networks}\\
\hline
macaque visual cortex & 14.0416 & 7.3329 & 5.7037 \\
\hline
macaque visual and sensorimotor area & 16.8302 & 8.6048 & 10.0539\\
\hline
macaque cortical connectivity & 16.33 & 11.0694 & 8.8560\\
\hline
cat cortex (complete) & 31.9566 & 13.6595 & 8.5597\\
\hline
cat cortex connectivity & 22.9151 & 10.6677 & 9.7824\\
\hline
{\it C. elegan} & 24.3655 & 14.2428 & 11.9718\\
\hline
\multicolumn{4}{|c|}{Food webs}\\
\hline
Ythan Estuary & 17.0246 & 7.3913 & 9.6333\\
\hline
Little Rock Lake & 41.0126 & 10.7348 & 30.2778 \\
\hline

         Grassland & 5.6437 & 4.4565 & 1.1872\\
\hline
         Silwood & 14.7225 & 9.7215 & 5.001\\
\hline
        St Marks Seagrass & 11.8536 & 6.4522 & 5.4014\\
\hline
         St Martin & 12.5528 & 7.1137 & 5.4391\\
\hline 
\multicolumn{4}{|c|}{Protein-protein interaction network}\\
\hline
 {\it E. coli} & 15.9311 & 12.2921 & 3.639\\
\hline
      {\it S. cerevisiae } & 7.5350 & 7.5163 & 0.0187\\
\hline
      {\it H. pylori}  & 10.4658 & 9.1747 & 1.2911  \\
\hline
\multicolumn{4}{|c|}{Metabolic networks}\\
\hline
   {\it A. pernix} & 12.6330 & 7.7995 & 4.8335\\
\hline
   {\it A. fulgidus} & 17.4103 & 11.6872 &  5.7231\\
\hline
{\it C. pneumoniae} & 11.2525 & 7.0190 & 4.2335 \\
\hline
    {\it N. gonorrhoeae }    & 17.0745 & 10.6622 & 6.4123 \\
\hline
  {\it E. nidulans } & 16.2199 & 10.5369 & 5.683 \\
\hline
  {\it S. cerevisiae }  & 19.8917 & 12.2595 &  7.6322\\
\hline
\multicolumn{4}{|c|}{Gene regulatory networks}\\
\hline
 {\it E. coli} & 9.0636 & 8.5917 & 0.4719 \\
\hline
        {\it S. cerevisiae } & 9.9761 & 9.9648 & 0.0113 \\
\hline 
\end{tabular}
\label{spectralgap}
}
\end{table}

\begin{table}[h]
\centering
{\scriptsize
\caption{{\bf Z-score of small-world-ness of a network}.  Here, $n$, $m$ are the number of nodes and edges in a network $G$ respectively. $L_G$ is the 
average shortest path length and $T_G$ is the transitivity of $G$. Small-world-ness is given by
$SW_G$. Z-score of small-world-ness is represented by $Z_G$, which is computed over a set of networks have the same number of nodes and degree sequence as $G$ has.}
\begin{tabular}{|c|c|c|c|c|c|c|}
\hline 
 Network ${G}$ & $\,\,\,\, n\,\,\,\,$ & $\,\,\,\, m\,\,\,\,$ & $\,\,\,\,
 L_{G}\,\,\,\,$ & $\,\,\,\, T_{G}\,\,\,\,$ & $\,\,\,\, SW_{G}\,\,\,\,$ & $\,\,\,\, Z_{G}\,\,\,\,$\\
\hline
\hline 
\multicolumn{7}{|c|}{\bf{Neuronal networks}}\\
\hline 
 macaque visual cortex & 32 & 194 & 1.6593 & 0.5812 & 1.4715 & 10.2856\\
\hline 
macaque visual and sensorimotor area & 47 & 313 & 1.8501 & 0.5472 & 1.7914 & 14.2038 \\
\hline 
 macaque cortical connectivity & 71 & 438 & 2.2447 & 0.4418 & 2.1872 & 16.9169\\
\hline 
cat cortex (complete) & 95 & 1170 & 1.8645 & 0.4891 & 1.7389 & 15.7985 \\
\hline 
  cat cortex connectivity & 52 & 515 & 1.6357 & 0.5850 & 1.4940 & 18.1573\\
\hline 
{\it C. elegans}  neuronal network & 297 & 2418 & 2.4553 & 0.1807 & 3.6338 & 19.3799\\
\hline 
\multicolumn{7}{|c|}{\bf{Food webs}}\\
\hline
Ythan Estuary & 135 & 596 & 2.4135 &  0.1420 & 3.0227 & -0.1597\\
\hline 
 Little Rock Lake & 183 & 2434 & 2.1466 & 0.3323  & 2.1125 & -1.0949 \\
\hline 
 Grassland & 88 & 137 & 3.9924 & 0.1664 & 4.5413 & 4.3507 \\  
 \hline
Silwood Park & 135 & 365 & 3.3887 & 0.0314 & 6.5290 & 0.2847\\
\hline 
 St Marks Seagrass & 49 & 223 & 2.0876 & 0.1896 & 1.4011 & 0.5451 \\
 \hline
St Martin & 45 & 224 & 1.9333 & 0.2263  & 1.3833 & 0.4440\\
\hline
\multicolumn{7}{|c|}{\bf{Protein-protein interaction networks}}\\
\hline 
 {\it E. coli} & 270 & 716 & 2.7450 & 0.1552 & 10.0457 & 8.7941\\
\hline
 {\it S. cerevisiae } & 1846 & 2203 & 4.2494 & 0.0550 & 100.5855 & 37.1999\\
\hline 
 {\it H. Pylori} & 724 & 1403 & 3.9931 & 0.0152 & 3.5092 & -5.3138\\
\hline
\multicolumn{7}{|c|}{\bf{Metabolic networks}}\\
\hline
{\it A. pernix} & 201 & 548 & 2.9597 & 0.1005 & 4.0823 & 2.9362\\
\hline 
  {\it A. fulgidus} & 493 & 1402 & 3.1839 & 0.0668 & 6.7729 & 2.7412\\
\hline
 {\it C. pneumoniae} & 187 & 435 & 3.2643 & 0.1131 & 4.8712 & 4.9490\\
\hline 
 {\it N. gonorrhoeae} & 399 & 1185 & 2.8778  & 0.0737 & 6.0026 & 1.9721\\
\hline
 {\it E. nidulans} & 377 & 1074 & 3.0405 & 0.0789 & 6.1203 & 4.6917\\
\hline 
 {\it C. elegans} & 452 & 1332 & 3.1013 & 0.0710 & 7.8039 & 5.0742\\
\hline 
\multicolumn{7}{|c|}{\bf{Gene regulatory networks}}\\
\hline
{\it E. coli} & 328 & 456 & 4.8337 & 0.0243 & 3.3048 & -3.4238\\
\hline
{\it S. cerevisiae } & 662 & 1062 & 5.1995 & 0.0163 & 3.5882 & -6.0691\\
\hline
\end{tabular}
\label{swness_table}
}
\end{table}

\end{document}